\documentclass[twocolumn]{aastex62}
\usepackage{amsmath}
\usepackage{color}
\usepackage{bm}
\usepackage{natbib}
\definecolor{redgray}{rgb}{0.2,0.0,0.0}
\definecolor{red}{rgb}{0.8,0.0,0.0}

\newcommand{\catalog}{the KeLP catalog}
\newcommand{\dbic}{\Delta\mathrm{BIC}}
\newcommand{\kepler}{\textit{Kepler}}
\newcommand{\gaia}{\textit{Gaia}}
\graphicspath{{./fig/},{./large/}}

\shortauthors{Kawahara and Masuda}
\shorttitle{Kepler Long-period Planet Catalog}

\begin{document}
\title{Transiting Planets near the Snow Line from {\it Kepler}. I. Catalog\footnote{Based in part on data collected at Subaru Telescope, which is operated by the National Astronomical Observatory of Japan.}}

\correspondingauthor{Hajime Kawahara}
\email{kawahara@eps.s.u-tokyo.ac.jp}

\author{Hajime Kawahara}
\affil{Department of Earth and Planetary Science, The University of Tokyo, Tokyo 113-0033, Japan}
\affil{Research Center for the Early Universe, 
School of Science, The University of Tokyo, Tokyo 113-0033, Japan}

\author{Kento Masuda}
\altaffiliation{NASA Sagan Fellow}
\affil{Department of Astrophysical Sciences, Princeton University, Princeton, NJ 08544, USA}

\begin{abstract}

We present a comprehensive catalog of cool (period $P\gtrsim 2\,\mathrm{yr}$) transiting planet candidates in the four-year light curves from the prime \kepler\ mission. Most of the candidates show only one or two transits and have largely been missed in the original \kepler\ Object of Interest catalog. 
Our catalog is based on all known such candidates in the literature as well as new candidates from the search in this paper, and provides a resource 
to explore the planet population near the snow line of Sun-like stars. 
We homogeneously performed pixel-level vetting, stellar characterization with \gaia\ parallax and archival/Subaru spectroscopy, and light-curve modeling to derive planet parameters and to eliminate stellar binaries. The resulting clean sample consists of 67 planet candidates 
whose radii are typically constrained to 5\%,
in which 23 are newly reported. 
The number of Jupiter-sized candidates (29 with $r>8\,R_\oplus$) in the sample is consistent with the Doppler occurrence.
The smaller candidates are more prevalent (23 with $4<r/R_\oplus<8$, 15 with $r/R_\oplus<4$) and suggest that long-period Neptune-sized planets are at least as common as the Jupiter-sized ones, although our sample is yet to be corrected for detection completeness. If the sample is assumed to be complete, these numbers imply the occurrence rate of $0.39\pm0.07$ planets with $4<r/R_\oplus<14$ and $2<P/\mathrm{yr}<20$ per FGK dwarf. 
The stars hosting candidates with $r>4\,R_\oplus$ have systematically higher [Fe/H] than the \kepler\ field stars, providing evidence that giant planet--metallicity correlation extends to $P>2\,\mathrm{yr}$. 

\end{abstract}

\section{Introduction}\label{sec:intro}

Planets near the snow line, the freezing point of water, are expected to provide critical information on the core accretion theory. 
The expected location of the snow line around a Sun-like star is about 3~au (or 5~yr) in the traditional planet formation theory \citep{1981PThPS..70...35H} and may vary by a factor of a few depending on the stellar type, mass accretion rate, dust opacity, and evolutionary stage of the disk \citep[e.g.][]{2011ApJ...738..141O,2012MNRAS.425L...6M,2015ApJ...807....9M}. 
Those distant planets may also dynamically affect the architecture of their inner planetary systems \citep{2017AJ....153..210H,2017AJ....153...42L,2017MNRAS.467.1531H,2017MNRAS.468..549B,2017MNRAS.468.3000M,2018MNRAS.478..197P}, which almost always exist when the outer planet is Jupiter-sized \citep{2018AJ....156...92Z, 2019AJ....157...52B}. 
While the population of long-period ($>$ a few yr) planets has been probed with the Doppler method for Jupiter-mass ones around Sun-like stars \citep{2008PASP..120..531C, 2011arXiv1109.2497M} and with microlensing for planets mainly around late-type dwarfs \citep[e.g.][]{2010ApJ...720.1073G, 2012Natur.481..167C, 2016ApJ...833..145S}, the transit search has been less complete. The \kepler\ mission \citep{2011ApJ...736...19B} provided a detailed view of planets within $1\,\mathrm{au}$ of Sun-like stars, but those with periods longer than two years were originally missed. 

Several recent works have demonstrated the potential of \kepler\ to probe the population of long-period giant planets by identifying single and double transiting events (STEs and DTEs) in the four-year data. These events were out of the scope of the original \kepler\ pipeline that required three transits, and so dedicated searches have been performed via visual inspection \citep{2015ApJ...815..127W,2016ApJ...822....2U} and automated pipelines \citep{2016MNRAS.457.2273O,2016AJ....152..206F,2017AJ....153..180S,2018A&A...615L..13G}, both in the {\it Kepler} and \textit{K2} \citep{2014PASP..126..398H} data sets. 
These searches have revealed long-period transiting planets more than anticipated before the launch \citep{2008ApJ...688..616Y}, some of which have been successfully followed-up \citep{2016ApJ...826L...7D,2019ApJ...873L..17D},
and even provided a quantitative occurrence rate of such planets around Sun-like stars \citep{2016AJ....152..206F}.
It is particularly noteworthy that the sensitivity of the \kepler\ data goes down to Neptune-sized planets around Sun-like stars in this period range, which are currently out of the scope of the Doppler (in terms of mass) and microlensing (in terms of host stars) methods.

In this work, we present a comprehensive catalog of long-period transiting planets from the prime \kepler\ mission, the \kepler\ long-period planet (KeLP) catalog, as a resource complementary to the \kepler\ object of interest (KOI) catalog \citep{2016AJ....152..158T}. This paper improves upon the previous works in the following three aspects. In terms of targets, we compile all the known events and add new targets from our additional search (Section \ref{sec:ste}); we independently vet all of them based on the centroid analysis and visual inspection of the pixel data, matching epochs, and detector positions of the events, and remove obvious false positives (Section \ref{sec:fp}).
For stars, we perform precise characterization combining \gaia\ astrometry, archival spectroscopy, and our follow-up spectroscopy using the Subaru 8.2~m telescope (Section \ref{sec:stars}).
For planets, we perform homogeneous light-curve modeling to derive precise planet parameters and to remove stellar binaries as possible on the basis of the inferred parameters (Section \ref{sec:planets}). 

The properties of the promising candidates are discussed in Section \ref{sec:planets_clean}. Their properties are used to evaluate the detectability of long-period planets as in this catalog
with future direct imaging instruments in Section \ref{sec:imaging}.

\section{Input Catalog}\label{sec:ste}

We first created an input catalog of long-period planet candidates by compiling known events from the literature (Section \ref{ss:prc}) and by performing a new dedicated search (Sections \ref{ss:cvi} and \ref{ss:tls}). Table \ref{tab:source} summarizes the sources of the input catalog, as well as the outputs after vetting processes in Sections \ref{sec:fp}--\ref{sec:planets}. The ``clean sample" column corresponds to the final product in \catalog.
\begin{table*}[!tbh]
\begin{center}
\caption{Breakdown of the sources in the \catalog. S/D/T stand for single/double/triple transit events. \label{tab:source}}
\begin{tabular}{ccccc}
  \hline\hline
  method & input & after FP test in Section \ref{sec:fp} (S/D/T) & clean sample (S/D/T) & reference \\
  \hline
  Planet Hunter& 1 &  1 (0/1/0) & 1 (0/1/0) & \cite{2014AJ....148...28S}  \\
  Planet Hunter& 30 &  27 (13/11/3)& 24 (11/10/3) & \cite{2015ApJ...815..127W}  \\
  - & 1 & 1 (0/1/0) & 1 (0/1/0) & \cite{2016ApJ...820..112K}\\  
  \kepler\ eclipsing binary & 3 &  2 (1/1/0) & 0 & \cite{2016AJ....151...68K}\\
  VI of KOI (Jun 4, 2015)& 19 & 15 (14/1/0) & 13 (12/1/0) &\cite{2016ApJ...822....2U}  \\
  automated search & 7 & 5 (4/1/0) & 2 (1/1/0) & \cite{2016AJ....152..206F} \\
  Swiss cheese effect & 1 & 1 (0/1/0) & 1 (0/1/0) & \cite{2017AJ....153..180S}\\
  \hline
  clipping+VI & 43 & 28 (24/4/0) & 16 (13/3/0) & this paper\\
  trapezoid least square+VI & 16 & 13 (10/3/0) & 9 (7/2/0) & this paper\\
  \hline
  total & 121 & 93 (66/24/3) & 67 (44/19/3) &
\end{tabular}
\end{center}
\end{table*}

\subsection{Previously Reported Candidates}\label{ss:prc}

We collected long-period planet candidates from the following literature.\footnote{During the preparation of this manuscript,  \cite{2019arXiv190101974H} was posted on arXiv, who reported 12 long-period planet candidates using the code of \cite{2016AJ....152..206F}, refined stellar radii from \gaia\ DR2, and their own light-curve detrending. The two new events reported in their paper (KIC 6186417 and 7906827) were also detected in our analysis in Section \ref{ss:cvi}, and so do not affect the present paper.}
\begin{itemize}
\item Those from visual inspection by citizen scientists, namely {\it the Planet Hunters}, are described in \citet{2014AJ....148...28S} and \citet{2015ApJ...815..127W}. We included one DTE from the former, and 27 STEs and DTEs from the latter in the input catalog. We decided to included three triple transit events (TTEs) in \citet{2015ApJ...815..127W} as well, because two (KIC 10024862 and KIC 9413313) are not in the KOI catalog and one (KIC 8012732, KOI-8151) is classified as a false positive presumably due to transit timing variation of the second event. 
\item \cite{2016ApJ...822....2U} also performed a visual search for STEs/DTEs around KOI stars. We included 19 targets from this work.  
\item \cite{2017AJ....153..180S} found a DTE in KIC 5351520, which was overlooked by \cite{2016ApJ...822....2U} due to the ``Swiss-cheesing" effect of the light curve caused by inner transiting planets. This target was added.
\item \cite{2016AJ....152..206F} performed an automated search focusing on $\sim 40,000$ bright and quiet Sun-like stars. Seven new STEs/DTEs from this search were added. 
\item We realized that some targets detected in Section \ref{ss:cvi} have already been reported by \cite{2016ApJ...820..112K} and in the {\it Kepler} Eclipsing Binary (KEB) catalog by \cite{2016AJ....151...68K}. We classify them as inputs from the literature, rather than new candidates from our search.
\end{itemize}

\subsection{Sigma Clipping and Visual Inspection}\label{ss:cvi}

As the first simple search for deep transit events, we performed a search based on sigma-clipping and subsequent visual inspection. 
We identified dips as $>2\sigma$ flux deviation lasting longer than 2.5h, where $\sigma$ is the standard deviation of the light curve detrended with the second-order spline interpolation. Each candidate was visually inspected and the obvious stellar eclipses were removed; they are listed in Table \ref{tab:fp}.
We applied this procedure to all the long-cadence (29.4~min), pre-search data conditioning (PDC) light curves. As a result, we found 43 new STEs and DTEs, which are listed as ``clipping+VI" in Table \ref{tab:source}. We found no new TTE by this method. 
The sample of 102 STEs and DTEs both from this search and the previous literature (Section \ref{ss:prc}) was used to calibrate a systematic search by the trapezoid fitting in Section \ref{ss:tls}.

\subsection{Trapezoid Least Square}\label{ss:tls}

\begin{figure}[htbp]
\begin{center}
  \includegraphics[width=\linewidth]{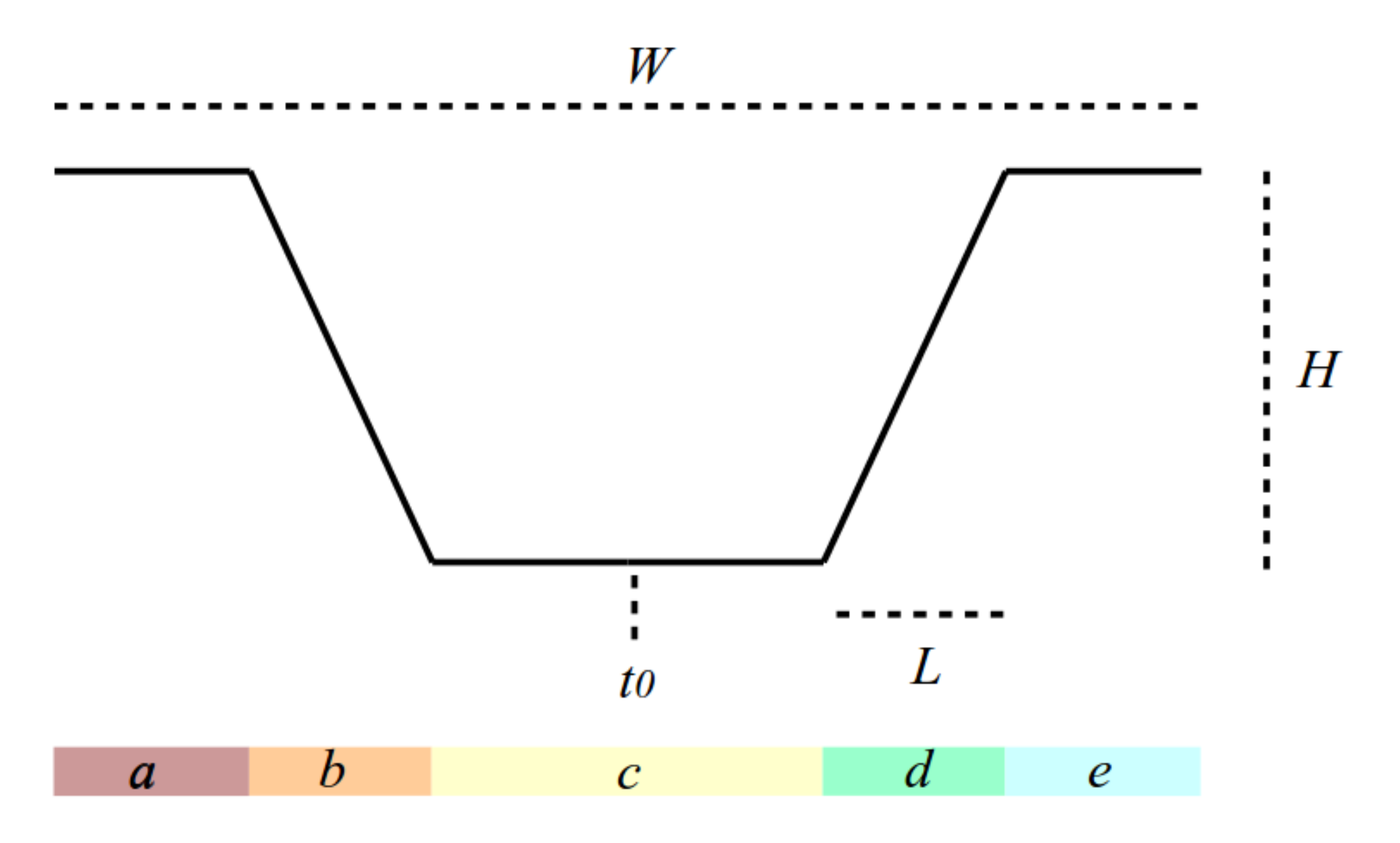}
\caption{Parameters of the trapezoid model. The central time, width, and height of the template are given by $t_0$, $W$, and $H$, respectively. $L$ is the length of $b$ and $d$, and $c$ is twice as long as $a$ and $e$; so $c/2=a=e=W/2-L$.\label{fig:trapezoid}}
\end{center}
\end{figure}

We then performed a systematic search by matching a trapezoid template to the long-cadence PDC light curves (trapezoid least square; TLS). As shown in Figure \ref{fig:trapezoid}, the fitting parameters are the central time $t_0$, height $H$, gradient $H/L$, and total width $W$. 
For each $t=t_0$, we minimize the $\chi^2$  defined by 
\begin{eqnarray}
  \label{eq:residHLchi}
  &\,&\chi^2 = \sum_{t_i \in a,e} \frac{(x_i + H/2)^2}{\sigma_i^2} + \sum_{t_i \in b} \frac{[x_i - H L^{-1} (t_i + W/4)]^2}{\sigma_i^2} \nonumber \\
  &+& \sum_{t_i \in c} \frac{(x_i - H/2)^2}{\sigma_i^2} +  \sum_{t_i \in d} \frac{[x_i + H L^{-1} (t_i - W/4)]^2}{\sigma_i^2},
\end{eqnarray}
where $a$ to $e$ are the regions shown in Figure \ref{fig:trapezoid} and $x_i = x(t_i)$ is the light curve normalized to zero within the fitting region.
Because minimization of equation (\ref{eq:residHLchi}) is time consuming, we implement the process using the graphics processing unit (GPU) and pycuda \citep{kloeckner_pycuda_2012}, as described in Appendix \ref{sec:tlsalg}. Before the trapezoid fitting, we detrended the light curve using a median filter with a timescale of 64~h. Because the median filter with a larger window size is also time consuming, we also implemented a GPU-based median filter, the kernel\_MedianV\_sh algorithm proposed by \cite{couturier2013designing}. We executed the code using GeForce Titan X and GTX 1080Ti. The resulting computing time for a full long cadence time series is less than a second. To increase the precision of the best-fit parameters, we refit the trapezoid model with higher precision using CPU.

\begin{figure}[htbp]
\begin{center}
  \includegraphics[width=\linewidth]{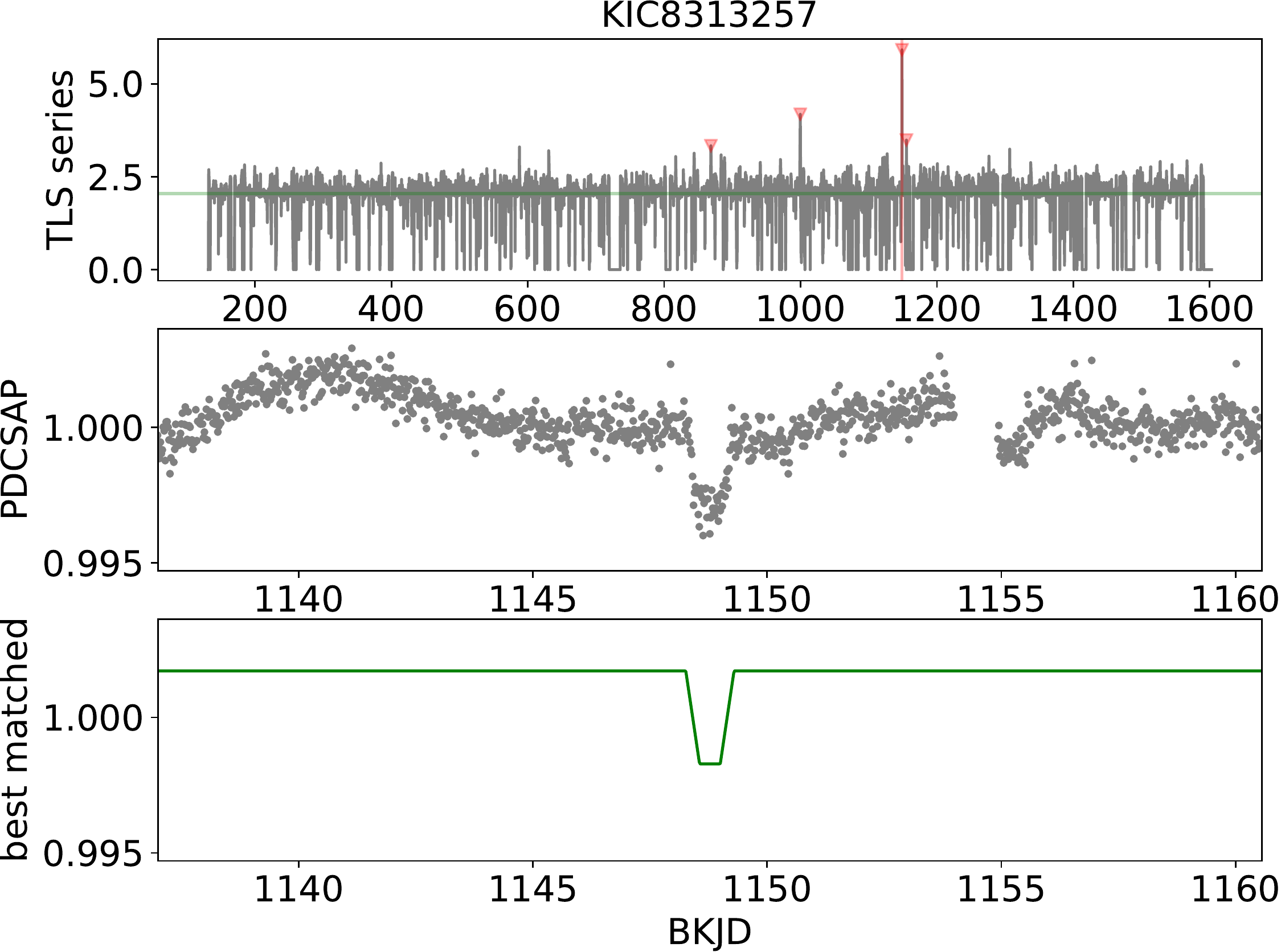}
\caption{
One of the STE detected by the trapezoid least square (TLS) algorithm (Section \ref{ss:tls}). The top panel shows the TLS time series for the four-year data defined by Equation (\ref{eq:snx}). The strongest peak is located at the Barycentric \kepler\ Julian Day (BKJD) $\sim$ 1149. The PDC flux and the best-match trapezoid of the corresponding STE candidate are shown in the middle and bottom panels, respectively. \label{fig:TLSex}}
\end{center}
\end{figure}

Figure \ref{fig:TLSex} shows an example of an STE in KIC 8313257 detected by the TLS algorithm. We compute the signal to noise ratio for each $t_0$ by
\begin{eqnarray}
  \label{eq:snx}
  s(t_0) \equiv \frac{\tilde{H}}{\sqrt{\mathrm{res}\cdot (n - 3)}},
\end{eqnarray}
where $\tilde{H}$ is the maximum value of $H$ among the grids of $L$ and $W$ for a given $t_0$, $\mathrm{res}$ is the squared norm of the residual of the trapezoid fit, $n$ is the number of data points in the fitting region, and $n-3$ is the degrees of freedom of the fit. We call $s(t_0)$ the TLS series (the top panel in Figure \ref{fig:TLSex}). 
We pick up the dip at $t_0=\tilde t_0$ that maximizes $s(t_0)$ as an STE candidate, and the value of $s(\tilde t_0)$ is defined as the S/N of the candidate dip.

The fitting procedure described above always yields one dip candidate in each light curve. However, many of the dip candidates, even with high S/N values, are false positives. Most of the false dips turned out to be due to stellar variability producing many peaks and valleys in the light curve. The trapezoid fitting detects one of those valleys, and the resulting TLS series exhibits a series of false peaks with comparable S/N to $s(\tilde t_0)$. Similarly, short-period transiting planets exhibit multiple high peaks in the TLS series. In contrast, typical STEs/DTEs exhibit only one or two high peaks in the TLS series, as shown in the top panel in Figure \ref{fig:TLSex}.

To characterize the difference between these two cases, we introduce another empirical metric to describe the significance of the fourth peak in the TLS series: 
\begin{eqnarray}
\Delta \equiv \frac{s(t_{0,4}) - \overline{s(t_0)}}{\sigma_s},
\end{eqnarray}
where $t_{0,4}$ is the time at the fourth highest peak in the TLS series, and $\overline{s(t_0)}$ and $\sigma_s$ are the median and standard deviation of the TLS series, respectively. The false positive cases with multiple high peaks are expected to have higher $\Delta$ compared to true STEs/DTEs with a small number of high peaks. The choice of the fourth peak is motivated by the fact that the second or third highest peaks occasionally become high due to a gap in the light curve, rather than stellar variability or short-period transiting planets; the choice of the fourth peak thus helps to suppress the rate of true negatives caused by the gap.

\begin{figure}[htbp]
\begin{center}
  \includegraphics[width=\linewidth]{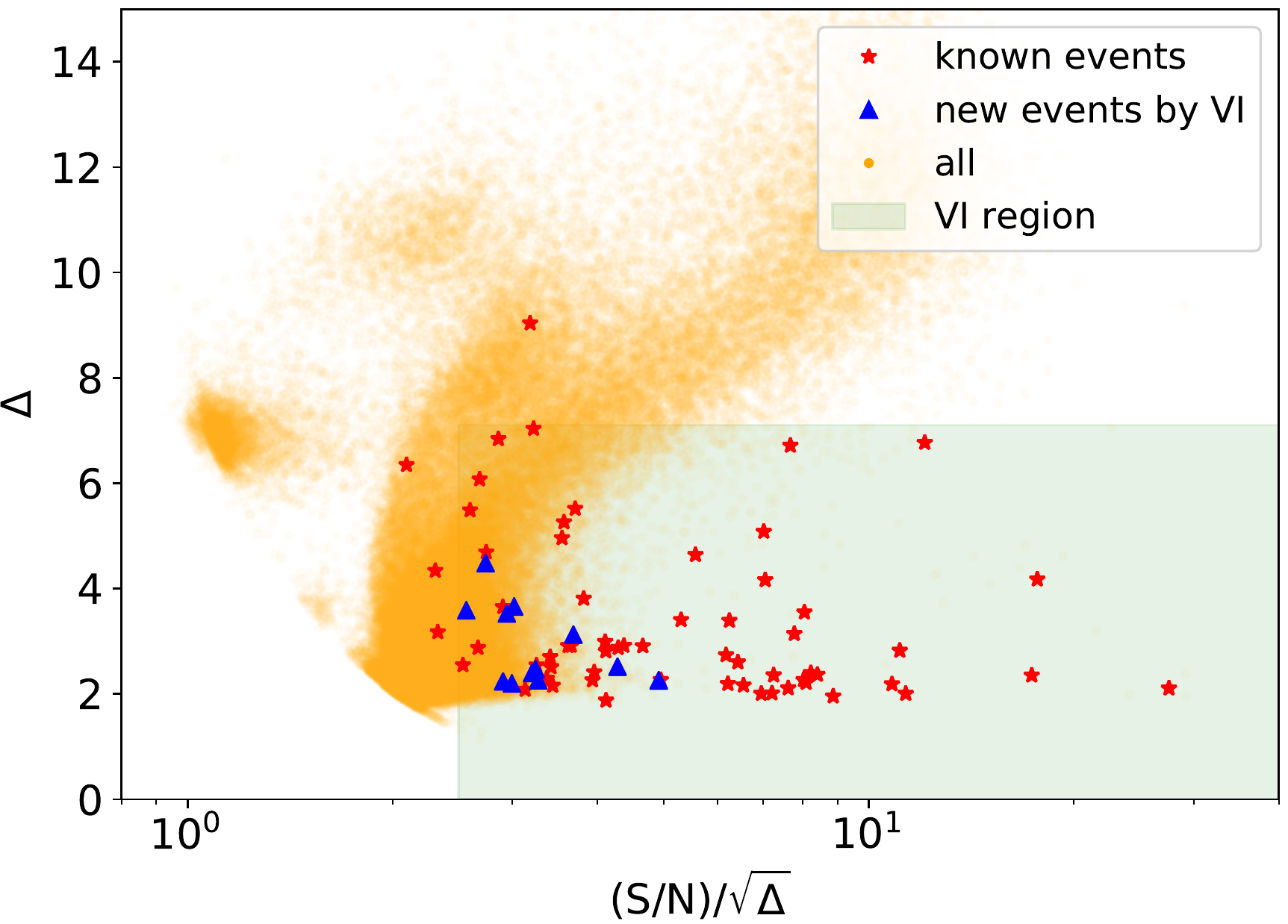}
\caption{
Classification of the STE candidates from the TLS algorithm on the (S/N)/$\sqrt{\Delta}$--$\Delta$ plane. Orange dots are all the candidates. The known events from the literature (Section \ref{ss:prc}) and our search in Section \ref{ss:cvi} are shown with red stars; they were used as the training set to determine the region for visual inspection ($\Delta < 7$ and $\mathrm{S/N} > 2.5 \sqrt{\Delta}$; green square). From the visual inspection of the orange candidates in this region, we found additional 13 candidates that passed the vetting in Section \ref{sec:fp}, shown by blue triangles \label{fig:tlscrit}.}
\end{center}
\end{figure}

Figure \ref{fig:tlscrit} shows $\Delta$ and (S/N)/$\sqrt{\Delta}$ of all the events detected by the TLS algorithm (orange dots). The choice of the $x$-axis is motivated by the empirically-found scaling $\mathrm{S/N}\propto \sqrt{\Delta}$ for small S/N; this normalization aligns the orange dots vertically and makes the following classification simpler. We also ran the TLS algorithm for the systems known to have STEs/DTEs in Sections \ref{ss:prc} and \ref{ss:cvi}, and recovered 61 of them; their positions are marked by red stars in Figure \ref{fig:tlscrit}. As expected, these events are clustered in the region with high S/N and low $\Delta$. Based on these training set results, we chose the region defined by $\Delta < 7$ and $\mathrm{S/N} > 2.5 \sqrt{\Delta}$ (green square region in Figure \ref{fig:tlscrit}) and one of us (KH) visually inspected all 20816 events in this region. From this search, we identified 10 STEs and 3 DTEs that have not been reported in the literature. We added these 13 systems into our input catalog. We also found three TTEs; one of them, KIC 6681473 (KOI-5312) was a KOI false positive and included in the input list for further vetting. The other two were not included, because they were either a KOI candidate (KIC 3634051, KOI-6103) or already in the input catalog (KIC 8012732).
We also found four obvious stellar primary and secondary (flat-bottomed) eclipses. These events are listed in Table \ref{tab:fp}.

\section{Eliminating Suspicious Signals}\label{sec:fp}

\subsection{First Screening of the Targets}

From the STEs/DTEs in \cite{2015ApJ...815..127W}, we excluded the following two targets. For KIC 5522786, candidate transits are reported at $\mathrm{BKJD}=283$ and 1040 in the PDC light curve \citep{2014AJ....148...28S, 2015ApJ...815..127W}, but the latter is not seen in the simple aperture photometry (SAP) light curve. Moreover, there are other suspicious dips at $\mathrm{BKJD} = 268.5$ and 872.5 in the SAP light curve. Thus the reliability of the reported signal appears to be low. KIC 8540376 exhibits DTEs, but the separation is $31.8\,\mathrm{days}$ because only two quarters are available. The period is too short for our purpose.
Similarly, we removed the STE in KIC 8489948 (the sigma clipping+VI in Section \ref{ss:cvi}) from the list, because only two quarters (Q16 and 17) are available. These three objects are listed as ``SC" in Table \ref{tab:fp} in Appendix \ref{sec:fplist}.

\subsection{Events Sharing Similar Epochs}

\begin{figure}[htbp]
  \begin{center}
  \includegraphics[width=\linewidth]{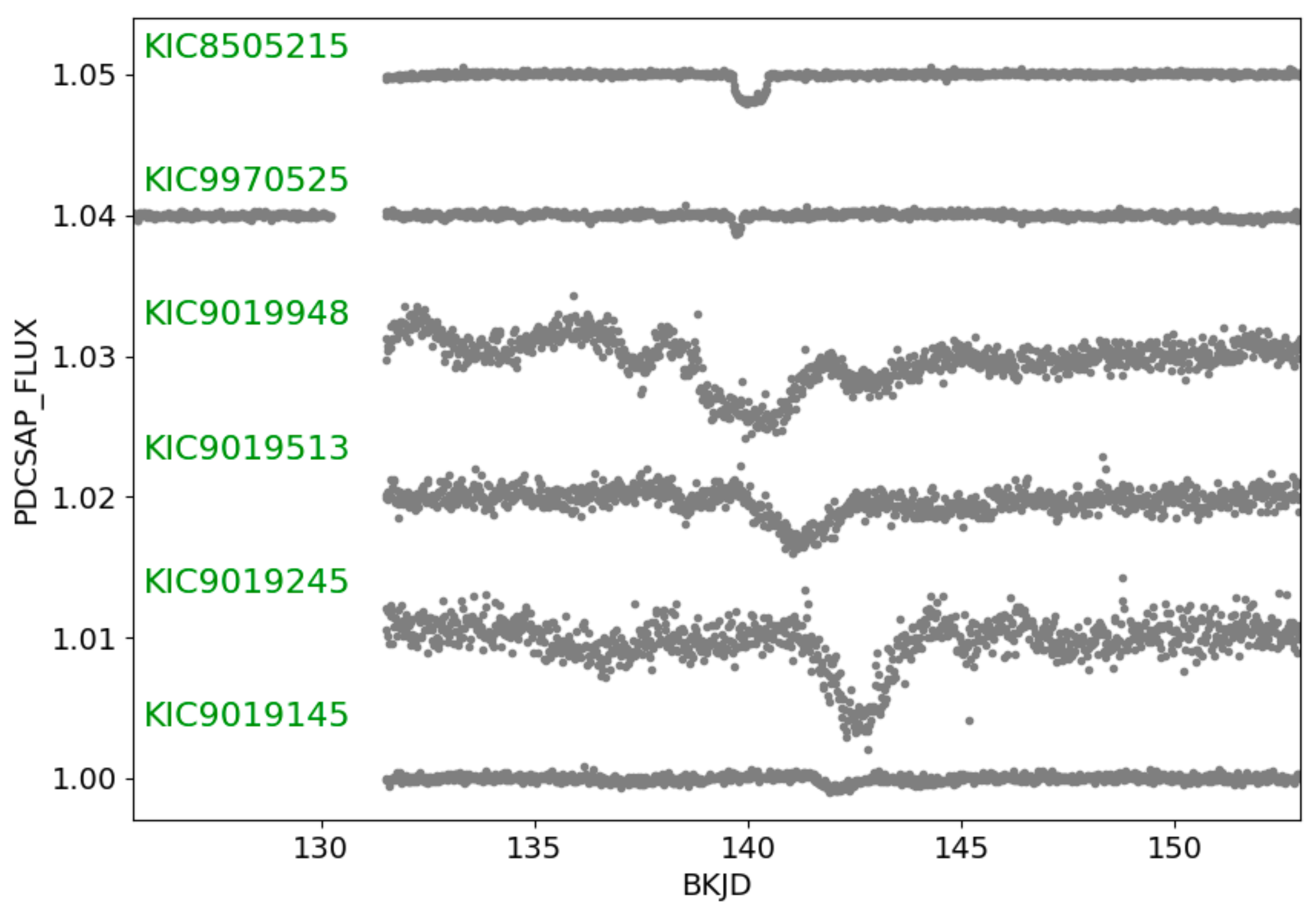}
\caption{
Suspicious dips around $\mathrm{BKJD}=140$ identified from the similar transit times. The bottom four events with similar non-transit-like shapes were excluded, and KIC 8505215 and KIC 9970525 were kept in the candidate list. \label{fig:cross}}
\end{center}
\end{figure}

Matching ephemerides in two different targets is known to be a strong indicator of false positives \footnote{As an example for long-period signals, the transit times of the triple dips with an interval of $\sim 375 $d in KIC 8622875 (KOI-5551) matches those of pulse-like signals in KIC 8557406 and KIC 8622134, therefore, those are considered as false positives \citep{2018AJ....155..144K}.  }. Although the period is not precisely constrained for most of our sample by construction, signals originating from contamination may still share similar transit times, which may be used to flag spurious signals.  
We therefore checked the events with similar central times among all the input candidates. 
We found that STEs in KIC 8505215, 9970525, 9019948, 9019513, 9019245, and 9019145 occurred within a few days (Figure \ref{fig:cross}), and some of them do have spurious shapes.
We excluded the latter four (listed as ``SE" in Table \ref{tab:fp}), but decided to keep the events in KIC 8505215 and KIC 9970525 (top two rows) in the list because they have distinct transit-like shapes unlike the removed four.

\begin{figure}[htbp]
  \begin{center}
  \includegraphics[width=\linewidth]{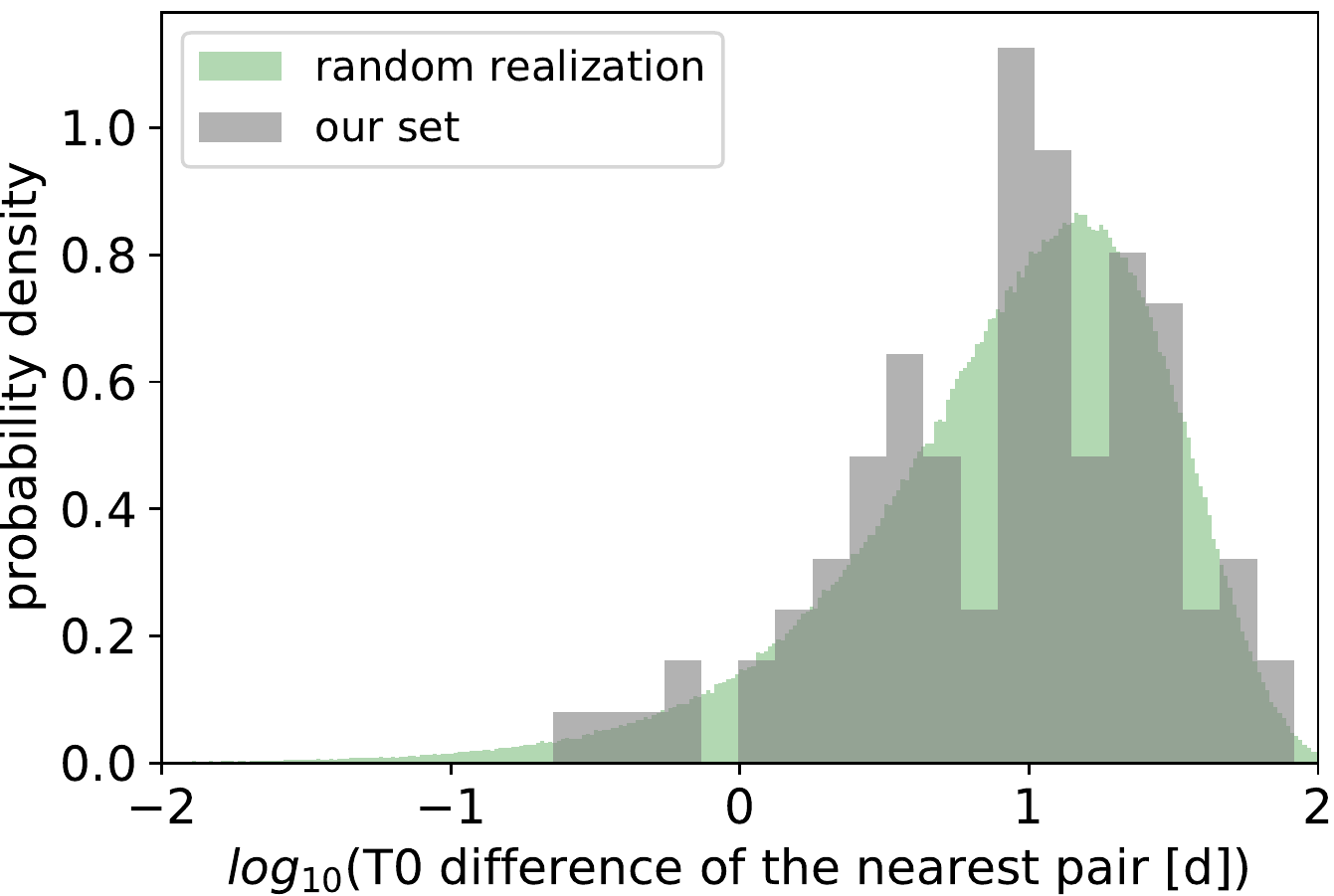}
\caption{The distribution of the time intervals between the two nearest STE pairs. The gray histogram is the distribution from the input catalog, and the green histogram is a simulated probability density function for a random realization.\label{fig:em}}
\end{center}
\end{figure}

In addition to KIC 8505215 and KIC 9970525, several other STEs were found to be separated by less than a day. To check if this is a statistically natural outcome, we compared the distribution of the time intervals of the nearest STEs/DTEs against that of a completely random realization of the transit events (Figure \ref{fig:em}). The resulting distributions are consistent with each other, with the Kolmogorov-Smirnov (KS) $p$-value of 0.98. 

\subsection{Visual Inspection of the Pixel Data}

Here and in Section \ref{ss:pldi}, we describe the pixel-level vetting of the signals. We first visually checked the pixel-level light curves around the detected events. We excluded the events that exhibit either one of the following two features:
\begin{itemize}
\item The signal originates from a single pixel. This is likely due to a sudden drop in the pixel sensitivity and produces a box-shaped false positive. The ``transit depth" in the single pixel is deeper than that seen in the SAP flux, because the contribution comes from that one pixel. For example, dips in KIC 7190443 at BKJD = 1201 and KIC 5621767 at BKJD = 839 belong to this category.
\item The signal is apparently from outside the aperture. In this case, the signal depths significantly vary over the pixels in the aperture because the amount of contamination depends on the distance from the nearby source. One obvious such case is KIC 5480825 at BKJD = 363.
\end{itemize}
The automated centroid analysis in Section \ref{ss:pldi} will anyway remove the latter class of false positives, but visual inspection is important for the former class that does not necessarily produce centroid shifts. The false positives identified here are listed in Table \ref{tab:fp} as ``VIP" (visual inspection of pixel data).

\subsection{Pixel-level Difference Imaging \label{ss:pldi}}

The centroid shift of the pixel-level difference image between in- and out-of-transit data is an excellent indicator of the contamination from a nearby star \citep{2013PASP..125..889B}. We define the difference image by
\begin{eqnarray}
  \label{eq:diffimage}
  \delta I(X,Y) \equiv \langle I(X,Y,t) \rangle_{\mathrm{out}} - \langle I(X,Y,t) \rangle_{\mathrm{in}},
\end{eqnarray}
where $X$ and $Y$ are the column and row positions of the pixel on image, $I(X,Y,t)$ is the flux in the pixel $(X,Y)$ at time $t$, and $\langle  I(X,Y,t) \rangle_{\mathrm{in}}$ and $\langle I(X,Y,t) \rangle_{\mathrm{out}}$ are the time-medians of the flux inside (i.e. between ingress and egress) and outside (i.e. both sides of) the dip. 

\begin{figure}[htbp]
\begin{center}
  \includegraphics[width=\linewidth]{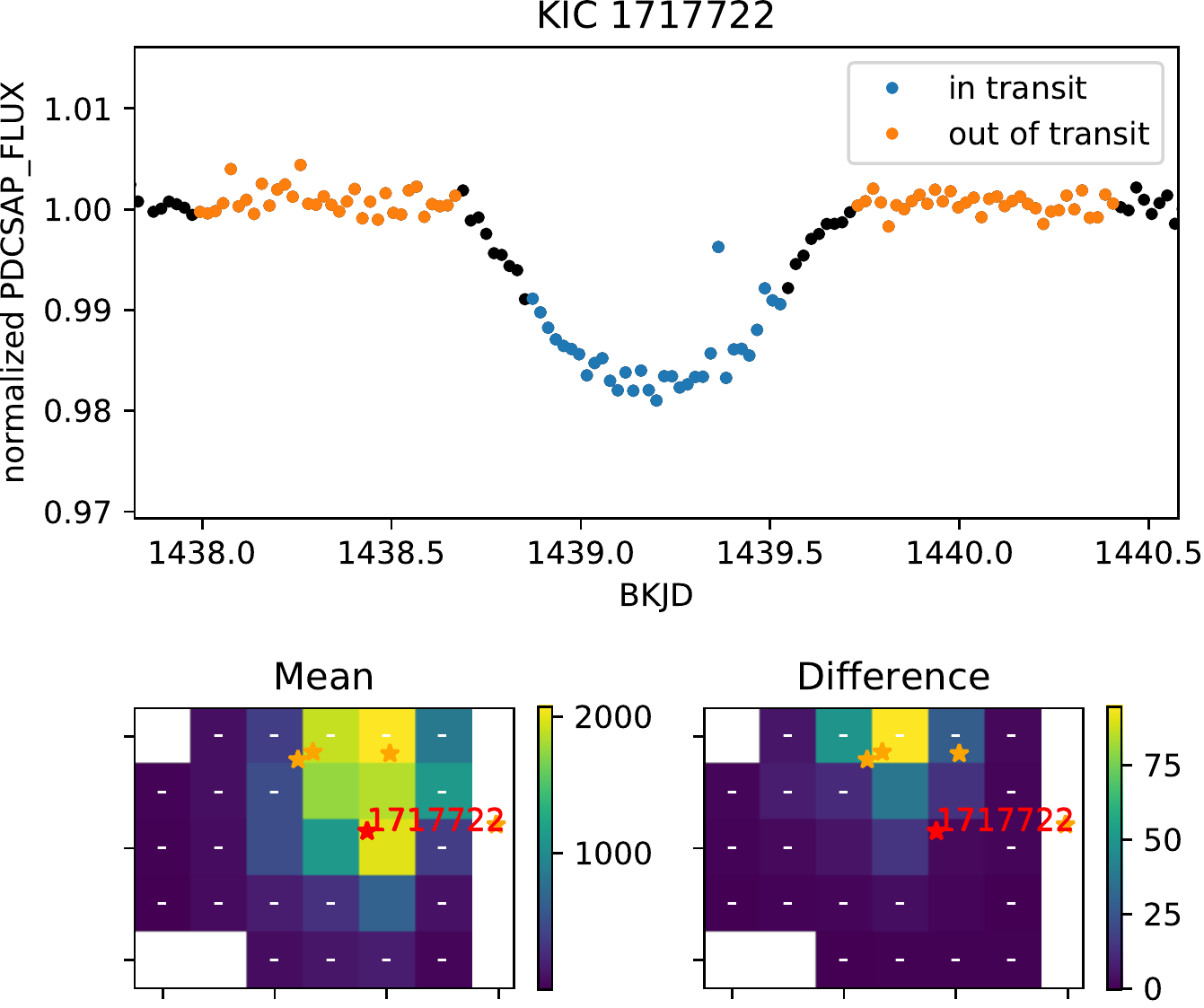}
  \caption{
  An example of a contaminated signal detected by the pixel-level difference imaging. The top panel shows the dip of KIC 1717722 and the in-transit (blue) and out-of-transit (orange) ranges we adopted. The bottom panels represent the mean image for the quarter in which the dip exists (left), and the difference image defined by Equation (\ref{eq:diffimage}) (right). The minus symbol (-) indicates the pixels outside the aperture used in the SAP/PDC light curves. The unit for pixel images is $e^-/\mathrm{s}$.  \label{fig:pdiff1717722}}
\end{center}
\end{figure}

Figure \ref{fig:pdiff1717722} shows a clear example of contamination from a nearby star in KIC 1717722. The top panel shows the in- and out-of-transit data. The lower left panel shows the mean pixel image of the quarter that includes the detected dip. Pixels labeled with ``-'' are outside the aperture used for SAP and PDC light curves. The difference image computed by Equation (\ref{eq:diffimage}) is shown in the bottom right panel, along with the position of KIC 1717722 (red star symbol) and nearby stars (orange star symbols) taken from the KIC stellar list provided by the \kepler\ team. During the dip, the center of the flux moves toward nearby stars in the upper left of the target star, suggesting that the dip originates from one of those nearby stars, and not on KIC 1717722. Two other targets were found to exhibit such a significant centroid shift greater than a pixel. 

To systematically identify such shifts including the smaller ones, we evaluated the centroid shift of the difference image:
\begin{eqnarray}
  \label{eq:PRFcoffset}
  {\bf \delta X} = {\bf X}_\mathrm{PRF} [\delta I(X,Y)] - {\bf X}_\mathrm{PRF} [\langle I(X,Y,t) \rangle_{\mathrm{out}}],
\end{eqnarray}
where ${\bf X}_\mathrm{PRF} [I(X,Y)]$ is the center of the best-fit pixel-response function (PRF) image to $I(X,Y)$. The PRF fitting in this paper was performed using the {\tt lightkurve} package \citep{lightkurve}.
The PRF fitting did not work well and returned unstable centroid positions when the total electron number of the difference image is below 30. We decided to just keep 19 such low-S/N targets in our list.

\begin{figure}[htbp]
\begin{center}
  \includegraphics[width=\linewidth]{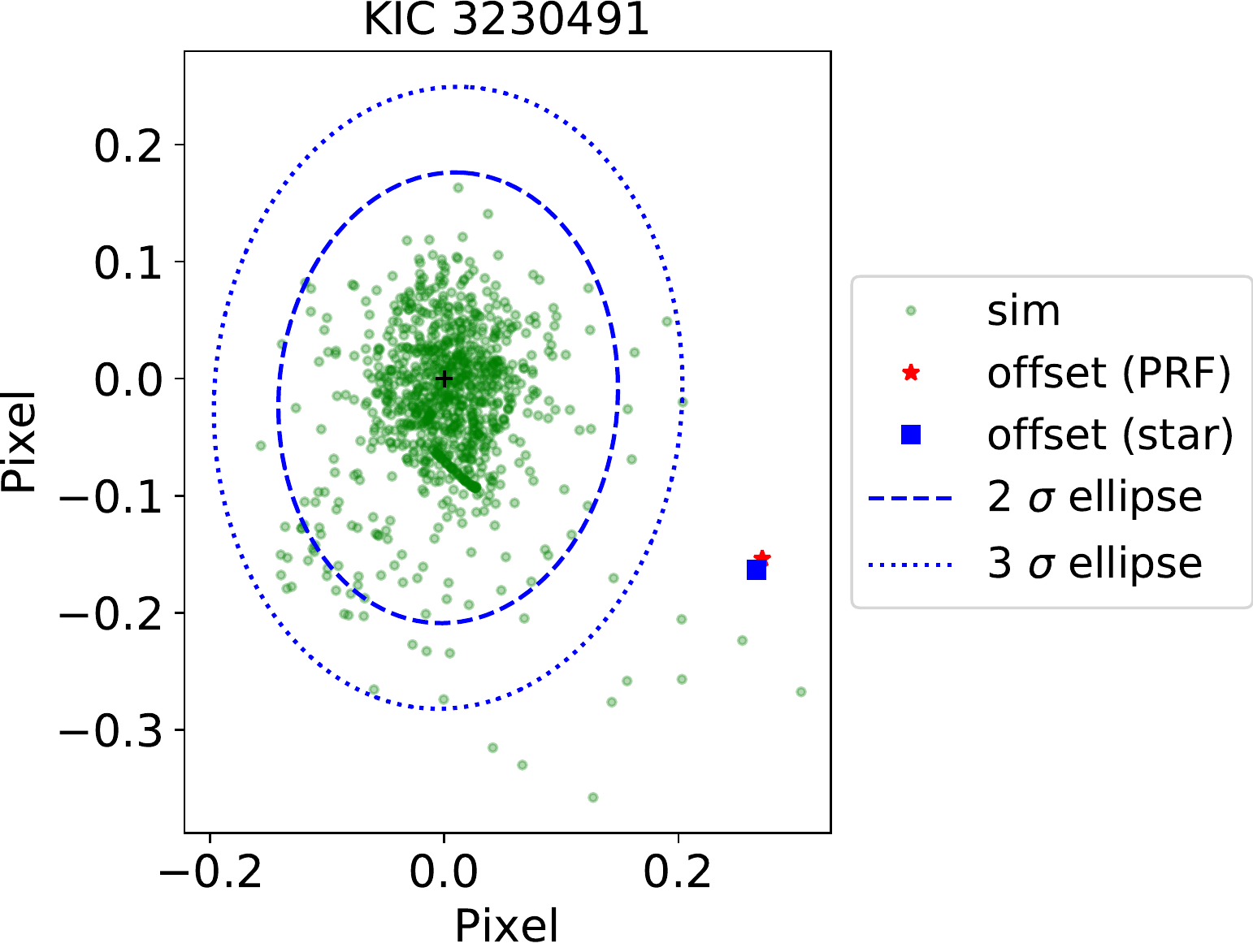}
  \caption{An example of a false positive showing a significant centroid shift (KIC 3230491). The red star is the shift of the difference image centroid from the out-of-transit one, ${\bf \delta X}$ (Equation \ref{eq:PRFcoffset}). The green dots show the distribution of the centroid shifts computed for injected dips ${\bf \delta X}^\ast$ (Equation \ref{eq:PRFcoffsetsim}). The dashed and dotted lines are the 95.5\% and 99.7\% ellipses computed from the green dots. The blue square is the shift of the difference image centroid from the catalog position of the target (Equation \ref{eq:PRFcoffsetartcat}). Both the red star and blue square show significant shifts, and so this case is a false positive.\label{fig:stat}}
\end{center}
\end{figure}

\cite{2013PASP..125..889B} proposed to use the mean and variance of the centroid shifts from all repeating transits to quantify the significance of the shift. In our case, however, there are only one or two dips in the light curve and this method is not applicable. Instead, we follow a method similar to the one proposed by \cite{2018AJ....155..144K}, who checked the centroid shift in double or triple pulses due to gravitational microlensing by a white-dwarf (WD) companion. First, we injected 1,000 simuated dips with the same depth as the detected one, at randomly selected times in the light curves from the same season. The images within each injected dip were given by
\begin{eqnarray}
  \label{eq:diffimageart}
  I^{\ast}(X,Y,t) = I(X,Y,t) - \alpha I_\mathrm{sim}(X,Y),
\end{eqnarray}
where $I_\mathrm{sim}(X,Y)$ is the best-fit PRF model for the median of $I(X,Y, t)$ in this range, and the scaling factor $\alpha$ is chosen so that the injected dip has the same depth as the original one.
Second, the difference image of the injected signal, 
\begin{eqnarray}
  \delta I^{\ast}(X,Y) \equiv \langle I(X,Y,t) \rangle_{\mathrm{out}} - \langle I^{\ast}(X,Y,t) \rangle_\mathrm{in},
\end{eqnarray}
was used to compute its centroid shift $\delta {\bf X}^\ast$ 
in the same way as for the detected signal:
\begin{eqnarray}
    \label{eq:PRFcoffsetsim}
  {\bf \delta X}^\ast = {\bf X}_\mathrm{PRF} [\delta I^\ast (X,Y)] - {\bf X}_\mathrm{PRF} [\langle I(X,Y,t) \rangle_{\mathrm{out}}].
\end{eqnarray}
Finally, the distribution of $\delta {\bf X}^*$ was used to estimate statistical uncertainty of the centroid shift determination. 
We computed ellipses that include 95.5\%/99.7\% of $\delta {\bf X}^*$ (dashed/dotted lines in Figure \ref{fig:stat}), which we conventionally call $2\sigma/3\sigma$ ellipses.\footnote{The ellipses were computed as follows. We performed the principle component analysis (PCA) whitening of the dataset of ${\bf \delta X}^\ast$ and determine the center and the radius that surround 95.5\% of the distribution. The radius that encircles 99.7\% was estimated by multiplying the 95.5\% radius by 1.38, assuming the two-dimensional normal distribution. Then we performed the inverse PCA transform whitening of those radii.} 
We flagged 10 events outside the $3\sigma$ ellipse and discuss them further below. We decided to keep six events between $2\sigma$ and $3\sigma$ ellipses, considering that the number ($6/93\sim6\%$) is consistent with what we expect from statistical fluctuation. 

We further examined the 10 flagged events, because a true signal can still cause a large centroid shift as defined by Equation (\ref{eq:PRFcoffset}) due to a nearby unrelated star \citep{2013PASP..125..889B}. In this case, the out-of-transit centroid is shifted toward the nearby star, while the in-transit one remains close to the original target star. To avoid erroneous identification of this case as a false positive, \cite{2013PASP..125..889B} proposed another centroid test using the difference image and the catalog position of a star ${\bf X}_\mathrm{cat}$: 
\begin{eqnarray}
  \label{eq:PRFcoffsetartcat}
  {\bf \delta X}_{\rm cat} = {\bf X}_\mathrm{PRF} [\delta I (X,Y)] - {\bf X}_\mathrm{cat}.
\end{eqnarray}
Again, \cite{2013PASP..125..889B} estimated the significance of the shift from quarter-to-quarter variations but this is not possible in our case, and so we adopted the same $3\sigma$ ellipse constructed from injected signals. Among the 10 flagged events, ${\bf \delta X}_{\rm cat}$ for the event in KIC 11709124 was found to be within $3\sigma$, and so this target was moved back to the list. In addition to the nearby star, a true signal can exhibit a larger centroid shift when the signal is located near the gap and the out-of-transit data are partially unavailable \citep{2018AJ....155..144K}. This was the case for KIC 6191521, both of whose DTEs are located near data gaps. One shows the centroid shift beyond the 3$\sigma$ ellipse, the other is between the $2\sigma$ and $3\sigma$ ones, and their values are all near the outliers of the shifts from simulated dips. These imply that the large offset is due to the data gaps, and we decided to keep KIC 6191521 in our list as well. 

\begin{figure}[htbp]
\begin{center}
  \includegraphics[width=\linewidth]{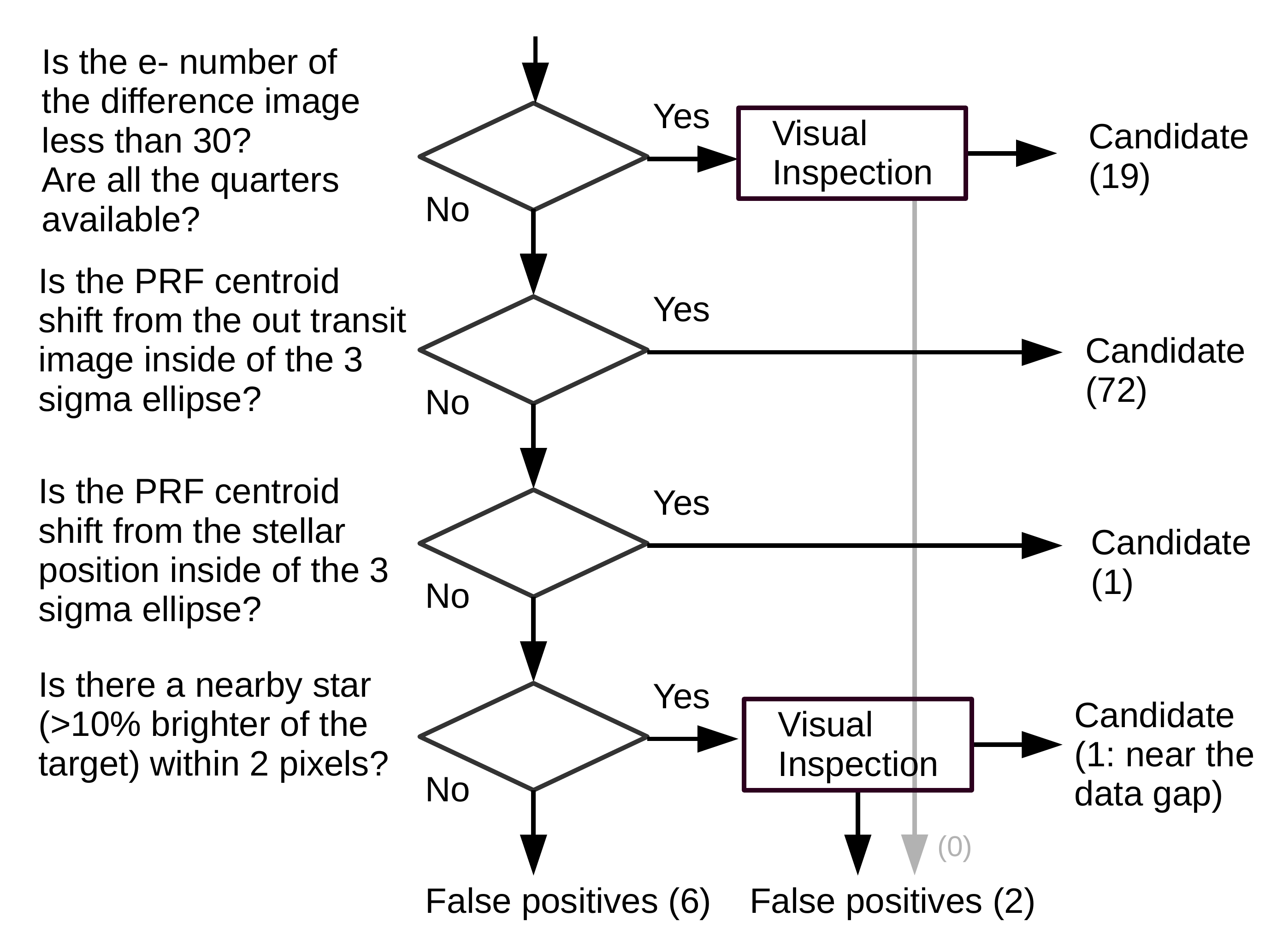}
  \caption{Flow chart of the centroid test using the pixel-level difference image.\label{fig:flowchart}}
\end{center}
\end{figure}

Figure \ref{fig:flowchart} summarizes the pixel-data examination described in this subsection. 
Here eight targets were classified as false positives and eliminated from the list (``CS" in Table \ref{tab:fp}).

\subsection{Detector Positions of the Transit Events}

\begin{figure}[htbp]
\begin{center}
  \includegraphics[width=\linewidth]{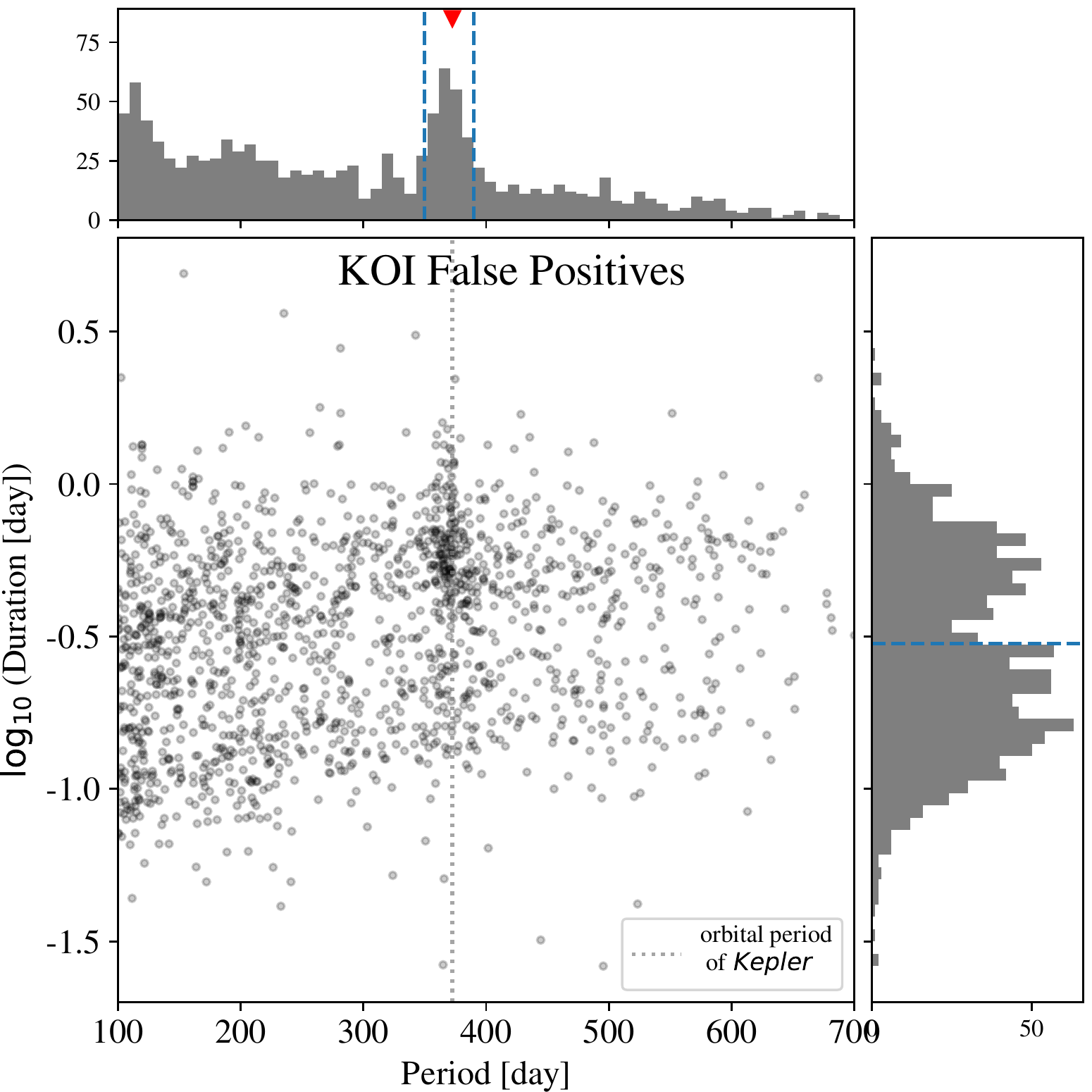}
  \caption{KOI false positives on the orbital period--transit duration plane. The dashed line and red triangle indicate the orbital period of {\it Kepler} (372 d). The top and left panels are the histograms of the period and duration. The clustering around $P = 372\,\mathrm{d}$ and duration of 0.6~d is due to instrumental artifacts. We call the region with periods 350--390~d (blue dashed lines in the histograms) and duration longer than 0.3~d as an ``FP bump."\label{fig:KOIpdFP}}
\end{center}
\end{figure}

\cite{2015ApJS..217...31M} pointed out two types of instrumental effects causing false positives. One is the edge effect around a data gap; we already excluded suspicious dips around the gaps during the visual inspection in Sections \ref{ss:cvi} and \ref{ss:tls}. The other is an artifact with the period close to the orbital period of the {\it Kepler} spacecraft (372~days), which may contaminate long-period candidates discussed here. 
As shown in Figure \ref{fig:KOIpdFP}, such artifacts are clustered around the orbital period of {\it Kepler} and have durations typically longer than 0.3~days. 

\begin{figure*}[htbp]
\begin{center}
  \includegraphics[width=0.49\linewidth]{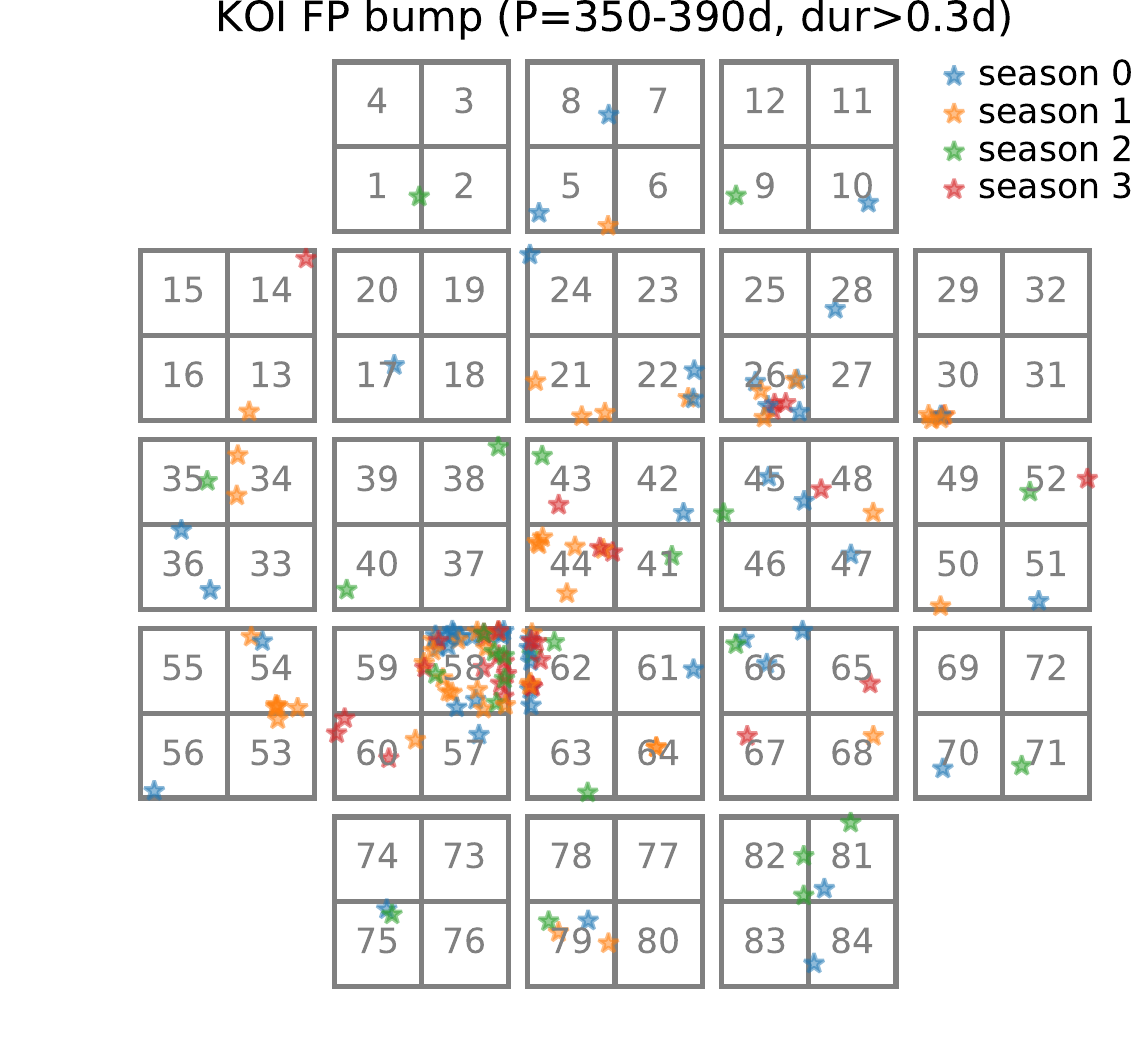}
  \includegraphics[width=0.49\linewidth]{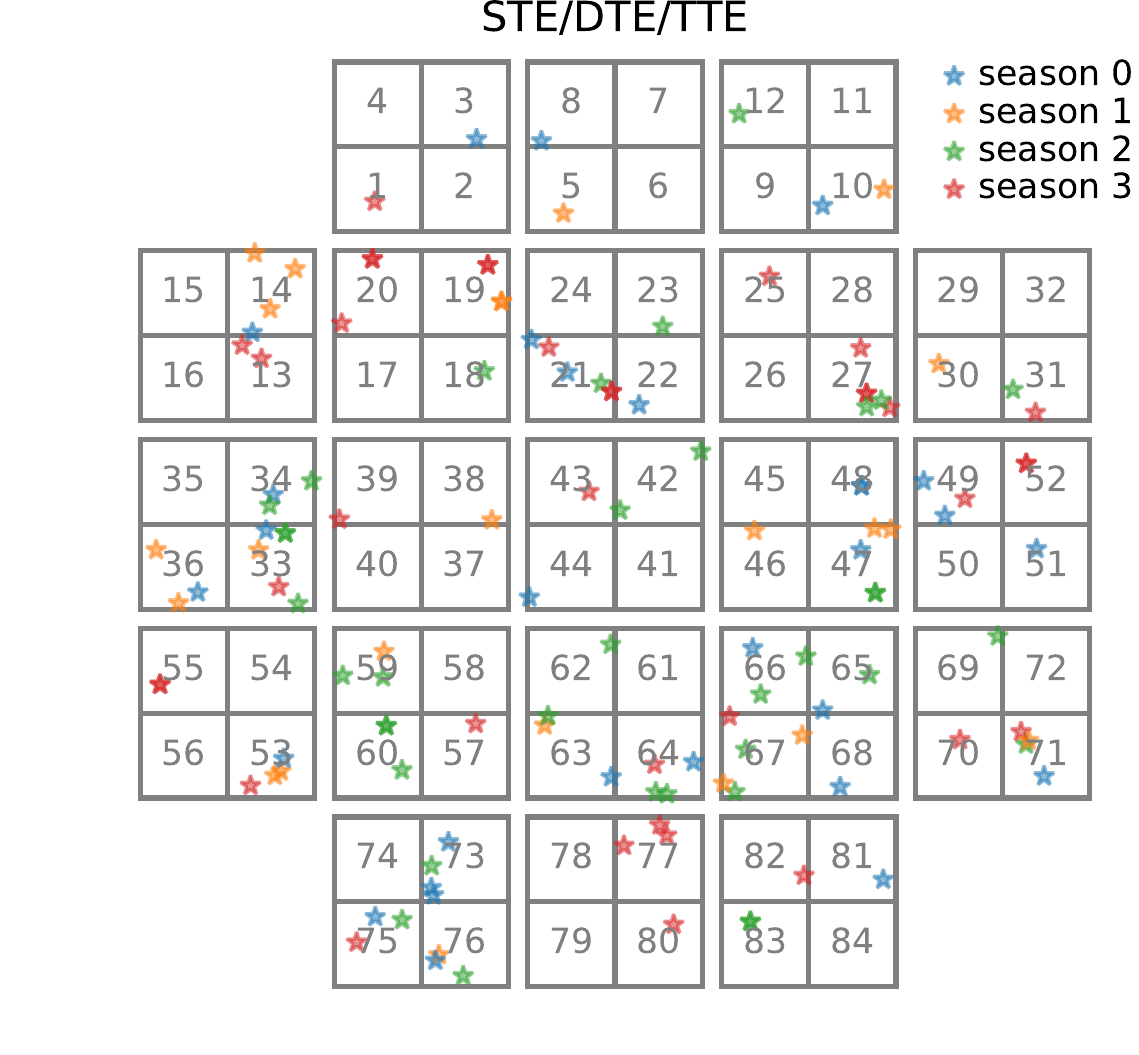}
  \caption{The CCD positions of the events (chips 1--84). The left panel shows the distribution for those in the KOI FP bump, defined by $P=350$--390~d and duration longer than 0.3~d (Figure \ref{fig:KOIpdFP}). The color indicates the season in which the events are observed: blue, yellow, green, and red stars correspond to seasons 0, 1, 2, and 3, respectively. Only the first transit event for each target is plotted, because the other transits occur mostly in the same season by construction of the FP bump (i.e. period close to one year). The right panel shows the same distribution for all the long-period planet candidates in our list. 
   \label{fig:longP_d}}
\end{center}
\end{figure*}

To identify the nature of the artifact, we checked the CCD positions of the KOI false positives with a period between 350--390~d and a duration longer than 0.3~d, which we define as an ``FP bump," taking into account the rotation of the spacecraft every three months. As shown in the left panel of Figure \ref{fig:longP_d}, the transits in the FP bump are clustered around the CCD chip 58 and the left edge of the chip 62. This clustering strongly indicates the instrumental origin; the FP bump appears to be produced by repeating signals from the same CCD chips. On the other hand, we did not find any such clustering for long-period planet candidates in our list (right panel of Figure \ref{fig:longP_d}). Thus we find no evidence that a majority of these events originate from similar artifacts as producing the FP bump.

\section{The Stars}\label{sec:stars}

Precise stellar parameters, especially radii, are important for characterization and vetting of the detected planet candidates. Here we characterize the host stars in the input catalog by fitting stellar evolutionary models to spectroscopic atmospheric parameters or broad-band photometry, and parallax from \gaia\ DR2. The stars analyzed here include not only those hosting candidates that passed the FP test in Section \ref{sec:fp}, but also some that failed.

\subsection{Spectroscopy}

First, we collected archival spectroscopic parameters from the California \kepler\ Survey \citep[CKS,][]{2017AJ....154..107P}, the Spectral Properties of Cool Stars (SPOCS) catalog by \citet{2016ApJS..225...32B}, the Large Sky Area Multi-Object Fiber Spectroscopic Telescope \citep[LAMOST;][]{2012RAA....12.1197C, 2015RAA....15.1095L}/LASP stellar parameters \citep{2014IAUS..306..340W} from DR4, {\it The Payne} \citep{2018arXiv180401530T} parameters based on APOGEE DR14 \citep{2017AJ....154...94M}, and the \kepler\ input catalog DR25 \citep{2017ApJS..229...30M} with spectroscopic provenance (KIC 6191521 alone, originally from \citealt{2017AJ....153...71F}). For these stars, we adopted the literature values of effective temperature $T_{\rm eff}$, surface gravity $\log g$, and iron abundance [Fe/H] with preferences given in this order. Note that 14 of the stars analyzed here are in the CKS sample and have known inner transiting planets and planet candidates (see also Tables \ref{tab:clean} and \ref{tab:flagged}, and Section \ref{ssec:planets_clean_koi}).

\begin{figure*}[htbp]
\begin{center}
  \includegraphics[width=0.49\linewidth]{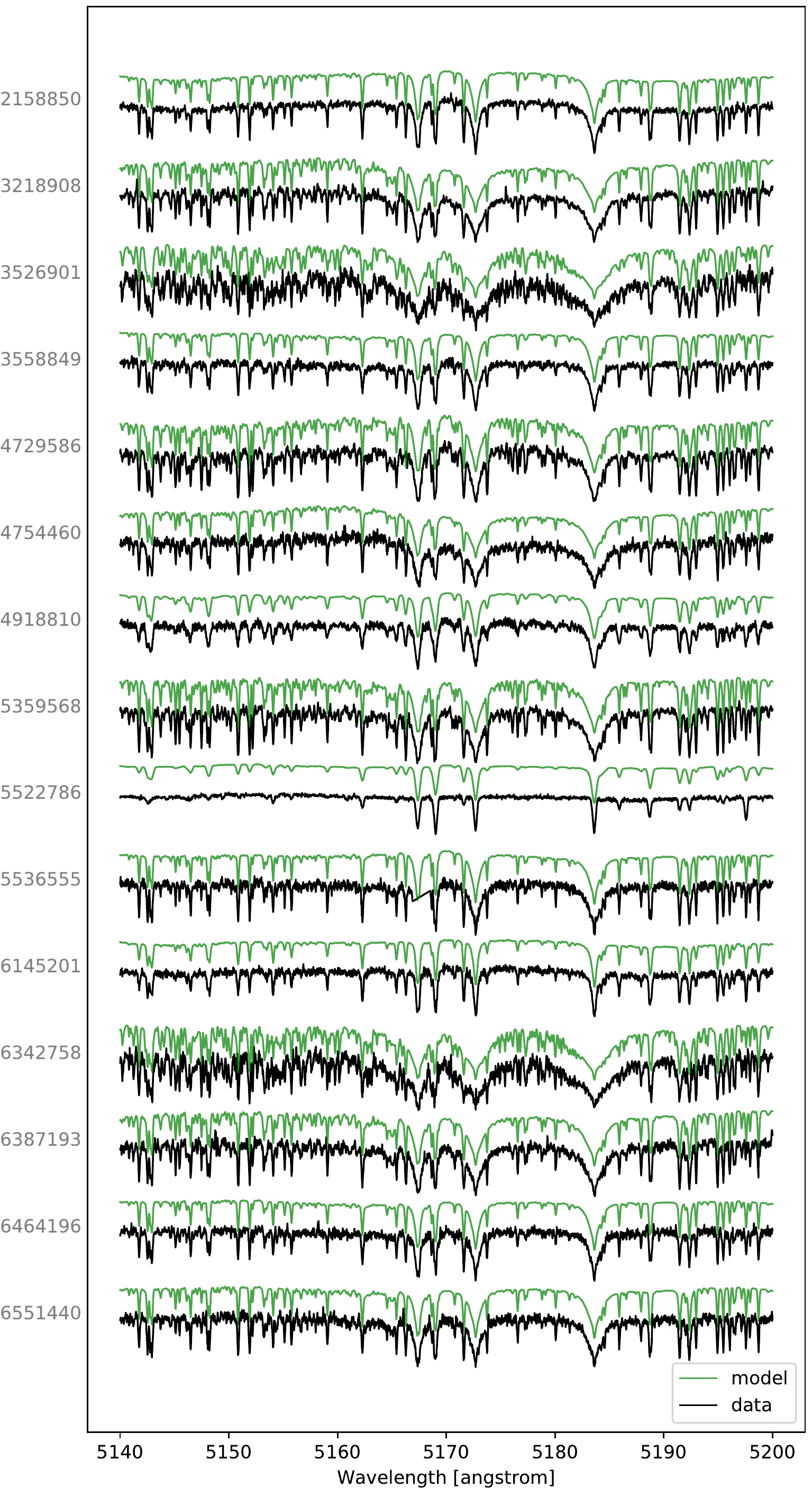}
  \includegraphics[width=0.49\linewidth]{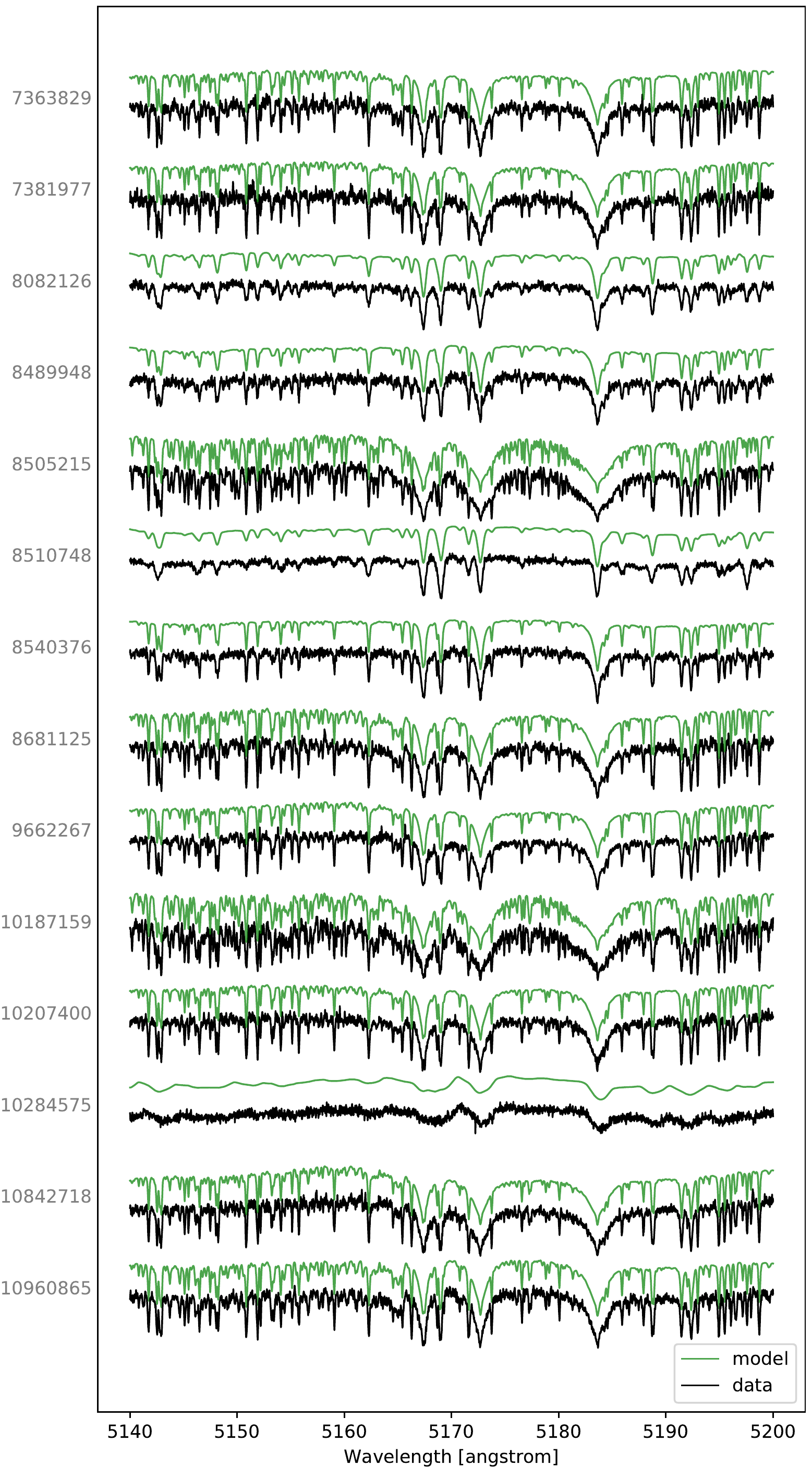}
  \caption{Subaru/HDS high resolution spectra near the Mg triplet (black). KIC numbers are shown at the left of the corresponding spectrum. Green curves are the best-match model returned by the {\tt Specmatch-Emp} code \citep{2017ApJ...836...77Y}. \label{fig:allspectra}}
\end{center}
\end{figure*}

For stars without available archival spectra, we obtained high-resolution spectra using the high dispersion spectrograph \citep[HDS;][]{2002PASJ...54..855N} installed on the Subaru 8.2~m telescope. We mainly observed targets brighter than $r\mathchar`-\mathrm{mag}=15$ and without V-shaped dips.
The observations were performed on UT July 28, August 5, and September 6 in 2018 (proposal IDs S18A-044 and S18B-062, PI: Kawahara) with the standard I2a setup and $2\times1$ binning without an image rotator. Image slicer \#2 \citep{2012PASJ...64...77T} was used for targets with $r\mathchar`-\mathrm{mag}\lesssim14.5$ to increase the signal-to-noise ratio. The resulting spectral resolution was typically 60,000--87,000. We then fit the spectra around the Mg triplet using the {\tt Specmatch-Emp} code \citep{2017ApJ...836...77Y} to derive $T_{\rm eff}$, $\log g$, and [Fe/H] via empirical matching to the spectra of the touchstone stars with well-determined parameters. Figure \ref{fig:allspectra} shows the observed spectra (black) along with the best matches from the code (green) for FGK stars. We cross-checked our estimates for KIC 8505215 ($T_\mathrm{eff}=4954 \pm 110\,$K, $\log g=4.54 \pm 0.12$, [Fe/H]$=-0.45 \pm 0.09$) against the CKS values ($T_\mathrm{eff} = 4935 \pm 116$ K, $\log g=4.42 \pm 0.10$, [Fe/H]$=-0.28 \pm 0.07$, including systematic uncertainties), and found a reasonable agreement. 

The atmospheric parameters directly derived from spectroscopy are summarized in the last three columns of Table \ref{tab:starparams} with the source in the fourth column from the last. These parameters are collected from multiple sources including the present paper, and their accuracy and precision are not uniform. For the following isochrone modeling, we adopt the same uncertainties of $110\,\mathrm{K}$ for $T_\mathrm{eff}$, 0.12 for $\log g$, and 0.1 for [Fe/H] for all stars, consdering the typical accuracy of these parameters \citep[cf.][]{2017AJ....154..107P}. In practice, the stellar radius is mainly determined by $T_{\rm eff}$ and \gaia\ parallax, and the other parameters play minor roles. 

\subsection{Photometry}

We also collected \textit{JHK} photometry from the Two Micron All Sky Survey \citep[2MASS;][]{2006AJ....131.1163S} for all the stars in the sample. For those without spectroscopic observations, the colors are used to estimate $T_{\rm eff}$; otherwise only the $K$-band magnitude was used so that the extinction does not introduce systematic bias on $T_{\rm eff}$. We decided not to use \textit{griz} colors considering the systematic trend reported in \citet{2012ApJS..199...30P}.

\subsection{Isochrone Modeling}

For the stars with spectroscopic parameters, we fitted stellar evolutionary models to the spectroscopic $T_{\rm eff}$, $\log g$, and [Fe/H], along with the \gaia\ DR2 parallax \citep{2018A&A...616A...1G, 2018AJ....156...58B} and the 2MASS $K$-magnitude. The $K$-magnitude was corrected for the extinction $A_K$ using the value of $E(B-V)$ and extinction vectors from the dust map \citep[Bayestar17;][]{2018MNRAS.478..651G}. The $30\%$ fractional uncertainty was assumed for $A_K$ following \citet{2018AJ....156..264F}. The stellar-evolutionary models were from the Dartmouth Stellar Evolution Database \citep{2008ApJS..178...89D} and were compared to the data using the {\tt isochrones} package by \citet{2015ascl.soft03010M}.

For the stars without available spectroscopy, we fitted stellar models to the 2MASS \textit{JHK} colors instead of $T_{\rm eff}$, $\log g$, and [Fe/H]. We also floated $A_V$ as a free parameter rather than using a value from the extinction map. This is because the model sometimes failed to fit the \textit{JHK} colors corrected for the map-based extinction and yielded unreasonably small formal uncertainties for $T_{\rm eff}$ when combined with the precise \gaia\ parallax. Here $A_V$ is introduced to mitigate such incompleteness of the model.

\subsection{Results}

\begin{figure*}[htbp]
    \epsscale{0.95}
    \plotone{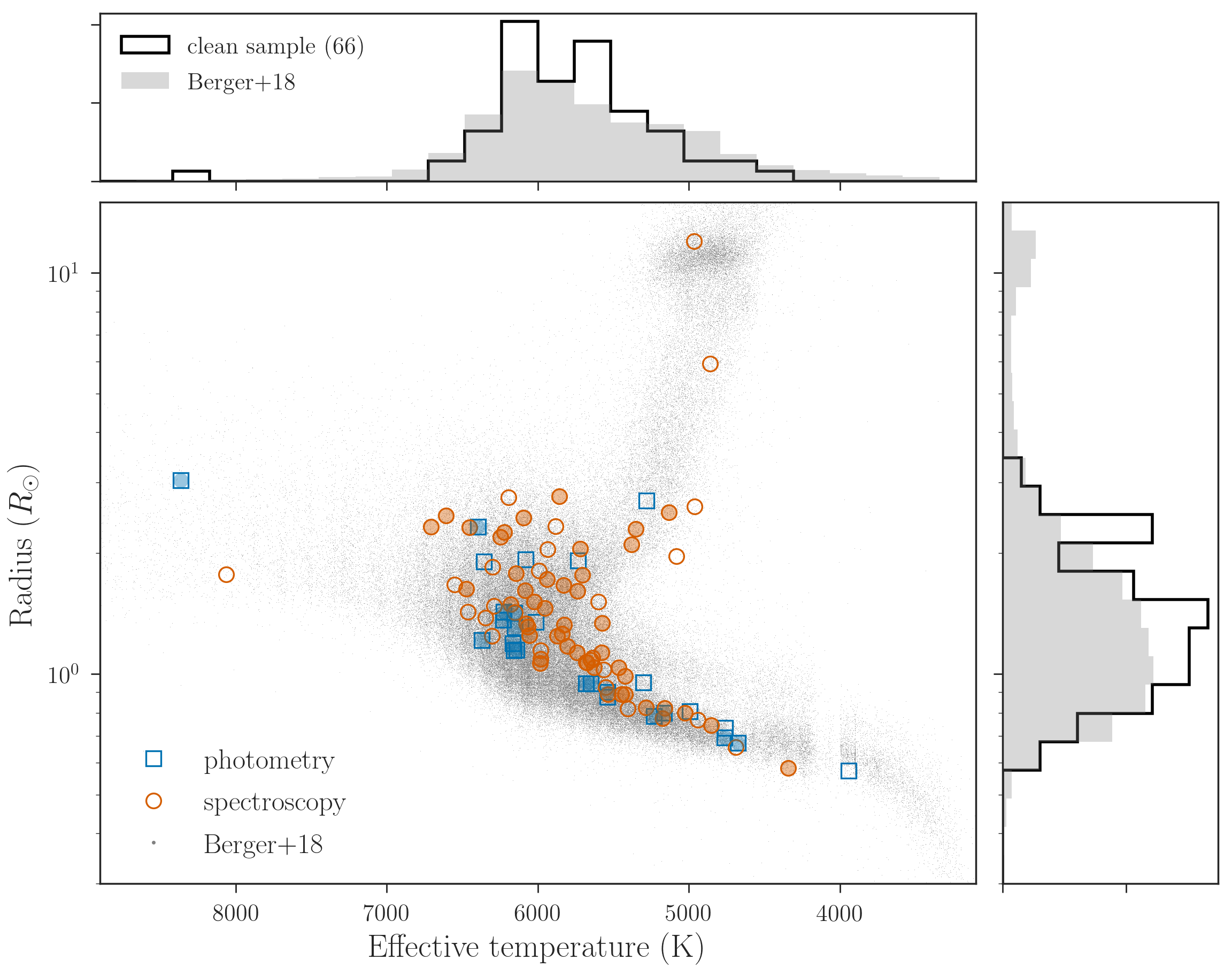}
    \caption{The host stars in the HR diagram (open circles/squares) and the \kepler\ stars from \citet{2018ApJ...866...99B} (gray dots). The filled circles/squares correspond to the host stars in the clean sample (Section \ref{sec:planets_clean}), whose corresponding histograms are shown with the thick black lines in the top and right panels. The gray histograms are for the \kepler\ field stars. There are 66 stars in the clean sample (compared to 67 candidates) because KIC 10024862 has two long-period transiting planet candidates.}
    \label{fig:radius_teff}
\end{figure*}

The results of the isochrone modeling for all the stars are summarized in Table \ref{tab:starparams}. Here the second to seventh columns report constraints from the isochrone model (median and $68\%$ credible interval of the marginal posterior), and the remaining columns show the values from raw spectroscopy when available. We cross-checked the resulting stellar radii against the values in \citet{2018ApJ...866...99B} and found a good agreement without any apparent systematic trend, although the radius precision was improved to a few percent for the sample with spectroscopic information added. Figure \ref{fig:radius_teff} shows the radii and effective temperatures of the spectroscopic (orange circles) and photometric (blue squares) samples along with all the \kepler\ field stars (gray dots) from \citet{2018ApJ...866...99B}. 

The distribution of [Fe/H] from the final isochrone fit has the mean of $0.00$ and the standard deviation of $0.15$. These values agree with the typical metallicity of the \kepler\ field stars \citep{2014ApJ...789L...3D, 2017ApJ...838...25G}. That said, the [Fe/H] distribution of our sample is skewed toward higher values (see Section \ref{ssec:planets_clean_feh}).
This explains the lack of smaller stars along the main-sequence in our sample (Figure \ref{fig:radius_teff}).

\startlongtable
\begin{deluxetable*}{lcccccccccccccc}
\tablecaption{Parameters of the Host Stars.\label{tab:starparams}}
\tablehead{
\colhead{KIC} & \colhead{$T_\mathrm{eff}$ (K)} & \colhead{$\log g$ (cgs)} & \colhead{[Fe/H]} & \colhead{$M_\star$ ($M_\odot$)} & \colhead{$R_\star$ ($R_\odot$)} & \colhead{$\log_{10}(\mathrm{age}/\mathrm{yr})$} & \colhead{Source$^\dagger$} & \colhead{$T_\mathrm{eff}$ (K)} & \colhead{$\log g$ (cgs)} & \colhead{[Fe/H]}
}
\startdata
1717722 & $4689^{+80}_{-75}$ & $4.63^{+0.02}_{-0.02}$ & $-0.28^{+0.08}_{-0.08}$ & $0.68^{+0.02}_{-0.02}$ & $0.66^{+0.01}_{-0.01}$ & $9.9^{+0.2}_{-0.4}$ & APOGEE & $4669$ & $4.48$ & $-0.38$\\
2158850 & $5980^{+96}_{-98}$ & $4.34^{+0.02}_{-0.03}$ & $-0.04^{+0.08}_{-0.08}$ & $1.05^{+0.05}_{-0.04}$ & $1.15^{+0.02}_{-0.02}$ & $9.7^{+0.1}_{-0.2}$ & HDS & $5976$ & $4.28$ & $-0.06$\\
2162635 & $4964^{+73}_{-60}$ & $3.62^{+0.03}_{-0.03}$ & $0.05^{+0.09}_{-0.04}$ & $1.03^{+0.06}_{-0.04}$ & $2.61^{+0.05}_{-0.05}$ & $10.02^{+0.07}_{-0.09}$ & CKS & $4921$ & $3.62$ & $0.08$\\
3111510 & $6230^{+260}_{-265}$ & $4.24^{+0.05}_{-0.05}$ & $0.1^{+0.1}_{-0.2}$ & $1.19^{+0.08}_{-0.08}$ & $1.36^{+0.06}_{-0.05}$ & $9.5^{+0.2}_{-0.3}$ & \nodata & \nodata & \nodata & \nodata\\
3218908$^b$ & $5578^{+91}_{-77}$ & $4.31^{+0.03}_{-0.03}$ & $0.18^{+0.08}_{-0.08}$ & $0.96^{+0.04}_{-0.03}$ & $1.13^{+0.03}_{-0.03}$ & $10.01^{+0.08}_{-0.12}$ & HDS & $5561$ & $4.25$ & $0.24$\\
3230491 & $5406^{+86}_{-90}$ & $4.53^{+0.03}_{-0.03}$ & $-0.21^{+0.08}_{-0.08}$ & $0.83^{+0.04}_{-0.04}$ & $0.82^{+0.02}_{-0.02}$ & $9.8^{+0.2}_{-0.3}$ & LAMOST & $5405$ & $4.51$ & $-0.28$\\
3239945 & $4851^{+63}_{-64}$ & $4.59^{+0.02}_{-0.02}$ & $0.02^{+0.08}_{-0.08}$ & $0.78^{+0.03}_{-0.03}$ & $0.745^{+0.008}_{-0.008}$ & $9.8^{+0.2}_{-0.4}$ & CKS & $4868$ & $4.57$ & $0.03$\\
3241604 & $6358^{+353}_{-250}$ & $4.03^{+0.06}_{-0.06}$ & $0.1^{+0.1}_{-0.1}$ & $1.42^{+0.08}_{-0.06}$ & $1.9^{+0.1}_{-0.1}$ & $9.4^{+0.1}_{-0.1}$ & \nodata & \nodata & \nodata & \nodata\\
3346436 & $5882^{+100}_{-104}$ & $3.80^{+0.06}_{-0.04}$ & $-0.03^{+0.08}_{-0.01}$ & $1.23^{+0.16}_{-0.06}$ & $2.33^{+0.07}_{-0.07}$ & $9.66^{+0.07}_{-0.11}$ & LAMOST & $5887$ & $3.92$ & $0.00$\\
3351971 & $5539^{+172}_{-155}$ & $4.49^{+0.03}_{-0.04}$ & $-0.0^{+0.2}_{-0.2}$ & $0.92^{+0.05}_{-0.06}$ & $0.90^{+0.02}_{-0.02}$ & $9.7^{+0.3}_{-0.5}$ & \nodata & \nodata & \nodata & \nodata\\
3526901 & $5082^{+85}_{-68}$ & $3.85^{+0.05}_{-0.05}$ & $0.01^{+0.02}_{-0.01}$ & $0.98^{+0.04}_{-0.02}$ & $2.0^{+0.1}_{-0.1}$ & $10.08^{+0.03}_{-0.06}$ & HDS & $4881$ & $4.49$ & $-0.17$\\
3558849 & $6026^{+111}_{-104}$ & $4.14^{+0.03}_{-0.03}$ & $0.01^{+0.06}_{-0.08}$ & $1.14^{+0.05}_{-0.04}$ & $1.51^{+0.05}_{-0.04}$ & $9.72^{+0.09}_{-0.10}$ & HDS & $5976$ & $4.25$ & $-0.06$\\
3756801 & $5857^{+91}_{-87}$ & $3.68^{+0.07}_{-0.04}$ & $-0.04^{+0.15}_{-0.01}$ & $1.3^{+0.2}_{-0.1}$ & $2.77^{+0.08}_{-0.08}$ & $9.5^{+0.1}_{-0.1}$ & CKS & $5834$ & $3.82$ & $0.12$\\
3962440 & $6453^{+104}_{-101}$ & $3.90^{+0.06}_{-0.05}$ & $0.03^{+0.06}_{-0.04}$ & $1.55^{+0.07}_{-0.07}$ & $2.3^{+0.2}_{-0.2}$ & $9.33^{+0.06}_{-0.05}$ & CKS & $6417$ & $4.16$ & $-0.07$\\
4042088 & $6463^{+100}_{-109}$ & $4.23^{+0.02}_{-0.03}$ & $-0.04^{+0.08}_{-0.08}$ & $1.25^{+0.04}_{-0.05}$ & $1.43^{+0.03}_{-0.03}$ & $9.4^{+0.1}_{-0.1}$ & LAMOST & $6473$ & $4.09$ & $-0.05$\\
4729586 & $4966^{+36}_{-43}$ & $2.71^{+0.04}_{-0.04}$ & $0.02^{+0.08}_{-0.04}$ & $2.65^{+0.08}_{-0.04}$ & $12.0^{+0.6}_{-0.6}$ & $8.76^{+0.02}_{-0.03}$ & HDS & $4812$ & $2.96$ & $-0.06$\\
4754460 & $5637^{+91}_{-85}$ & $4.30^{+0.03}_{-0.03}$ & $-0.22^{+0.08}_{-0.07}$ & $0.87^{+0.03}_{-0.02}$ & $1.10^{+0.03}_{-0.03}$ & $10.07^{+0.04}_{-0.07}$ & HDS & $5514$ & $4.56$ & $-0.35$\\
4754691$^b$ & $5303^{+71}_{-96}$ & $4.421^{+0.012}_{-0.009}$ & $0.2^{+0.1}_{-0.2}$ & $0.88^{+0.02}_{-0.02}$ & $0.95^{+0.01}_{-0.01}$ & $10.10^{+0.02}_{-0.05}$ & \nodata & \nodata & \nodata & \nodata\\
4772953 & $9431^{+681}_{-554}$ & $4.16^{+0.05}_{-0.06}$ & $-0.0^{+0.2}_{-0.2}$ & $2.2^{+0.1}_{-0.1}$ & $2.04^{+0.10}_{-0.09}$ & $8.6^{+0.2}_{-0.1}$ & \nodata & \nodata & \nodata & \nodata\\
4918810 & $6144^{+80}_{-87}$ & $4.05^{+0.02}_{-0.02}$ & $0.09^{+0.08}_{-0.06}$ & $1.31^{+0.03}_{-0.07}$ & $1.78^{+0.04}_{-0.03}$ & $9.56^{+0.07}_{-0.05}$ & HDS & $6106$ & $4.12$ & $0.12$\\
5010054 & $5940^{+100}_{-93}$ & $4.03^{+0.02}_{-0.02}$ & $0.00^{+0.06}_{-0.13}$ & $1.16^{+0.03}_{-0.03}$ & $1.72^{+0.04}_{-0.04}$ & $9.72^{+0.08}_{-0.09}$ & LAMOST & $5898$ & $4.01$ & $-0.11$\\
5184479 & $5953^{+110}_{-94}$ & $4.17^{+0.03}_{-0.03}$ & $0.10^{+0.07}_{-0.07}$ & $1.13^{+0.05}_{-0.05}$ & $1.46^{+0.04}_{-0.04}$ & $9.8^{+0.1}_{-0.1}$ & LAMOST & $5954$ & $4.15$ & $0.11$\\
5351250 & $5534^{+75}_{-82}$ & $4.52^{+0.02}_{-0.03}$ & $0.07^{+0.07}_{-0.08}$ & $0.96^{+0.03}_{-0.04}$ & $0.89^{+0.02}_{-0.02}$ & $9.4^{+0.3}_{-0.4}$ & CKS & $5588$ & $4.58$ & $0.12$\\
5359568 & $4858^{+108}_{-98}$ & $2.96^{+0.05}_{-0.07}$ & $-0.12^{+0.04}_{-0.04}$ & $1.2^{+0.1}_{-0.2}$ & $5.9^{+0.2}_{-0.2}$ & $9.8^{+0.2}_{-0.2}$ & HDS & $4841$ & $3.16$ & $0.08$\\
5536555 & $5984^{+91}_{-101}$ & $4.41^{+0.03}_{-0.03}$ & $-0.02^{+0.08}_{-0.08}$ & $1.06^{+0.05}_{-0.05}$ & $1.06^{+0.02}_{-0.02}$ & $9.5^{+0.2}_{-0.3}$ & HDS & $5992$ & $4.40$ & $-0.03$\\
5623581 & $6228^{+297}_{-316}$ & $4.21^{+0.06}_{-0.06}$ & $0.1^{+0.1}_{-0.1}$ & $1.21^{+0.08}_{-0.10}$ & $1.43^{+0.07}_{-0.07}$ & $9.6^{+0.2}_{-0.3}$ & \nodata & \nodata & \nodata & \nodata\\
5732155 & $6395^{+222}_{-183}$ & $3.90^{+0.05}_{-0.04}$ & $0.11^{+0.13}_{-0.08}$ & $1.57^{+0.06}_{-0.06}$ & $2.3^{+0.1}_{-0.1}$ & $9.33^{+0.08}_{-0.08}$ & \nodata & \nodata & \nodata & \nodata\\
5871088 & $4675^{+167}_{-148}$ & $4.63^{+0.02}_{-0.02}$ & $-0.2^{+0.2}_{-0.2}$ & $0.71^{+0.04}_{-0.04}$ & $0.67^{+0.03}_{-0.03}$ & $9.7^{+0.3}_{-0.6}$ & \nodata & \nodata & \nodata & \nodata\\
5942949 & $4765^{+169}_{-136}$ & $4.62^{+0.02}_{-0.02}$ & $-0.2^{+0.2}_{-0.2}$ & $0.73^{+0.04}_{-0.03}$ & $0.69^{+0.02}_{-0.02}$ & $9.7^{+0.3}_{-0.6}$ & \nodata & \nodata & \nodata & \nodata\\
5951458 & $5993^{+99}_{-88}$ & $4.00^{+0.02}_{-0.02}$ & $-0.05^{+0.09}_{-0.09}$ & $1.20^{+0.05}_{-0.03}$ & $1.81^{+0.03}_{-0.04}$ & $9.65^{+0.09}_{-0.04}$ & LAMOST & $5948$ & $4.02$ & $-0.11$\\
6145201 & $6302^{+133}_{-119}$ & $4.01^{+0.03}_{-0.03}$ & $-0.19^{+0.14}_{-0.07}$ & $1.26^{+0.08}_{-0.06}$ & $1.85^{+0.05}_{-0.05}$ & $9.52^{+0.09}_{-0.06}$ & HDS & $6302$ & $4.05$ & $-0.24$\\
6186417 & $6151^{+270}_{-280}$ & $4.26^{+0.06}_{-0.06}$ & $0.0^{+0.1}_{-0.2}$ & $1.15^{+0.08}_{-0.09}$ & $1.31^{+0.09}_{-0.07}$ & $9.6^{+0.2}_{-0.3}$ & \nodata & \nodata & \nodata & \nodata\\
6191521 & $5574^{+97}_{-75}$ & $4.18^{+0.04}_{-0.04}$ & $0.16^{+0.09}_{-0.08}$ & $0.98^{+0.04}_{-0.03}$ & $1.34^{+0.06}_{-0.06}$ & $10.06^{+0.05}_{-0.08}$ & M17 & $5512$ & $4.22$ & $0.20$\\
6203563 & $6370^{+205}_{-185}$ & $4.35^{+0.03}_{-0.04}$ & $-0.1^{+0.2}_{-0.2}$ & $1.19^{+0.06}_{-0.07}$ & $1.21^{+0.03}_{-0.02}$ & $9.3^{+0.3}_{-0.4}$ & \nodata & \nodata & \nodata & \nodata\\
6342758 & $4942^{+67}_{-64}$ & $4.57^{+0.02}_{-0.02}$ & $0.07^{+0.08}_{-0.08}$ & $0.81^{+0.03}_{-0.03}$ & $0.769^{+0.010}_{-0.009}$ & $9.8^{+0.2}_{-0.4}$ & HDS & $4974$ & $4.51$ & $0.11$\\
6387193 & $5566^{+97}_{-93}$ & $4.40^{+0.04}_{-0.03}$ & $0.18^{+0.08}_{-0.08}$ & $0.96^{+0.05}_{-0.04}$ & $1.03^{+0.02}_{-0.02}$ & $9.9^{+0.1}_{-0.2}$ & HDS & $5581$ & $4.28$ & $0.26$\\
6464196 & $5870^{+106}_{-113}$ & $4.25^{+0.03}_{-0.03}$ & $-0.07^{+0.09}_{-0.10}$ & $1.01^{+0.04}_{-0.05}$ & $1.24^{+0.04}_{-0.03}$ & $9.86^{+0.08}_{-0.08}$ & HDS & $5836$ & $4.27$ & $-0.14$\\
6510758 & $5422^{+75}_{-59}$ & $4.41^{+0.02}_{-0.02}$ & $0.17^{+0.08}_{-0.08}$ & $0.91^{+0.04}_{-0.02}$ & $0.99^{+0.01}_{-0.01}$ & $10.04^{+0.07}_{-0.13}$ & LAMOST & $5354$ & $4.35$ & $0.21$\\
6551440 & $5804^{+102}_{-101}$ & $4.31^{+0.03}_{-0.03}$ & $0.07^{+0.08}_{-0.06}$ & $1.02^{+0.04}_{-0.04}$ & $1.17^{+0.02}_{-0.02}$ & $9.9^{+0.1}_{-0.1}$ & HDS & $5795$ & $4.29$ & $0.07$\\
6690896 & $6181^{+96}_{-99}$ & $4.09^{+0.03}_{-0.02}$ & $-0.41^{+0.09}_{-0.07}$ & $1.01^{+0.05}_{-0.04}$ & $1.49^{+0.03}_{-0.03}$ & $9.83^{+0.06}_{-0.08}$ & LAMOST & $6132$ & $4.25$ & $-0.50$\\
6804821 & $8366^{+1344}_{-796}$ & $3.83^{+0.10}_{-0.07}$ & $0.1^{+0.1}_{-0.1}$ & $2.3^{+0.3}_{-0.2}$ & $3.0^{+0.1}_{-0.2}$ & $8.8^{+0.1}_{-0.2}$ & \nodata & \nodata & \nodata & \nodata\\
7040629 & $6055^{+105}_{-110}$ & $4.29^{+0.03}_{-0.02}$ & $0.03^{+0.07}_{-0.08}$ & $1.11^{+0.04}_{-0.04}$ & $1.25^{+0.02}_{-0.02}$ & $9.7^{+0.1}_{-0.1}$ & CKS & $6053$ & $4.24$ & $0.02$\\
7176219 & $5280^{+93}_{-79}$ & $3.70^{+0.02}_{-0.02}$ & $0.04^{+0.06}_{-0.03}$ & $1.34^{+0.03}_{-0.03}$ & $2.70^{+0.05}_{-0.06}$ & $9.61^{+0.03}_{-0.03}$ & \nodata & \nodata & \nodata & \nodata\\
7363829 & $5705^{+110}_{-109}$ & $4.01^{+0.04}_{-0.04}$ & $0.17^{+0.08}_{-0.08}$ & $1.15^{+0.09}_{-0.05}$ & $1.76^{+0.08}_{-0.08}$ & $9.82^{+0.07}_{-0.11}$ & HDS & $5668$ & $4.06$ & $0.21$\\
7381977 & $5551^{+96}_{-95}$ & $4.46^{+0.04}_{-0.04}$ & $-0.05^{+0.08}_{-0.08}$ & $0.91^{+0.05}_{-0.04}$ & $0.93^{+0.02}_{-0.02}$ & $9.9^{+0.2}_{-0.3}$ & HDS & $5557$ & $4.40$ & $-0.08$\\
7447005 & $5649^{+180}_{-168}$ & $4.47^{+0.04}_{-0.04}$ & $0.0^{+0.1}_{-0.2}$ & $0.95^{+0.06}_{-0.06}$ & $0.95^{+0.03}_{-0.03}$ & $9.7^{+0.3}_{-0.5}$ & \nodata & \nodata & \nodata & \nodata\\
7672940 & $6553^{+101}_{-107}$ & $4.13^{+0.03}_{-0.03}$ & $0.01^{+0.07}_{-0.08}$ & $1.36^{+0.04}_{-0.04}$ & $1.67^{+0.06}_{-0.06}$ & $9.39^{+0.05}_{-0.06}$ & CKS & $6532$ & $4.19$ & $-0.04$\\
7875441 & $6082^{+267}_{-326}$ & $3.99^{+0.06}_{-0.05}$ & $0.10^{+0.12}_{-0.08}$ & $1.35^{+0.08}_{-0.14}$ & $1.9^{+0.1}_{-0.1}$ & $9.5^{+0.2}_{-0.1}$ & \nodata & \nodata & \nodata & \nodata\\
7906827 & $6157^{+231}_{-202}$ & $4.37^{+0.04}_{-0.05}$ & $-0.0^{+0.2}_{-0.2}$ & $1.12^{+0.07}_{-0.08}$ & $1.14^{+0.06}_{-0.06}$ & $9.4^{+0.3}_{-0.5}$ & \nodata & \nodata & \nodata & \nodata\\
7947784 & $6012^{+325}_{-279}$ & $4.23^{+0.07}_{-0.07}$ & $0.1^{+0.1}_{-0.1}$ & $1.12^{+0.10}_{-0.09}$ & $1.35^{+0.08}_{-0.07}$ & $9.7^{+0.2}_{-0.3}$ & \nodata & \nodata & \nodata & \nodata\\
8012732 & $5841^{+98}_{-101}$ & $4.26^{+0.03}_{-0.03}$ & $0.09^{+0.07}_{-0.07}$ & $1.05^{+0.04}_{-0.04}$ & $1.26^{+0.03}_{-0.03}$ & $9.8^{+0.1}_{-0.1}$ & SPOCS & $5846$ & $4.21$ & $0.11$\\
8082126 & $6289^{+96}_{-110}$ & $4.18^{+0.03}_{-0.03}$ & $-0.08^{+0.08}_{-0.09}$ & $1.20^{+0.04}_{-0.05}$ & $1.48^{+0.03}_{-0.03}$ & $9.57^{+0.08}_{-0.07}$ & HDS & $6274$ & $4.12$ & $-0.14$\\
8168680 & $6347^{+104}_{-102}$ & $4.17^{+0.03}_{-0.03}$ & $-0.41^{+0.09}_{-0.08}$ & $1.03^{+0.05}_{-0.05}$ & $1.38^{+0.02}_{-0.02}$ & $9.76^{+0.08}_{-0.09}$ & LAMOST & $6338$ & $4.17$ & $-0.49$\\
8313257 & $5164^{+150}_{-131}$ & $4.56^{+0.03}_{-0.03}$ & $-0.0^{+0.2}_{-0.2}$ & $0.84^{+0.04}_{-0.04}$ & $0.80^{+0.02}_{-0.02}$ & $9.8^{+0.2}_{-0.5}$ & \nodata & \nodata & \nodata & \nodata\\
8410697 & $5648^{+98}_{-85}$ & $4.33^{+0.03}_{-0.02}$ & $-0.07^{+0.10}_{-0.09}$ & $0.93^{+0.04}_{-0.04}$ & $1.08^{+0.01}_{-0.01}$ & $10.00^{+0.07}_{-0.09}$ & LAMOST & $5617$ & $4.26$ & $-0.11$\\
8426957 & $5936^{+102}_{-99}$ & $3.94^{+0.03}_{-0.03}$ & $0.11^{+0.10}_{-0.07}$ & $1.33^{+0.06}_{-0.07}$ & $2.05^{+0.04}_{-0.04}$ & $9.59^{+0.05}_{-0.05}$ & LAMOST & $5922$ & $3.79$ & $0.15$\\
8489948 & $6152^{+94}_{-94}$ & $4.20^{+0.03}_{-0.03}$ & $0.00^{+0.06}_{-0.07}$ & $1.17^{+0.04}_{-0.04}$ & $1.42^{+0.05}_{-0.05}$ & $9.63^{+0.09}_{-0.08}$ & HDS & $6123$ & $4.28$ & $-0.05$\\
8505215$^{ab}$ & $5176^{+73}_{-66}$ & $4.536^{+0.015}_{-0.009}$ & $-0.25^{+0.07}_{-0.07}$ & $0.75^{+0.02}_{-0.01}$ & $0.775^{+0.007}_{-0.007}$ & $10.07^{+0.05}_{-0.11}$ & HDS & $4954$ & $4.54$ & $-0.49$\\
8510748 & $6608^{+105}_{-102}$ & $3.87^{+0.03}_{-0.03}$ & $0.10^{+0.08}_{-0.07}$ & $1.66^{+0.04}_{-0.04}$ & $2.48^{+0.10}_{-0.09}$ & $9.23^{+0.04}_{-0.04}$ & HDS & $6604$ & $3.96$ & $0.11$\\
8636333 & $6166^{+250}_{-234}$ & $4.34^{+0.05}_{-0.05}$ & $-0.0^{+0.2}_{-0.2}$ & $1.13^{+0.08}_{-0.08}$ & $1.19^{+0.06}_{-0.05}$ & $9.5^{+0.3}_{-0.4}$ & \nodata & \nodata & \nodata & \nodata\\
8648356 & $8062^{+107}_{-108}$ & $4.17^{+0.02}_{-0.02}$ & $-0.13^{+0.08}_{-0.09}$ & $1.70^{+0.05}_{-0.05}$ & $1.77^{+0.03}_{-0.03}$ & $8.93^{+0.08}_{-0.09}$ & LAMOST & $8080$ & $3.98$ & $-0.17$\\
8681125 & $5466^{+79}_{-68}$ & $4.37^{+0.03}_{-0.03}$ & $0.20^{+0.08}_{-0.08}$ & $0.93^{+0.04}_{-0.02}$ & $1.04^{+0.03}_{-0.03}$ & $10.04^{+0.06}_{-0.12}$ & HDS & $5435$ & $4.16$ & $0.26$\\
8738735 & $6085^{+89}_{-108}$ & $4.10^{+0.03}_{-0.03}$ & $0.02^{+0.06}_{-0.07}$ & $1.19^{+0.06}_{-0.04}$ & $1.61^{+0.04}_{-0.04}$ & $9.67^{+0.09}_{-0.08}$ & CKS & $6052$ & $4.12$ & $-0.02$\\
8800954$^{ab}$ & $5285^{+81}_{-74}$ & $4.52^{+0.03}_{-0.02}$ & $-0.13^{+0.08}_{-0.08}$ & $0.82^{+0.04}_{-0.03}$ & $0.82^{+0.01}_{-0.01}$ & $10.0^{+0.1}_{-0.2}$ & CKS & $5223$ & $4.57$ & $-0.23$\\
9388752 & $6195^{+118}_{-94}$ & $3.74^{+0.03}_{-0.02}$ & $-0.03^{+0.03}_{-0.02}$ & $1.51^{+0.09}_{-0.05}$ & $2.75^{+0.04}_{-0.05}$ & $9.37^{+0.05}_{-0.07}$ & LAMOST & $6146$ & $3.88$ & $-0.24$\\
9413313 & $5160^{+73}_{-74}$ & $4.54^{+0.03}_{-0.03}$ & $0.11^{+0.08}_{-0.08}$ & $0.86^{+0.04}_{-0.03}$ & $0.82^{+0.01}_{-0.01}$ & $9.8^{+0.2}_{-0.4}$ & SPOCS & $5196$ & $4.49$ & $0.16$\\
9419047 & $5721^{+99}_{-99}$ & $3.92^{+0.03}_{-0.02}$ & $0.10^{+0.06}_{-0.05}$ & $1.27^{+0.05}_{-0.05}$ & $2.05^{+0.04}_{-0.04}$ & $9.66^{+0.07}_{-0.05}$ & LAMOST & $5691$ & $3.99$ & $0.14$\\
9581498 & $6155^{+220}_{-264}$ & $4.20^{+0.07}_{-0.06}$ & $0.1^{+0.1}_{-0.1}$ & $1.20^{+0.08}_{-0.11}$ & $1.4^{+0.1}_{-0.1}$ & $9.6^{+0.2}_{-0.2}$ & \nodata & \nodata & \nodata & \nodata\\
9662267 & $5742^{+103}_{-100}$ & $4.33^{+0.03}_{-0.03}$ & $0.07^{+0.08}_{-0.07}$ & $0.99^{+0.04}_{-0.04}$ & $1.13^{+0.03}_{-0.03}$ & $9.9^{+0.1}_{-0.1}$ & HDS & $5747$ & $4.25$ & $0.08$\\
9663113 & $6247^{+99}_{-99}$ & $3.93^{+0.02}_{-0.02}$ & $0.19^{+0.08}_{-0.08}$ & $1.51^{+0.04}_{-0.04}$ & $2.20^{+0.07}_{-0.07}$ & $9.40^{+0.04}_{-0.04}$ & CKS & $6217$ & $4.07$ & $0.25$\\
9704149 & $5679^{+178}_{-168}$ & $4.47^{+0.03}_{-0.04}$ & $-0.0^{+0.1}_{-0.2}$ & $0.96^{+0.06}_{-0.07}$ & $0.94^{+0.03}_{-0.03}$ & $9.7^{+0.3}_{-0.5}$ & \nodata & \nodata & \nodata & \nodata\\
9822143 & $5379^{+100}_{-80}$ & $3.75^{+0.03}_{-0.02}$ & $-0.47^{+0.04}_{-0.02}$ & $0.91^{+0.04}_{-0.03}$ & $2.10^{+0.04}_{-0.04}$ & $10.04^{+0.06}_{-0.08}$ & LAMOST & $5392$ & $3.63$ & $-0.67$\\
9838291 & $6094^{+102}_{-109}$ & $3.84^{+0.02}_{-0.03}$ & $0.10^{+0.07}_{-0.06}$ & $1.51^{+0.04}_{-0.04}$ & $2.45^{+0.06}_{-0.05}$ & $9.44^{+0.04}_{-0.04}$ & LAMOST & $6024$ & $3.83$ & $0.10$\\
9970525 & $6305^{+103}_{-103}$ & $4.27^{+0.03}_{-0.03}$ & $-0.3^{+0.1}_{-0.1}$ & $1.04^{+0.05}_{-0.05}$ & $1.24^{+0.02}_{-0.02}$ & $9.7^{+0.1}_{-0.1}$ & LAMOST & $6301$ & $4.23$ & $-0.39$\\
10024862 & $5983^{+106}_{-103}$ & $4.38^{+0.06}_{-0.06}$ & $-0.08^{+0.08}_{-0.09}$ & $1.04^{+0.05}_{-0.05}$ & $1.09^{+0.07}_{-0.07}$ & $9.7^{+0.2}_{-0.3}$ & SPOCS & $5979$ & $4.38$ & $-0.12$\\
10058021 & $4761^{+113}_{-102}$ & $4.59^{+0.02}_{-0.02}$ & $0.0^{+0.1}_{-0.1}$ & $0.76^{+0.03}_{-0.03}$ & $0.73^{+0.01}_{-0.01}$ & $9.9^{+0.2}_{-0.4}$ & \nodata & \nodata & \nodata & \nodata\\
10187159 & $5026^{+72}_{-62}$ & $4.55^{+0.03}_{-0.02}$ & $0.10^{+0.08}_{-0.08}$ & $0.82^{+0.04}_{-0.03}$ & $0.80^{+0.01}_{-0.01}$ & $9.9^{+0.2}_{-0.3}$ & HDS & $5000$ & $4.48$ & $0.13$\\
10190048$^b$ & $4995^{+84}_{-89}$ & $4.53^{+0.02}_{-0.01}$ & $0.1^{+0.1}_{-0.1}$ & $0.81^{+0.03}_{-0.02}$ & $0.81^{+0.02}_{-0.02}$ & $10.04^{+0.06}_{-0.16}$ & \nodata & \nodata & \nodata & \nodata\\
10207400 & $5679^{+103}_{-98}$ & $4.37^{+0.04}_{-0.04}$ & $0.07^{+0.08}_{-0.07}$ & $0.97^{+0.05}_{-0.04}$ & $1.07^{+0.04}_{-0.03}$ & $9.9^{+0.1}_{-0.2}$ & HDS & $5693$ & $4.28$ & $0.10$\\
10255705 & $5133^{+103}_{-99}$ & $3.69^{+0.04}_{-0.04}$ & $0.01^{+0.03}_{-0.02}$ & $1.11^{+0.16}_{-0.06}$ & $2.5^{+0.1}_{-0.1}$ & $9.89^{+0.09}_{-0.22}$ & APOGEE & $5110$ & $3.71$ & $-0.14$\\
10284575 & $6473^{+106}_{-114}$ & $4.15^{+0.02}_{-0.02}$ & $0.09^{+0.08}_{-0.07}$ & $1.35^{+0.03}_{-0.03}$ & $1.63^{+0.03}_{-0.03}$ & $9.40^{+0.06}_{-0.07}$ & HDS & $6506$ & $3.83$ & $0.13$\\
10287723 & $4343^{+49}_{-46}$ & $4.68^{+0.01}_{-0.01}$ & $-0.43^{+0.09}_{-0.08}$ & $0.59^{+0.02}_{-0.01}$ & $0.582^{+0.006}_{-0.006}$ & $9.9^{+0.2}_{-0.4}$ & APOGEE & $4227$ & $4.60$ & $-0.56$\\
10321319 & $5598^{+101}_{-92}$ & $4.08^{+0.02}_{-0.02}$ & $0.02^{+0.05}_{-0.04}$ & $1.00^{+0.03}_{-0.03}$ & $1.51^{+0.02}_{-0.03}$ & $10.02^{+0.05}_{-0.07}$ & LAMOST & $5595$ & $3.93$ & $-0.10$\\
10384911 & $6065^{+100}_{-106}$ & $4.20^{+0.03}_{-0.03}$ & $-0.30^{+0.09}_{-0.10}$ & $0.98^{+0.05}_{-0.05}$ & $1.31^{+0.03}_{-0.03}$ & $9.86^{+0.08}_{-0.10}$ & LAMOST & $6031$ & $4.24$ & $-0.37$\\
10403228 & $3942^{+81}_{-55}$ & $4.69^{+0.01}_{-0.01}$ & $0.0^{+0.1}_{-0.2}$ & $0.59^{+0.01}_{-0.01}$ & $0.574^{+0.009}_{-0.010}$ & $9.9^{+0.2}_{-0.4}$ & \nodata & \nodata & \nodata & \nodata\\
10460629 & $6222^{+80}_{-74}$ & $3.90^{+0.02}_{-0.02}$ & $0.08^{+0.08}_{-0.06}$ & $1.49^{+0.03}_{-0.04}$ & $2.26^{+0.06}_{-0.06}$ & $9.43^{+0.04}_{-0.03}$ & CKS & $6140$ & $4.12$ & $0.06$\\
10525077 & $6140^{+207}_{-235}$ & $4.37^{+0.04}_{-0.05}$ & $-0.0^{+0.2}_{-0.2}$ & $1.11^{+0.07}_{-0.09}$ & $1.15^{+0.05}_{-0.04}$ & $9.5^{+0.3}_{-0.5}$ & \nodata & \nodata & \nodata & \nodata\\
10602068 & $5540^{+158}_{-133}$ & $4.52^{+0.03}_{-0.04}$ & $-0.0^{+0.2}_{-0.2}$ & $0.92^{+0.04}_{-0.06}$ & $0.87^{+0.02}_{-0.02}$ & $9.6^{+0.3}_{-0.5}$ & \nodata & \nodata & \nodata & \nodata\\
10683701 & $5736^{+99}_{-93}$ & $3.98^{+0.02}_{-0.02}$ & $-0.46^{+0.05}_{-0.03}$ & $0.90^{+0.03}_{-0.02}$ & $1.61^{+0.03}_{-0.03}$ & $10.03^{+0.05}_{-0.06}$ & LAMOST & $5708$ & $4.00$ & $-0.63$\\
10724544 & $5735^{+210}_{-208}$ & $3.95^{+0.05}_{-0.04}$ & $0.10^{+0.11}_{-0.07}$ & $1.21^{+0.11}_{-0.07}$ & $1.92^{+0.08}_{-0.08}$ & $9.7^{+0.1}_{-0.1}$ & \nodata & \nodata & \nodata & \nodata\\
10842718 & $5446^{+89}_{-86}$ & $4.49^{+0.03}_{-0.03}$ & $0.01^{+0.08}_{-0.08}$ & $0.90^{+0.04}_{-0.04}$ & $0.89^{+0.02}_{-0.02}$ & $9.8^{+0.2}_{-0.3}$ & HDS & $5463$ & $4.45$ & $0.02$\\
10960865 & $5350^{+100}_{-87}$ & $3.80^{+0.03}_{-0.03}$ & $0.18^{+0.07}_{-0.07}$ & $1.22^{+0.07}_{-0.05}$ & $2.30^{+0.06}_{-0.06}$ & $9.76^{+0.06}_{-0.08}$ & HDS & $5322$ & $3.93$ & $0.26$\\
10976409 & $6708^{+103}_{-104}$ & $3.93^{+0.03}_{-0.03}$ & $0.17^{+0.08}_{-0.08}$ & $1.66^{+0.04}_{-0.04}$ & $2.33^{+0.08}_{-0.08}$ & $9.21^{+0.04}_{-0.05}$ & LAMOST & $6717$ & $3.95$ & $0.24$\\
11342550 & $5628^{+99}_{-95}$ & $4.38^{+0.04}_{-0.04}$ & $0.04^{+0.08}_{-0.08}$ & $0.95^{+0.04}_{-0.04}$ & $1.04^{+0.04}_{-0.04}$ & $9.9^{+0.1}_{-0.2}$ & CKS & $5623$ & $4.33$ & $0.04$\\
11558724 & $6079^{+102}_{-117}$ & $4.24^{+0.03}_{-0.03}$ & $0.06^{+0.07}_{-0.07}$ & $1.14^{+0.04}_{-0.04}$ & $1.34^{+0.04}_{-0.04}$ & $9.7^{+0.1}_{-0.1}$ & LAMOST & $6065$ & $4.26$ & $0.07$\\
11709124 & $5675^{+97}_{-100}$ & $4.36^{+0.03}_{-0.03}$ & $0.04^{+0.07}_{-0.08}$ & $0.96^{+0.04}_{-0.04}$ & $1.07^{+0.03}_{-0.03}$ & $9.9^{+0.1}_{-0.1}$ & CKS & $5666$ & $4.35$ & $0.02$\\
12066509 & $5826^{+93}_{-99}$ & $4.22^{+0.04}_{-0.04}$ & $0.13^{+0.08}_{-0.08}$ & $1.06^{+0.05}_{-0.04}$ & $1.33^{+0.05}_{-0.05}$ & $9.86^{+0.09}_{-0.12}$ & LAMOST & $5844$ & $4.10$ & $0.18$\\
12266600 & $5231^{+179}_{-139}$ & $4.57^{+0.02}_{-0.03}$ & $-0.1^{+0.2}_{-0.2}$ & $0.83^{+0.04}_{-0.04}$ & $0.78^{+0.02}_{-0.02}$ & $9.6^{+0.3}_{-0.5}$ & \nodata & \nodata & \nodata & \nodata\\
12356617 & $5829^{+84}_{-104}$ & $4.06^{+0.03}_{-0.02}$ & $0.17^{+0.07}_{-0.08}$ & $1.15^{+0.09}_{-0.04}$ & $1.66^{+0.03}_{-0.03}$ & $9.80^{+0.06}_{-0.12}$ & CKS & $5804$ & $4.03$ & $0.21$\\
12454613 & $5424^{+80}_{-89}$ & $4.51^{+0.02}_{-0.03}$ & $0.16^{+0.08}_{-0.08}$ & $0.94^{+0.04}_{-0.04}$ & $0.89^{+0.01}_{-0.01}$ & $9.6^{+0.3}_{-0.4}$ & LAMOST & $5448$ & $4.53$ & $0.24$ \\
\enddata
\tablenotetext{\dagger}{Sources of spectroscopic parameters: HDS --- this paper, CKS --- \citet{2017AJ....154..107P}, SPOCS --- \citet{2016ApJS..225...32B}, LAMOST --- LAMOST DR4 \citep{2014IAUS..306..340W}, APOGEE --- \citet{2018arXiv180401530T}, M17 --- \citet{2017ApJS..229...30M}.}
\tablenotetext{\textit{b}}{\ Flagged as photometric binary by \citet{2018ApJ...866...99B}.}
\tablenotetext{\textit{ab}}{\ Flagged as AO binary by \citet{2018ApJ...866...99B}.}
\tablecomments{The reported values and errors are medians and $15.87$th/$84.13$th percentiles of the marginal posterior.}
\end{deluxetable*}

\section{The Planet Candidates}\label{sec:planets}

Here we model the light curves of 93 candidates that passed the vetting in Section \ref{sec:fp}. We derive the parameters of the transiting objects including the radius and orbital period based on the stellar parameters determined in Section \ref{sec:stars}. The results are then used to further filter out potential stellar binaries to obtain the clean sample of planet candidates. 


\begin{figure*}
    \epsscale{1.2}
    \plotone{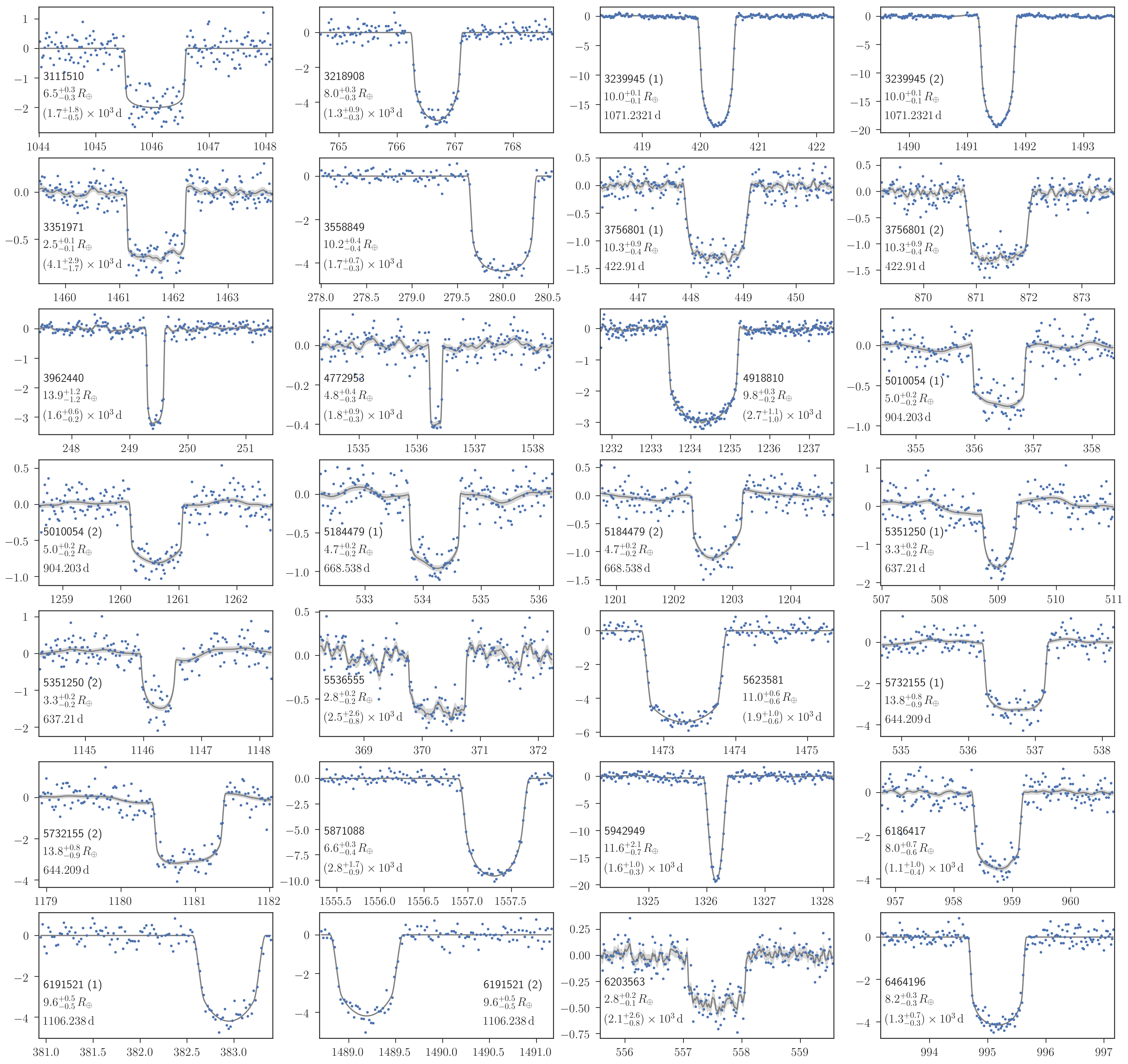}
    \caption{Transit light curves of the candidates in the clean sample (1/4). The vertical axis shows the relative flux deviation from unity, in units of $10^{-3}$. The horizontal axis shows the time $\mathrm{BJD}_{\rm TDB}-2454833$. The blue dots are the PDC data, the gray solid line shows the maximum a posteriori model, and the gray shaded region shows the $1\sigma$ prediction. The best-fit polynomial trend (see Section \ref{ssec:planets_model}) is removed here. When two transits are observed, the transit number is shown in the parentheses following the KIC names; here ``3" means that there is a gap in the middle of the two observed transits and that the second observed transit was assumed to be the third one. Shown below the KIC names are the radius and period estimated from the light curve modeling.}
    \label{fig:clean1}
\end{figure*}
\begin{figure*}
    \epsscale{1.2}
    \plotone{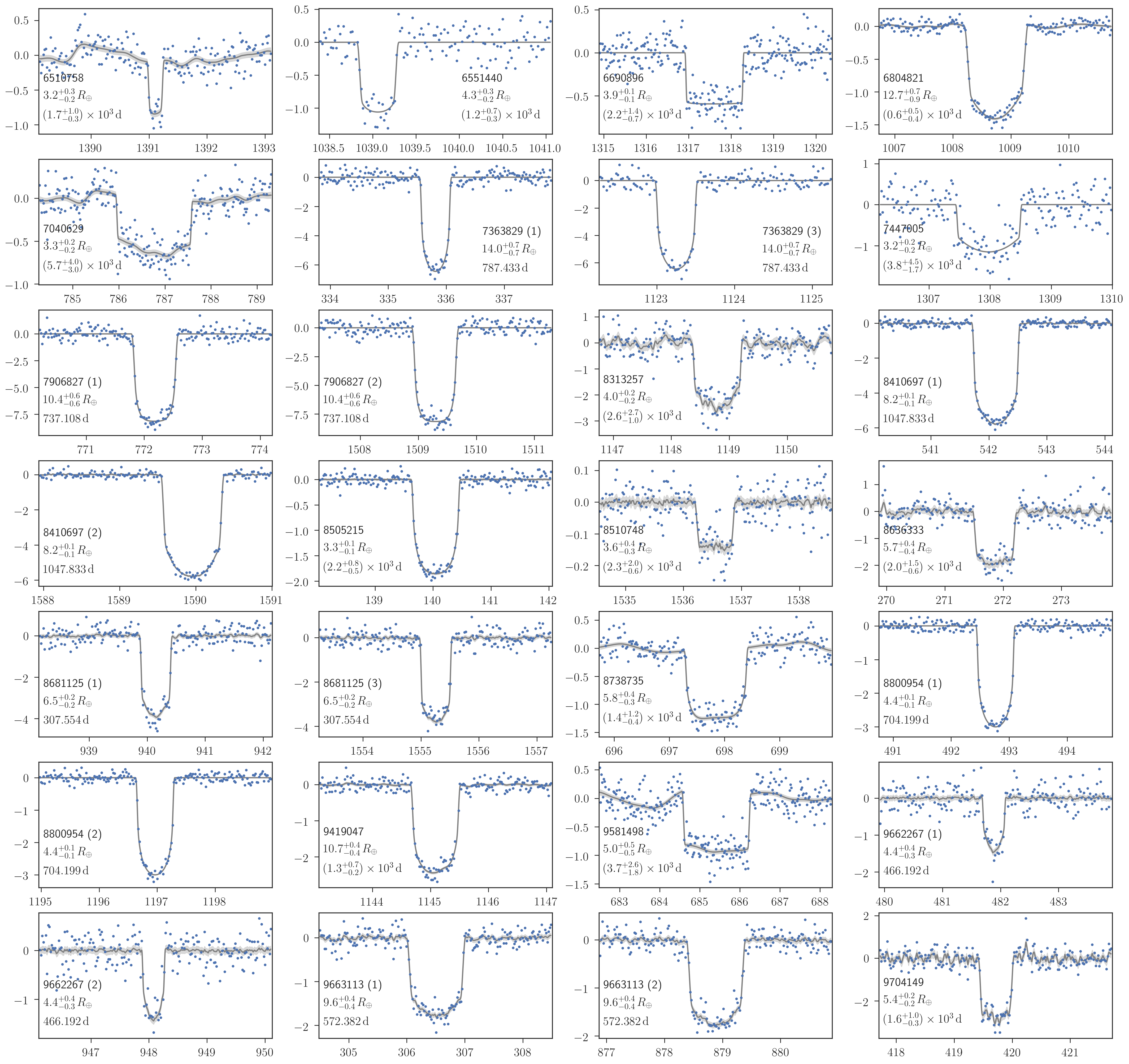}
    \caption{Transit light curves of the candidates in the clean sample (2/4). Same as Figure \ref{fig:clean1}.}
    \label{fig:clean2}
\end{figure*}
\begin{figure*}
    \epsscale{1.2}
    \plotone{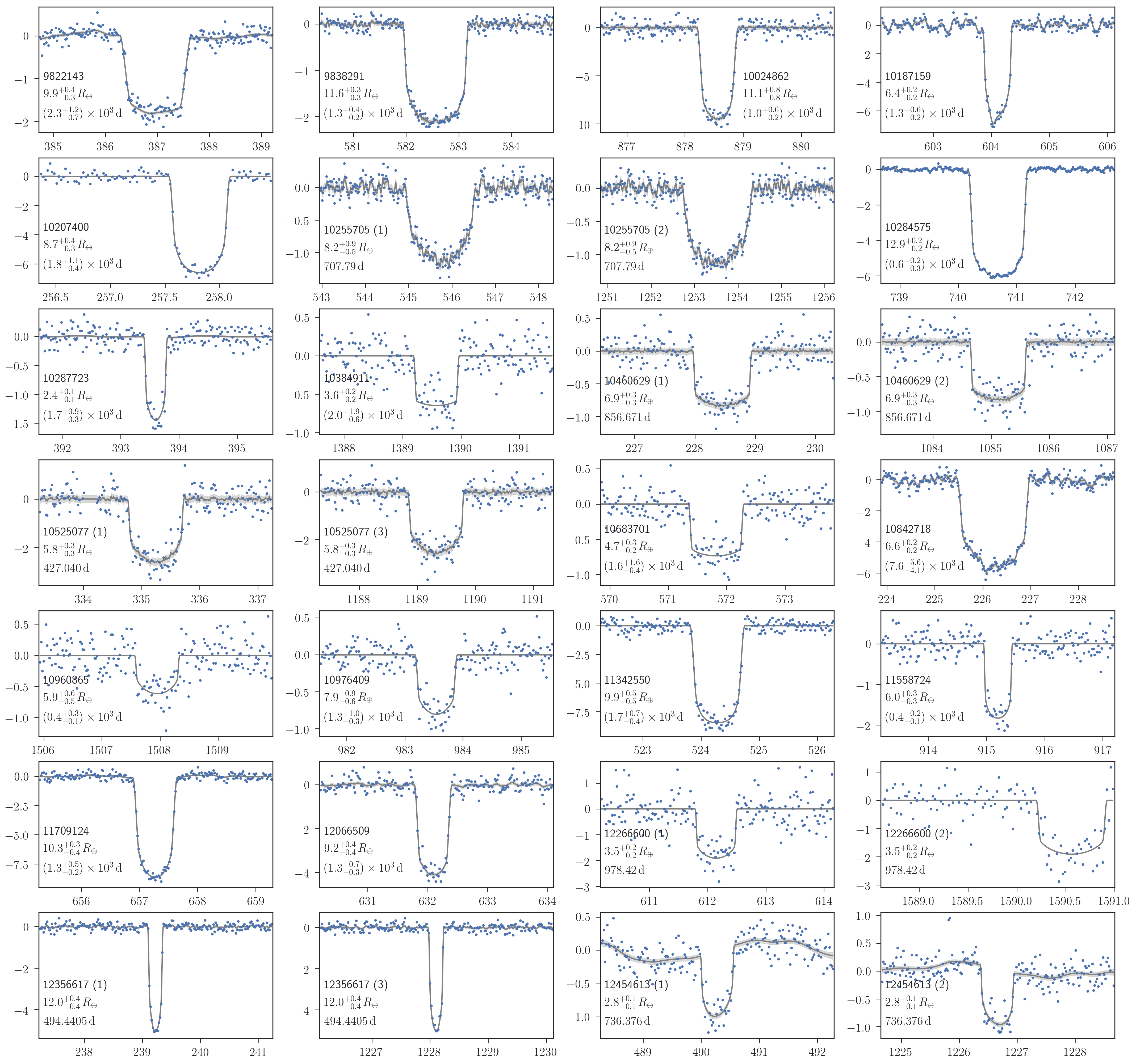}
    \caption{Transit light curves of the candidates in the clean sample (3/4). Same as Figure \ref{fig:clean1}.}
    \label{fig:clean3}
\end{figure*}
\begin{figure*}
    \epsscale{1.1}
    \plotone{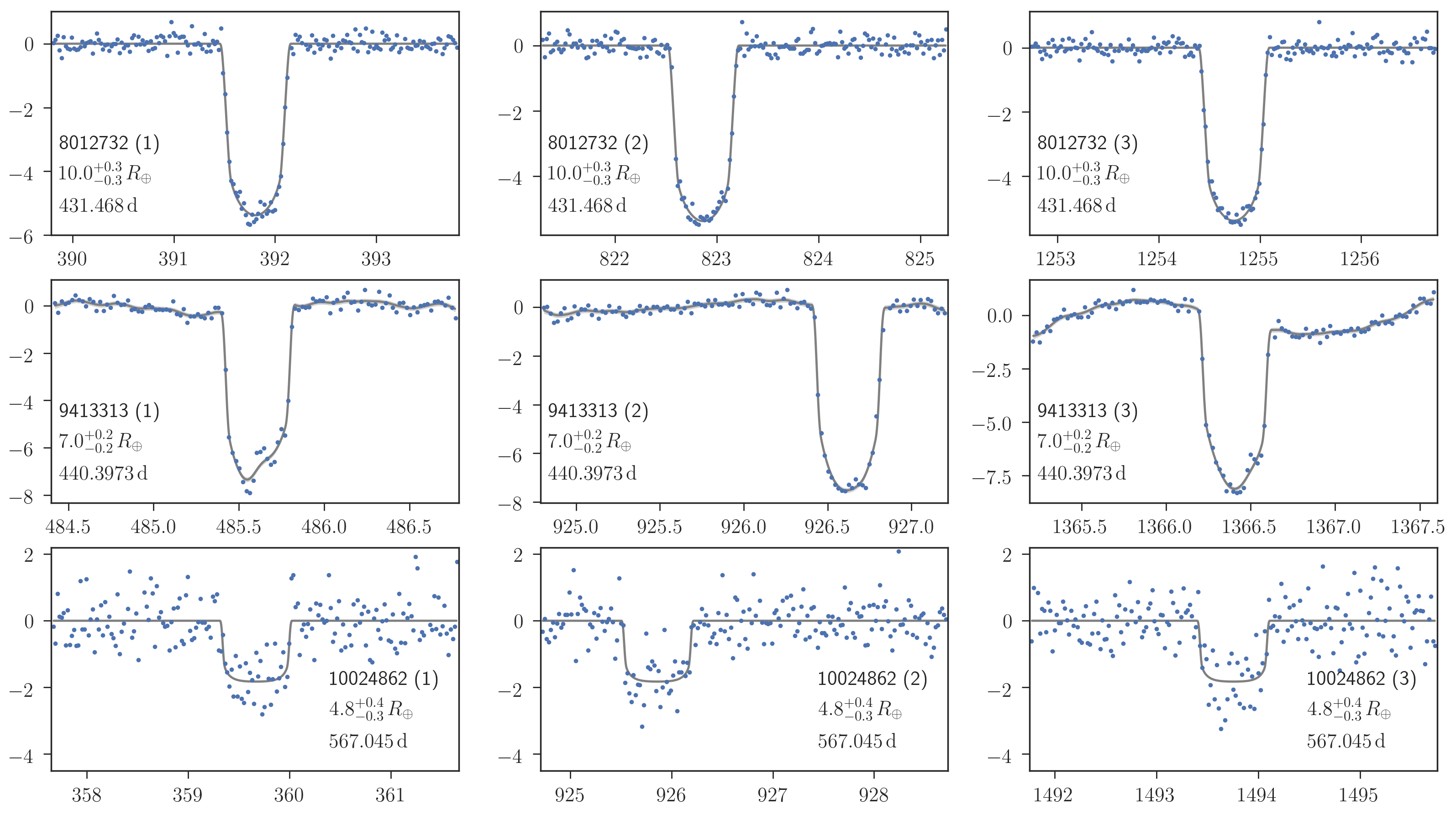}
    \caption{Transit light curves of the candidates in the clean sample (4/4). Same as Figure \ref{fig:clean1}, but here the candidates with three transits are shown. The second transits are all shifted from the center of the panels because of transit timing variations. The light curves of KIC 9413313 exhibits signatures of spot-induced modulations and spot crossing.}
    \label{fig:clean4}
\end{figure*}

\subsection{Modeling of the Light Curves}\label{ssec:planets_model}

We used the pre-search data conditioning (PDC) light curves downloaded from the Mikulski Archive for Space Telescopes.\footnote{\url{https://archive.stsci.edu}} For KIC 6804821 and KIC 10284575, the coherent pulsations were removed as described in Appendix \ref{sec:pulsation}. For each candidate, we extracted the data within $\mathrm{max}(2\,\mathrm{days},\ W)$ of the detected events (see Figure \ref{fig:trapezoid} for the definition of $W$), and removed $3\sigma$ outliers using a median filter. When any other shorter-period planet in the KOI catalog is transiting inside the extracted time window, that part was removed.

We modeled the light curves of each candidate as the sum of the mean model and the noise. The mean model $\bm{m}$ is a product of the transit model  assuming a quadratic limb-darkening law \citep{2002ApJ...580L.171M} computed with {\tt batman} \citep{ 2015PASP..127.1161K} and a second-order polynomial function of time to account for the longer-term trend. The noise was modeled as a Gaussian process whose covariance $K$ consists of a Mat\'ern-3/2 covariance and a white-noise term. The log-likelihood of the model $\ln\mathcal{L}$,
\begin{equation}
    \ln\mathcal{L} = -{1\over2}(\bm{f}-\bm{m})^T K^{-1} (\bm{f}-\bm{m})-{1\over 2}\ln\mathrm{det}\,K,
\end{equation}
was computed using {\tt celerite} \citep{celerite}.
Here $\bm{f}$ represents the observed PDC flux values and the covariance matrix $K$ is given by
\begin{equation}
    K_{ij}=(\sigma_i^2+\sigma_{\rm jit}^2)\delta_{ij}
    +\alpha^2 \left(1+{|t_i-t_j|\over 3\rho}\right)
    \exp\left(-{|t_i-t_j|\over 3\rho}\right),
\end{equation}
where $\delta_{ij}$ is the Kronecker delta, $\sigma_i$ is the PDC error of the $i$th measurement, and $\sigma_{\rm jit}$ models an additional white noise component that is not taken into account in $\sigma_i$.

The mean transit model has the following parameters: limb-darkening coefficients $q_1$ and $q_2$ as parametrized in \citet{2013MNRAS.435.2152K}, logarithm of planet-to-star radius ratio $\ln (r/R_\star)$, logarithm of mean stellar density $\ln \rho_\star$, transit impact parameter $b$, time of inferior conjunction $t_0$, logarithm of orbital period $\ln P$, eccentricity and argument of periastron $\sqrt{e}\cos\omega$ and $\sqrt{e}\sin\omega$, and the coefficients for the polynomial. 

We adopted uniform priors for the above parameters, except for the eccentricity (for which we adopted the beta distribution prior with $a=1.12$ and $b=3.09$; \citealt{2013MNRAS.434L..51K}), mean stellar density (for which we adopted a Gaussian prior based on the constraint in Section \ref{sec:stars}), and log orbital period (for which we adopted $-2\ln P/3$ so that the prior on $P$ is $\propto P^{-5/3}$).\footnote{The choice of $P^{-5/3}$ is motivated by the transit probability $\propto P^{-2/3}$ and the probability for a transit to occur within a finite baseline $\propto P^{-1}$ \citep[cf.][]{2018RNAAS...2d.223K}. So this analysis assumes the intrinsic prior uniform in $P$, and is not conditioned on the information that some of the candidates have inner transiting planets.} When only one transit is observed, the lower and upper bounds for the log period were set to be $\ln P_{\rm min}$ and $\ln(30000\,\mathrm{days})$, respectively, where $P_{\rm min}$ is the longer time interval between the observed event time and the two edges of the data. When two transits are observed, the limit was chosen to be $\pm0.5\,\mathrm{days}$ from the interval of the two transits, and we also fit additional three polynomial coefficients for the second event. If the data in the middle of the two transits are missing, we assumed that the true period is half the interval, because that case is a priori more likely. For TTEs, we fit yet another three polynomial coefficients, and floated the central time of the second transit to take into account transit timing variations, which turned out to be significant for all the TTEs analyzed here. The orbital period in this case corresponds to half the interval between the first and third observed transits.

The posterior samples for the transit and noise parameters $(\ln\alpha, \ln\rho, \ln\sigma_{\rm jit})$ were obtained simultaneously using a Markov chain Monte Carlo algorithm, as implemented in {\tt emcee} \citep{2013PASP..125..306F}. The resulting constraints are summarized in Tables \ref{tab:clean} and \ref{tab:flagged} separately for the clean and flagged candidates (see Section \ref{ssec:planets_prune}), respectively. We use the median and 15.87th/84.13th percentiles of marginal posteriors as the summary. The light curves and the models are shown in Figures \ref{fig:clean1}--\ref{fig:clean4} (for the clean sample) and Figures \ref{fig:lc_sec}--\ref{fig:lc_large} (for the flagged targets), after removing the polynomial trends.

In this analysis, 
we excluded STEs in KIC 2158850 and 6145201 and the second of the DTEs in KIC 8636333, based on the comparison of the models with and without a transit. For all the candidate events, we computed the Bayesian information criterion (BIC) 
\begin{equation}
    \mathrm{BIC}=-2\ln\mathcal{L}_{\rm max}+k\ln N,
\end{equation}
where $k$ and $N$ are the numbers of free parameters and data points, respectively, for the maximum likelihood models with and without a transit. We found that the polynomial model has smaller BIC values than the transit model for the above three events, and so they were excluded as insignificant events.

\subsection{Pruning the Sample}\label{ssec:planets_prune}

The combination of the stellar radius derived in Section \ref{sec:stars} and the modeling above revealed some candidates whose inferred radii are too large to be planets; these are obvious stellar binaries. In addition, our sample may be contaminated by secondary eclipses with depths similar to that of a planetary transit. 
Here we flag these events to obtain the cleaner sample of planet candidates, as described below in detail. The discussion in Section \ref{sec:planets_clean} focuses only on this sample (Table \ref{tab:clean}), although the parameters of the flagged systems are also reported in the catalog (Table \ref{tab:flagged}).

\begin{figure}
    \centering
    \epsscale{1.15}
    \plotone{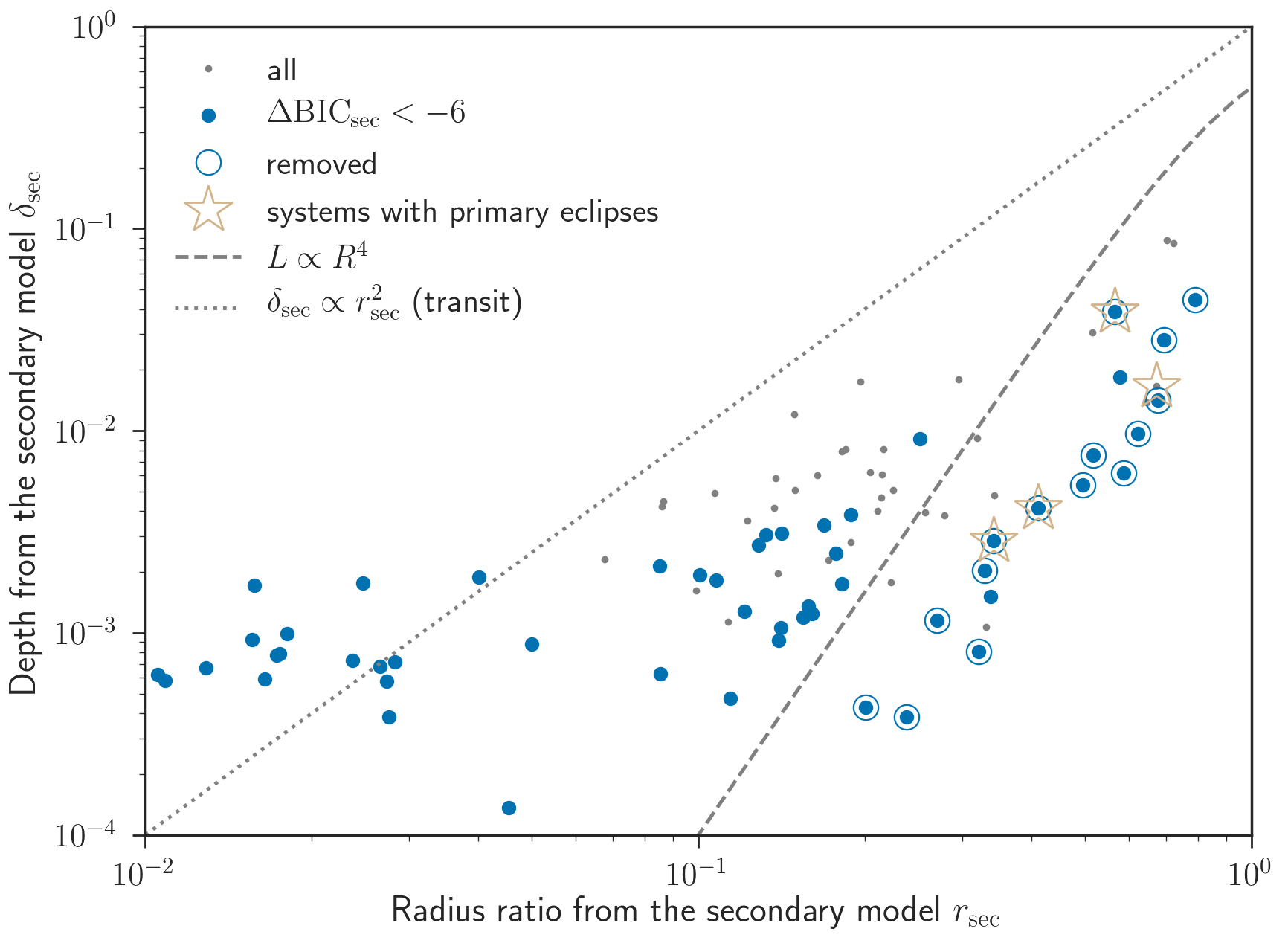}
    \plotone{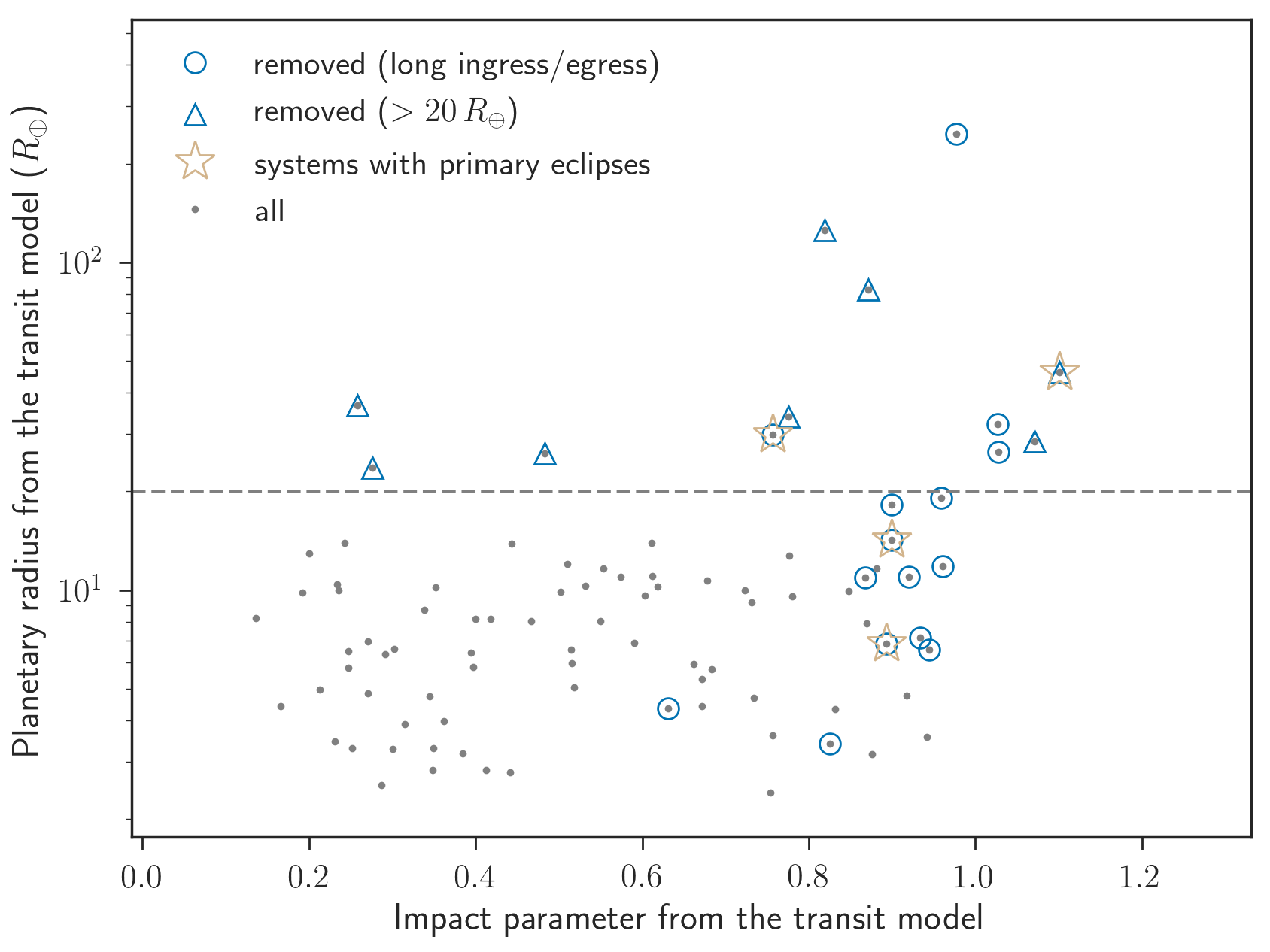}
    \caption{Two cuts to flag potential stellar binaries; see Section \ref{ssec:planets_prune} for details. {\it Upper panel} --- The depth $\delta_{\rm sec}$ and radius ratio $r_{\rm sec}$ from the secondary eclipse model. The gray dots show all the candidates and blue filled circles are the ones with $\Delta\mathrm{BIC}_{\rm sec}<-6$ (i.e. secondary model is at least as good as the transit model). The dashed line is the $\delta_{\rm sec}$--$r_{\rm sec}$ relation assuming $L_\star \sim R_\star^4$. The dotted line corresponds to $\delta_{\rm sec}=r_{\rm sec}^2$, which is expected for the transit case. The flagged systems are shown with outer circles, and the candidates with additional primary eclipses are marked by star symbols.
    {\it Bottom panel} --- Planetary radii and impact parameters based on the transit model and stellar radii derived in Section \ref{sec:stars}. The meaning of the gray dots, blue open circles, and stars are the same as in the upper panel. The triangles show the candidates that were not flagged in the upper panel but have radii greater than $20\,R_\oplus$ (i.e. stellar size).}
    \label{fig:secflag}
\end{figure}

To flag secondary eclipses with depths similar to that of a planetary transit, we use the relation between the ingress/egress duration and eclipse depth.
Considering that the luminosity $L_\star$ of main-sequence stars roughly scales as $L_\star\sim R_\star^4$, the secondary eclipses with depths $0.01$--$1\%$ roughly correspond to binaries with radius ratios of $0.1$--$0.3$. These ratios are larger than the corresponding radius ratios in the transit case ($0.01$--$0.1$), and so these events would have longer ingress/egress than expected for a planetary transit. We pick up such events as follows:
\begin{enumerate}
    \item We fit the light curves of each system using a model for secondary eclipses, where the limb-darkening coefficients are set to be zero and the depth $\delta_{\rm sec}$ is an additional free parameter. We also fixed the impact parameter to be zero to obtain an upper limit on the radius ratio allowed from ingress/egress duration, $r_{\rm sec}$. We then computed the BIC difference from the transit model, $\dbic_{\rm sec}$, for the maximum likelihood model. We picked up ``flat-bottomed" events as those with $\dbic_{\rm sec}<-6$. 
    \item A part of the ``flat-bottomed" events thus selected showed a distinct clustering in the $\delta_{\rm sec}$--$r_{\rm sec}$ plane (Figure \ref{fig:secflag}, upper panel). We removed all the events with $\delta_{\rm sec}>r_{\rm sec}^4/(1+r_{\rm sec}^4)$ in the lower right region. Because $r_{\rm sec}$ is an upper limit on the secondary radius, this is a conservative thresholding, and the other events are consistent with the planetary transit. As exceptions, we did not flag the candidates with inner confirmed KOIs \citep[KIC 5351250, KOI-408, Kepler-150;][]{2017AJ....153..180S} or with inner candidate KOIs with false-positive probabilities less than $10\%$ (KIC 5942949, KOI-2525), because these events in multis are less likely to be false positives \citep{2012ApJ...750..112L}.
\end{enumerate}
This cut removed $15$ systems. As shown in the bottom panel of Figure \ref{fig:secflag}, the cut turned out to be almost equivalent to excluding candidates with both large radii and impact parameters. This feature is consistent with what we expect for stellar eclipses.

As the second cut, we removed additional eight systems whose inferred radii based on the transit model exceed $20\,R_\oplus$, but whose light curves were not considered to be flat-bottomed in the first cut. 
These two cuts leaves us with $67$ ``clean" candidates in Table \ref{tab:clean}. Although the second cut on the radius is not so strict, other large candidates have already been removed in the first cut, and the largest object remaining in the clean sample has $r\approx14\,R_\oplus\approx1.25\,R_{\rm Jup}$. This value agrees well with the maximum radius of known planets with insolation flux lower than $10^7\,\mathrm{erg\,cm^{-2}\,s^{-1}}$ and radii measured to better than $10\%$.

The light curves of KIC 3526901, 4042088, 4754460, and 7947784 exhibit deeper, stellar eclipses in addition to the STEs in the input catalog, suggesting that the latter events are secondary eclipses. All of these systems, which are shown with star symbols in Figure \ref{fig:secflag}, have been successfully removed from the clean sample by the above criteria.

\subsubsection{How Clean is the Clean Sample?}

Our precise stellar radii allowed us to remove stellar eclipses caused by objects $>20\,R_\oplus$ reliably. We also attempted to flag possible secondary eclipses that produce shallower eclipses. Are the remaining candidates all likely to be planets?

First we examine the number of events flagged as potential secondary eclipses above. Considering that $L_\star\sim M_\star^4$ for main-sequence stars, companions with mass ratios $q\sim0.1$--$0.3$ can produce secondary eclipses with $0.01$--$1\%$. Assuming the log-normal period distribution in \cite{2010ApJS..190....1R}, binary fraction of $0.5$, and $1.3\,R_\star$ primary (median of KIC stars), EB occurrence in our search range is $\approx 2\times10^{-4}$. Thus the expected number of EBs with mass ratio $<0.3$ in the whole \kepler\ sample would be $\sim 2\times10^{-4}\times0.3\times 2\times10^{5}\sim10$ assuming flat mass-ratio distribution. This is comparable to the number of objects (15) flagged as potential secondary eclipses, serving as a sanity check of our procedure. 

Primary eclipses due to companions later than M7 ($\lesssim0.1\,M_\odot$) or brown-dwarfs can also be confused as Jupiter-sized planets, and they are basically indistinguishable based on the light curve alone. However, binaries with corresponding mass ratios ($\lesssim0.1$) or brown dwarfs \citep{2006ApJ...640.1051G} are even rarer than the above estimate. Thus the contaminations from these objects are likely $\mathcal{O}(1)$, if any. 

Another potential source of confusion is a secondary eclipse of a WD companion. Although the host stars in our sample are typically older than a few Gyr, some WD companions, if any, may still have luminosities of $\sim10^{-3}L_\odot$ and contaminate our sample. Since their physical radii are small, these cases should be at the leftward of the transit line in the upper panel of Figure \ref{fig:secflag}. \citet{2018MNRAS.474.4322M} estimated the occurrence of WD companions around A/F dwarfs to be $\sim3\%$ for the relevant period range, which is about one fourth of the occurrence of main-sequence companions estimated above. Thus, considering the luminosity function of WDs, some objects in the leftmost part of the upper panel of Figure \ref{fig:secflag} may be WDs. The objects in this area are marked with squares in Table \ref{tab:clean}, although they were not removed from the clean sample.

In summary, it is difficult to completely rule out planetary-depth eclipses caused by red-dwarf/WD companions, but we argue that their contamination is at most $\sim10\%$ in the clean sample. 

\startlongtable
\begin{deluxetable*}{lccccccccccccccccc}
\tablecaption{Parameters of the Systems in the Clean Sample.\label{tab:clean}}
\tablehead{
\colhead{KIC} & \colhead{KOI} & \colhead{$Kp$} & \colhead{$M_\star$ ($M_\odot$)} & \colhead{$R_\star$ ($R_\odot$)} & \colhead{$r$ ($R_\oplus$)} & \colhead{$t_0$ ($\mathrm{BKJD}$)} & \colhead{$P$ ($\mathrm{days}$)} & \colhead{$b$} & \colhead{$e$}
}
\startdata
3111510$^{}$ & \nodata & $14.8$ & $1.19^{+0.08}_{-0.08}$ & $1.36^{+0.06}_{-0.05}$ & $6.5^{+0.3}_{-0.3}$ & $1046.048^{+0.006}_{-0.005}$ & $(1.7^{+1.8}_{-0.5})\times10^3$ & $0.4^{+0.3}_{-0.3}$ & $0.2^{+0.2}_{-0.1}$\\
3218908$^{\dagger b}$ & 1108 (770) & $14.6$ & $0.96^{+0.04}_{-0.03}$ & $1.13^{+0.03}_{-0.03}$ & $8.0^{+0.3}_{-0.3}$ & $766.683^{+0.003}_{-0.003}$ & $(1.3^{+0.9}_{-0.3})\times10^3$ & $0.5^{+0.1}_{-0.3}$ & $0.1^{+0.2}_{-0.1}$\\
3239945$^{\dagger}$ & 490 (167) & $14.0$ & $0.78^{+0.03}_{-0.03}$ & $0.745^{+0.008}_{-0.008}$ & $10.0^{+0.1}_{-0.1}$ & $420.2869^{+0.0007}_{-0.0007}$ & $1071.2321^{+0.0010}_{-0.0009}$ & $0.24^{+0.06}_{-0.09}$ & $0.06^{+0.13}_{-0.04}$\\
3351971$^{\square}$ & \nodata & $12.7$ & $0.92^{+0.05}_{-0.06}$ & $0.90^{+0.02}_{-0.02}$ & $2.55^{+0.10}_{-0.08}$ & $1461.676^{+0.003}_{-0.004}$ & $(4^{+3}_{-2})\times10^3$ & $0.3^{+0.2}_{-0.2}$ & $0.2^{+0.2}_{-0.1}$\\
3558849$^{\dagger}$ & 4307 & $14.2$ & $1.14^{+0.05}_{-0.04}$ & $1.51^{+0.05}_{-0.04}$ & $10.2^{+0.4}_{-0.4}$ & $279.994^{+0.002}_{-0.002}$ & $(1.7^{+0.7}_{-0.3})\times10^3$ & $0.4^{+0.2}_{-0.2}$ & $0.54^{+0.09}_{-0.08}$\\
3756801$^{\dagger}$ & 1206$^*$ & $13.6$ & $1.3^{+0.2}_{-0.1}$ & $2.77^{+0.08}_{-0.08}$ & $10.3^{+0.9}_{-0.4}$ & $448.495^{+0.007}_{-0.007}$ & $422.91^{+0.01}_{-0.01}$ & $0.5^{+0.3}_{-0.3}$ & $0.2^{+0.2}_{-0.1}$\\
3962440$^{\dagger}$ & 1208$^*$ & $13.6$ & $1.55^{+0.07}_{-0.07}$ & $2.3^{+0.2}_{-0.2}$ & $14^{+1}_{-1}$ & $249.442^{+0.001}_{-0.002}$ & $(1.6^{+0.6}_{-0.2})\times10^3$ & $0.6^{+0.1}_{-0.2}$ & $0.92^{+0.02}_{-0.03}$\\
4772953$^{}$ & \nodata & $11.5$ & $2.2^{+0.1}_{-0.1}$ & $2.04^{+0.10}_{-0.09}$ & $4.8^{+0.4}_{-0.3}$ & $1536.321^{+0.003}_{-0.003}$ & $(1.8^{+0.9}_{-0.3})\times10^3$ & $0.92^{+0.03}_{-0.06}$ & $0.79^{+0.08}_{-0.10}$\\
4918810$^{\dagger}$ & \nodata & $13.4$ & $1.31^{+0.03}_{-0.07}$ & $1.78^{+0.04}_{-0.03}$ & $9.8^{+0.3}_{-0.2}$ & $1234.314^{+0.003}_{-0.003}$ & $(3^{+1}_{-1})\times10^3$ & $0.2^{+0.1}_{-0.1}$ & $0.2^{+0.2}_{-0.1}$\\
5010054$^{\dagger}$ & \nodata & $14.0$ & $1.16^{+0.03}_{-0.03}$ & $1.72^{+0.04}_{-0.04}$ & $5.0^{+0.2}_{-0.2}$ & $356.411^{+0.006}_{-0.005}$ & $904.203^{+0.008}_{-0.009}$ & $0.5^{+0.2}_{-0.3}$ & $0.2^{+0.1}_{-0.1}$\\
5184479$^{\dagger}$ & \nodata & $14.2$ & $1.13^{+0.05}_{-0.05}$ & $1.46^{+0.04}_{-0.04}$ & $4.7^{+0.2}_{-0.2}$ & $534.199^{+0.005}_{-0.005}$ & $668.538^{+0.008}_{-0.008}$ & $0.3^{+0.2}_{-0.2}$ & $0.11^{+0.15}_{-0.07}$\\
5351250$^{\dagger}$ & 408 (150) & $15.0$ & $0.96^{+0.03}_{-0.04}$ & $0.89^{+0.02}_{-0.02}$ & $3.3^{+0.2}_{-0.2}$ & $509.03^{+0.01}_{-0.01}$ & $637.21^{+0.02}_{-0.02}$ & $0.3^{+0.3}_{-0.2}$ & $0.1^{+0.2}_{-0.1}$\\
5536555$^{\dagger\square}$ & \nodata & $13.5$ & $1.06^{+0.05}_{-0.05}$ & $1.06^{+0.02}_{-0.02}$ & $2.8^{+0.2}_{-0.2}$ & $370.270^{+0.014}_{-0.008}$ & $(2.5^{+2.6}_{-0.8})\times10^3$ & $0.3^{+0.3}_{-0.2}$ & $0.2^{+0.2}_{-0.1}$\\
5623581$^{}$ & \nodata & $15.1$ & $1.21^{+0.08}_{-0.10}$ & $1.43^{+0.07}_{-0.07}$ & $11.0^{+0.6}_{-0.6}$ & $1473.287^{+0.003}_{-0.003}$ & $(1.9^{+1.0}_{-0.6})\times10^3$ & $0.57^{+0.08}_{-0.17}$ & $0.14^{+0.14}_{-0.09}$\\
5732155$^{}$ & \nodata & $15.2$ & $1.57^{+0.06}_{-0.06}$ & $2.3^{+0.1}_{-0.1}$ & $13.8^{+0.8}_{-0.9}$ & $536.704^{+0.005}_{-0.005}$ & $644.209^{+0.008}_{-0.008}$ & $0.4^{+0.2}_{-0.3}$ & $0.3^{+0.1}_{-0.1}$\\
5871088$^{}$ & \nodata & $15.6$ & $0.71^{+0.04}_{-0.04}$ & $0.67^{+0.03}_{-0.03}$ & $6.6^{+0.3}_{-0.4}$ & $1557.302^{+0.002}_{-0.002}$ & $(2.8^{+1.7}_{-0.9})\times10^3$ & $0.5^{+0.1}_{-0.2}$ & $0.2^{+0.1}_{-0.1}$\\
5942949$^{}$ & 2525 & $15.7$ & $0.73^{+0.04}_{-0.03}$ & $0.69^{+0.02}_{-0.02}$ & $11.6^{+2.1}_{-0.7}$ & $1326.160^{+0.001}_{-0.001}$ & $(1.6^{+1.0}_{-0.3})\times10^3$ & $0.88^{+0.07}_{-0.04}$ & $0.3^{+0.2}_{-0.1}$\\
6186417$^{}$ & \nodata & $15.4$ & $1.15^{+0.08}_{-0.09}$ & $1.31^{+0.09}_{-0.07}$ & $8.0^{+0.7}_{-0.6}$ & $958.751^{+0.006}_{-0.005}$ & $(1.1^{+1.0}_{-0.4})\times10^3$ & $0.5^{+0.2}_{-0.3}$ & $0.2^{+0.2}_{-0.1}$\\
6191521$^{\dagger}$ & 847 (700) & $15.2$ & $0.98^{+0.04}_{-0.03}$ & $1.34^{+0.06}_{-0.06}$ & $9.6^{+0.5}_{-0.5}$ & $382.951^{+0.004}_{-0.004}$ & $1106.238^{+0.006}_{-0.006}$ & $0.78^{+0.03}_{-0.04}$ & $0.10^{+0.15}_{-0.07}$\\
6203563$^{}$ & \nodata & $13.2$ & $1.19^{+0.06}_{-0.07}$ & $1.21^{+0.03}_{-0.02}$ & $2.8^{+0.2}_{-0.1}$ & $557.58^{+0.01}_{-0.01}$ & $(2.1^{+2.6}_{-0.8})\times10^3$ & $0.4^{+0.3}_{-0.3}$ & $0.2^{+0.2}_{-0.1}$\\
6464196$^{\dagger}$ & \nodata & $14.5$ & $1.01^{+0.04}_{-0.05}$ & $1.24^{+0.04}_{-0.03}$ & $8.2^{+0.3}_{-0.3}$ & $995.163^{+0.003}_{-0.003}$ & $(1.3^{+0.7}_{-0.3})\times10^3$ & $0.4^{+0.2}_{-0.2}$ & $0.1^{+0.1}_{-0.1}$\\
6510758$^{\dagger\square}$ & \nodata & $13.8$ & $0.91^{+0.04}_{-0.02}$ & $0.99^{+0.01}_{-0.01}$ & $3.2^{+0.3}_{-0.2}$ & $1391.113^{+0.005}_{-0.005}$ & $(1.7^{+1.0}_{-0.3})\times10^3$ & $0.88^{+0.05}_{-0.09}$ & $0.6^{+0.1}_{-0.2}$\\
6551440$^{\dagger}$ & \nodata & $13.6$ & $1.02^{+0.04}_{-0.04}$ & $1.17^{+0.02}_{-0.02}$ & $4.3^{+0.3}_{-0.2}$ & $1039.059^{+0.005}_{-0.005}$ & $(1.2^{+0.7}_{-0.3})\times10^3$ & $0.83^{+0.05}_{-0.11}$ & $0.3^{+0.2}_{-0.2}$\\
6690896$^{\dagger\square}$ & \nodata & $13.7$ & $1.01^{+0.05}_{-0.04}$ & $1.49^{+0.03}_{-0.03}$ & $3.9^{+0.1}_{-0.1}$ & $1317.612^{+0.007}_{-0.007}$ & $(2.2^{+1.4}_{-0.7})\times10^3$ & $0.3^{+0.3}_{-0.2}$ & $0.2^{+0.2}_{-0.1}$\\
6804821$^{}$ & \nodata & $10.6$ & $2.3^{+0.3}_{-0.2}$ & $3.0^{+0.1}_{-0.2}$ & $12.7^{+0.7}_{-0.9}$ & $1008.747^{+0.003}_{-0.004}$ & $(0.6^{+0.5}_{-0.4})\times10^3$ & $0.78^{+0.03}_{-0.04}$ & $0.2^{+0.3}_{-0.2}$\\
7040629$^{\dagger\square}$ & 671 (208) & $13.7$ & $1.11^{+0.04}_{-0.04}$ & $1.25^{+0.02}_{-0.02}$ & $3.3^{+0.2}_{-0.2}$ & $786.763^{+0.008}_{-0.007}$ & $(6^{+4}_{-3})\times10^3$ & $0.3^{+0.2}_{-0.2}$ & $0.2^{+0.2}_{-0.1}$\\
7363829$^{\dagger}$ & 1356$^*$ & $15.2$ & $1.15^{+0.09}_{-0.05}$ & $1.76^{+0.08}_{-0.08}$ & $14.0^{+0.7}_{-0.7}$ & $335.817^{+0.002}_{-0.002}$ & $787.433^{+0.003}_{-0.003}$ & $0.2^{+0.2}_{-0.2}$ & $0.68^{+0.03}_{-0.04}$\\
7447005$^{\square}$ & \nodata & $15.1$ & $0.95^{+0.06}_{-0.06}$ & $0.95^{+0.03}_{-0.03}$ & $3.2^{+0.2}_{-0.2}$ & $1307.98^{+0.01}_{-0.02}$ & $(4^{+4}_{-2})\times10^3$ & $0.4^{+0.3}_{-0.3}$ & $0.2^{+0.2}_{-0.1}$\\
7906827$^{}$ & \nodata & $15.7$ & $1.12^{+0.07}_{-0.08}$ & $1.14^{+0.06}_{-0.06}$ & $10.4^{+0.6}_{-0.6}$ & $772.185^{+0.002}_{-0.002}$ & $737.108^{+0.003}_{-0.003}$ & $0.2^{+0.2}_{-0.2}$ & $0.09^{+0.16}_{-0.06}$\\
8012732$^{\dagger}$ & 8151 (FP) & $13.9$ & $1.05^{+0.04}_{-0.04}$ & $1.26^{+0.03}_{-0.03}$ & $10.0^{+0.3}_{-0.3}$ & $391.807^{+0.002}_{-0.002}$ & $431.468^{+0.001}_{-0.001}$ & $0.72^{+0.02}_{-0.03}$ & $0.24^{+0.14}_{-0.06}$\\
8313257$^{}$ & \nodata & $15.4$ & $0.84^{+0.04}_{-0.04}$ & $0.80^{+0.02}_{-0.02}$ & $4.0^{+0.2}_{-0.2}$ & $1148.793^{+0.008}_{-0.008}$ & $(3^{+3}_{-1})\times10^3$ & $0.4^{+0.3}_{-0.2}$ & $0.2^{+0.2}_{-0.1}$\\
8410697$^{\dagger}$ & \nodata & $13.4$ & $0.93^{+0.04}_{-0.04}$ & $1.08^{+0.01}_{-0.01}$ & $8.2^{+0.1}_{-0.1}$ & $542.123^{+0.001}_{-0.001}$ & $1047.833^{+0.002}_{-0.002}$ & $0.14^{+0.12}_{-0.09}$ & $0.12^{+0.09}_{-0.04}$\\
8505215$^{\dagger ab}$ & 99 & $13.0$ & $0.75^{+0.02}_{-0.01}$ & $0.775^{+0.007}_{-0.007}$ & $3.28^{+0.09}_{-0.06}$ & $140.049^{+0.002}_{-0.002}$ & $(2.2^{+0.8}_{-0.5})\times10^3$ & $0.3^{+0.2}_{-0.2}$ & $0.11^{+0.12}_{-0.08}$\\
8510748$^{\dagger}$ & \nodata & $11.6$ & $1.66^{+0.04}_{-0.04}$ & $2.48^{+0.10}_{-0.09}$ & $3.6^{+0.4}_{-0.3}$ & $1536.55^{+0.01}_{-0.01}$ & $(2.3^{+2.0}_{-0.6})\times10^3$ & $0.94^{+0.02}_{-0.04}$ & $0.2^{+0.3}_{-0.2}$\\
8636333$^{\square}$ & 3349 (1475) & $15.3$ & $1.13^{+0.08}_{-0.08}$ & $1.19^{+0.06}_{-0.05}$ & $5.7^{+0.4}_{-0.4}$ & $271.87^{+0.01}_{-0.01}$ & $(2.0^{+1.5}_{-0.6})\times10^3$ & $0.7^{+0.1}_{-0.4}$ & $0.3^{+0.2}_{-0.2}$\\
8681125$^{\dagger}$ & \nodata & $15.0$ & $0.93^{+0.04}_{-0.02}$ & $1.04^{+0.03}_{-0.03}$ & $6.5^{+0.2}_{-0.2}$ & $940.149^{+0.003}_{-0.003}$ & $307.554^{+0.002}_{-0.002}$ & $0.2^{+0.2}_{-0.2}$ & $0.10^{+0.16}_{-0.06}$\\
8738735$^{\dagger}$ & 693 (214) & $13.9$ & $1.19^{+0.06}_{-0.04}$ & $1.61^{+0.04}_{-0.04}$ & $5.8^{+0.4}_{-0.3}$ & $697.856^{+0.006}_{-0.006}$ & $(1.4^{+1.2}_{-0.4})\times10^3$ & $0.4^{+0.3}_{-0.3}$ & $0.2^{+0.2}_{-0.1}$\\
8800954$^{\dagger ab}$ & 1274 (421)$^*$ & $13.4$ & $0.82^{+0.04}_{-0.03}$ & $0.82^{+0.01}_{-0.01}$ & $4.42^{+0.08}_{-0.07}$ & $492.767^{+0.001}_{-0.001}$ & $704.199^{+0.002}_{-0.002}$ & $0.2^{+0.2}_{-0.1}$ & $0.13^{+0.13}_{-0.07}$\\
9413313$^{\dagger}$ & \nodata & $14.1$ & $0.86^{+0.04}_{-0.03}$ & $0.82^{+0.01}_{-0.01}$ & $7.0^{+0.2}_{-0.2}$ & $485.611^{+0.001}_{-0.001}$ & $440.3973^{+0.0008}_{-0.0008}$ & $0.3^{+0.2}_{-0.2}$ & $0.25^{+0.07}_{-0.07}$\\
9419047$^{\dagger}$ & \nodata & $13.6$ & $1.27^{+0.05}_{-0.05}$ & $2.05^{+0.04}_{-0.04}$ & $10.7^{+0.4}_{-0.4}$ & $1145.086^{+0.003}_{-0.004}$ & $(1.3^{+0.7}_{-0.2})\times10^3$ & $0.68^{+0.09}_{-0.12}$ & $0.4^{+0.1}_{-0.1}$\\
9581498$^{\square}$ & 7194 (FP) & $14.2$ & $1.20^{+0.08}_{-0.11}$ & $1.4^{+0.1}_{-0.1}$ & $5.0^{+0.5}_{-0.5}$ & $685.417^{+0.005}_{-0.005}$ & $(4^{+3}_{-2})\times10^3$ & $0.2^{+0.2}_{-0.1}$ & $0.2^{+0.2}_{-0.1}$\\
9662267$^{\dagger}$ & \nodata & $14.9$ & $0.99^{+0.04}_{-0.04}$ & $1.13^{+0.03}_{-0.03}$ & $4.4^{+0.4}_{-0.3}$ & $481.886^{+0.006}_{-0.006}$ & $466.192^{+0.007}_{-0.008}$ & $0.7^{+0.1}_{-0.4}$ & $0.3^{+0.2}_{-0.2}$\\
9663113$^{\dagger}$ & 179 (458) & $14.0$ & $1.51^{+0.04}_{-0.04}$ & $2.20^{+0.07}_{-0.07}$ & $9.6^{+0.4}_{-0.4}$ & $306.507^{+0.004}_{-0.004}$ & $572.382^{+0.006}_{-0.006}$ & $0.60^{+0.08}_{-0.15}$ & $0.1^{+0.2}_{-0.1}$\\
9704149$^{}$ & \nodata & $15.1$ & $0.96^{+0.06}_{-0.07}$ & $0.94^{+0.03}_{-0.03}$ & $5.4^{+0.2}_{-0.2}$ & $419.720^{+0.004}_{-0.004}$ & $(1.6^{+1.0}_{-0.3})\times10^3$ & $0.67^{+0.09}_{-0.16}$ & $0.2^{+0.2}_{-0.2}$\\
9822143$^{\dagger}$ & \nodata & $13.8$ & $0.91^{+0.04}_{-0.03}$ & $2.10^{+0.04}_{-0.04}$ & $9.9^{+0.4}_{-0.3}$ & $386.963^{+0.005}_{-0.005}$ & $(2.3^{+1.2}_{-0.7})\times10^3$ & $0.85^{+0.02}_{-0.02}$ & $0.2^{+0.2}_{-0.1}$\\
9838291$^{\dagger}$ & \nodata & $12.9$ & $1.51^{+0.04}_{-0.04}$ & $2.45^{+0.06}_{-0.05}$ & $11.6^{+0.3}_{-0.3}$ & $582.562^{+0.002}_{-0.002}$ & $(1.3^{+0.4}_{-0.2})\times10^3$ & $0.55^{+0.08}_{-0.12}$ & $0.3^{+0.1}_{-0.1}$\\
10024862$^{\dagger}$ & \nodata & $15.9$ & $1.04^{+0.05}_{-0.05}$ & $1.09^{+0.07}_{-0.07}$ & $11.1^{+0.8}_{-0.8}$ & $878.561^{+0.003}_{-0.003}$ & $(1.0^{+0.6}_{-0.2})\times10^3$ & $0.61^{+0.08}_{-0.12}$ & $0.2^{+0.2}_{-0.1}$\\
10187159$^{\dagger}$ & 1870 (989) & $14.4$ & $0.82^{+0.04}_{-0.03}$ & $0.80^{+0.01}_{-0.01}$ & $6.4^{+0.2}_{-0.2}$ & $604.108^{+0.002}_{-0.002}$ & $(1.3^{+0.6}_{-0.2})\times10^3$ & $0.3^{+0.2}_{-0.2}$ & $0.4^{+0.1}_{-0.1}$\\
10207400$^{\dagger}$ & \nodata & $15.0$ & $0.97^{+0.05}_{-0.04}$ & $1.07^{+0.04}_{-0.03}$ & $8.7^{+0.4}_{-0.3}$ & $257.817^{+0.001}_{-0.001}$ & $(1.8^{+1.1}_{-0.4})\times10^3$ & $0.3^{+0.2}_{-0.2}$ & $0.56^{+0.09}_{-0.08}$\\
10255705$^{\dagger}$ & \nodata & $12.9$ & $1.11^{+0.16}_{-0.06}$ & $2.5^{+0.1}_{-0.1}$ & $8.2^{+0.9}_{-0.5}$ & $545.736^{+0.009}_{-0.008}$ & $707.79^{+0.01}_{-0.01}$ & $0.4^{+0.3}_{-0.2}$ & $0.2^{+0.2}_{-0.1}$\\
10284575$^{\dagger}$ & 3210 (FP) & $11.9$ & $1.35^{+0.03}_{-0.03}$ & $1.63^{+0.03}_{-0.03}$ & $12.9^{+0.2}_{-0.2}$ & $740.6801^{+0.0007}_{-0.0007}$ & $(0.6^{+0.2}_{-0.3})\times10^3$ & $0.2^{+0.1}_{-0.1}$ & $0.2^{+0.2}_{-0.1}$\\
10287723$^{\dagger}$ & 1174$^*$ & $13.4$ & $0.59^{+0.02}_{-0.01}$ & $0.582^{+0.006}_{-0.006}$ & $2.4^{+0.1}_{-0.1}$ & $393.598^{+0.003}_{-0.003}$ & $(1.7^{+0.9}_{-0.3})\times10^3$ & $0.75^{+0.07}_{-0.13}$ & $0.2^{+0.2}_{-0.2}$\\
10384911$^{\dagger}$ & \nodata & $14.0$ & $0.98^{+0.05}_{-0.05}$ & $1.31^{+0.03}_{-0.03}$ & $3.6^{+0.2}_{-0.2}$ & $1389.577^{+0.010}_{-0.009}$ & $(2.0^{+1.9}_{-0.6})\times10^3$ & $0.8^{+0.1}_{-0.3}$ & $0.3^{+0.2}_{-0.2}$\\
10460629$^{\dagger\square}$ & 1168$^*$ & $14.0$ & $1.49^{+0.03}_{-0.04}$ & $2.26^{+0.06}_{-0.06}$ & $6.9^{+0.3}_{-0.3}$ & $228.454^{+0.006}_{-0.006}$ & $856.671^{+0.008}_{-0.009}$ & $0.6^{+0.1}_{-0.2}$ & $0.3^{+0.1}_{-0.2}$\\
10525077$^{}$ & 5800 & $15.4$ & $1.11^{+0.07}_{-0.09}$ & $1.15^{+0.05}_{-0.04}$ & $5.8^{+0.3}_{-0.3}$ & $335.240^{+0.007}_{-0.008}$ & $427.040^{+0.005}_{-0.004}$ & $0.2^{+0.2}_{-0.2}$ & $0.46^{+0.14}_{-0.09}$\\
10683701$^{\dagger}$ & \nodata & $13.7$ & $0.90^{+0.03}_{-0.02}$ & $1.61^{+0.03}_{-0.03}$ & $4.7^{+0.3}_{-0.2}$ & $571.824^{+0.008}_{-0.007}$ & $(1.6^{+1.6}_{-0.4})\times10^3$ & $0.7^{+0.1}_{-0.3}$ & $0.3^{+0.2}_{-0.2}$\\
10842718$^{\dagger}$ & \nodata & $14.6$ & $0.90^{+0.04}_{-0.04}$ & $0.89^{+0.02}_{-0.02}$ & $6.6^{+0.2}_{-0.2}$ & $226.231^{+0.005}_{-0.005}$ & $(8^{+6}_{-4})\times10^3$ & $0.3^{+0.2}_{-0.2}$ & $0.2^{+0.2}_{-0.2}$\\
10960865$^{\dagger}$ & \nodata & $14.2$ & $1.22^{+0.07}_{-0.05}$ & $2.30^{+0.06}_{-0.06}$ & $5.9^{+0.6}_{-0.5}$ & $1507.95^{+0.02}_{-0.02}$ & $(0.35^{+0.29}_{-0.09})\times10^3$ & $0.7^{+0.2}_{-0.4}$ & $0.3^{+0.2}_{-0.2}$\\
10976409$^{\dagger\square}$ & \nodata & $13.9$ & $1.66^{+0.04}_{-0.04}$ & $2.33^{+0.08}_{-0.08}$ & $7.9^{+0.9}_{-0.6}$ & $983.539^{+0.008}_{-0.008}$ & $(1.3^{+1.0}_{-0.3})\times10^3$ & $0.87^{+0.05}_{-0.08}$ & $0.3^{+0.2}_{-0.2}$\\
11342550$^{\dagger}$ & 1421$^*$ & $15.3$ & $0.95^{+0.04}_{-0.04}$ & $1.04^{+0.04}_{-0.04}$ & $9.9^{+0.5}_{-0.5}$ & $524.281^{+0.002}_{-0.002}$ & $(1.7^{+0.7}_{-0.4})\times10^3$ & $0.5^{+0.1}_{-0.2}$ & $0.12^{+0.13}_{-0.08}$\\
11558724$^{\dagger\square}$ & \nodata & $14.7$ & $1.14^{+0.04}_{-0.04}$ & $1.34^{+0.04}_{-0.04}$ & $6.0^{+0.3}_{-0.3}$ & $915.196^{+0.003}_{-0.003}$ & $(0.37^{+0.22}_{-0.08})\times10^3$ & $0.5^{+0.2}_{-0.3}$ & $0.3^{+0.2}_{-0.2}$\\
11709124$^{\dagger}$ & 435 (154) & $14.5$ & $0.96^{+0.04}_{-0.04}$ & $1.07^{+0.03}_{-0.03}$ & $10.3^{+0.3}_{-0.4}$ & $657.268^{+0.001}_{-0.001}$ & $(1.3^{+0.5}_{-0.2})\times10^3$ & $0.62^{+0.04}_{-0.07}$ & $0.2^{+0.2}_{-0.1}$\\
12066509$^{\dagger}$ & \nodata & $14.7$ & $1.06^{+0.05}_{-0.04}$ & $1.33^{+0.05}_{-0.05}$ & $9.2^{+0.4}_{-0.4}$ & $632.092^{+0.002}_{-0.003}$ & $(1.3^{+0.7}_{-0.3})\times10^3$ & $0.73^{+0.06}_{-0.09}$ & $0.4^{+0.2}_{-0.1}$\\
12266600$^{\square}$ & \nodata & $15.4$ & $0.83^{+0.04}_{-0.04}$ & $0.78^{+0.02}_{-0.02}$ & $3.5^{+0.2}_{-0.2}$ & $612.145^{+0.008}_{-0.009}$ & $978.42^{+0.01}_{-0.01}$ & $0.2^{+0.2}_{-0.2}$ & $0.20^{+0.17}_{-0.09}$\\
12356617$^{\dagger}$ & 375$^*$ & $13.3$ & $1.15^{+0.09}_{-0.04}$ & $1.66^{+0.03}_{-0.03}$ & $12.0^{+0.4}_{-0.4}$ & $239.2243^{+0.0007}_{-0.0007}$ & $494.4405^{+0.0005}_{-0.0005}$ & $0.5^{+0.1}_{-0.2}$ & $0.85^{+0.02}_{-0.02}$\\
12454613$^{\dagger}$ & \nodata & $13.5$ & $0.94^{+0.04}_{-0.04}$ & $0.89^{+0.01}_{-0.01}$ & $2.8^{+0.1}_{-0.1}$ & $490.272^{+0.007}_{-0.007}$ & $736.376^{+0.008}_{-0.009}$ & $0.4^{+0.2}_{-0.3}$ & $0.2^{+0.2}_{-0.1}$\\
10024862$^{\dagger}$ & \nodata & $15.9$ & $1.04^{+0.05}_{-0.05}$ & $1.09^{+0.07}_{-0.07}$ & $4.8^{+0.4}_{-0.3}$ & $359.67^{+0.01}_{-0.01}$ & $567.045^{+0.007}_{-0.008}$ & $0.3^{+0.2}_{-0.2}$ & $0.13^{+0.21}_{-0.09}$\\
(triple) & & & & & & & &\\
\enddata
\tablenotetext{*}{These stars have KOI numbers because of the STEs/DTEs analyzed here; they are the only transiting planet candidates.}
\tablenotetext{\dagger}{\ Stars with spectra.}
\tablenotetext{\square}{\ Flat-bottomed transits with short ingress/egress; see Section \ref{ssec:planets_prune}.}
\tablenotetext{\textit{b}}{\ Flagged as photometric binary by \citet{2018ApJ...866...99B}.}
\tablenotetext{\textit{ab}}{\ Flagged as AO binary by \citet{2018ApJ...866...99B}.}
\tablecomments{The reported values and errors are medians and $15.87$th/$84.13$th percentiles of the marginal posterior.}
\end{deluxetable*}

\section{Properties of the Clean Sample}\label{sec:planets_clean}

\begin{figure*}[htbp]
    \epsscale{1.18}
    \plotone{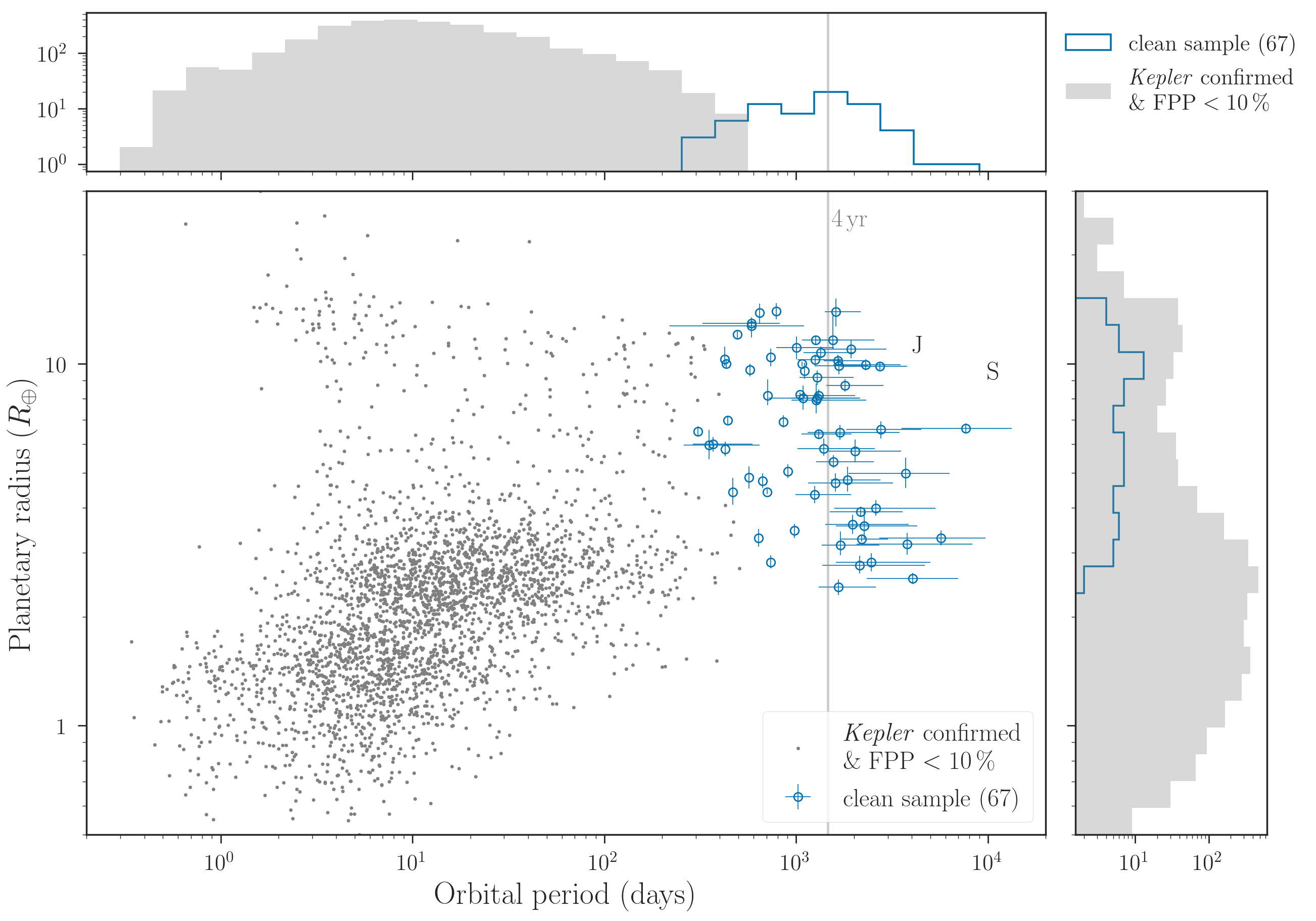}
    \caption{Radii and orbital periods of our clean sample defined in Section \ref{sec:planets} (open blue circles), along with confirmed {\it Kepler} planets and candidates with false positive probabilities \citep{2016ApJ...822...86M} less than $10\%$. The locations of Jupiter and Saturn are shown by ``J" and ``S," respectively.
    }
    \label{fig:rp}
\end{figure*}

Figure \ref{fig:rp} shows our clean sample in the radius--period plane, along with the ``clean KOI" sample constructed from confirmed planets and planet candidates with false positive probabilities calculated by \citet{2016ApJ...822...86M} less than $10\%$. For those without calculated values in \citet{2016ApJ...822...86M}, we adopted the values in the table available at the NASA exoplanet archive.\footnote{\url{https://exoplanetarchive.ipac.caltech.edu/cgi-bin/TblView/nph-tblView?app=ExoTbls&config=koifpp}} The current KOI catalog includes some of the STEs/DTEs in our clean sample, and they were excluded to avoid overlap. We updated the planetary radii in this clean KOI sample using the revised stellar radii in \citet{2018ApJ...866...99B}. 

Among the 67 candidates in our clean sample, 23 are newly reported in this paper (cf. Table \ref{tab:source}).\footnote{Two candidates reported in \citet{2019arXiv190101974H} are excluded in this counting.} In terms of planet size, 29 have $r>8R_\oplus$, 23 have $4\,R_\oplus<r<8\,R_\oplus$, and 15 have $r<4\,R_\oplus$. The number of Jupiter-sized planets with $r>8\,R_\oplus$ is found to be consistent with  the Doppler occurrence (see Section \ref{ssec:planets_clean_rv}). 
This implies that the catalog may have a high completeness for Jupiter-sized planets, although the result does not rule out the possibility that the occurrences of giant planets in the two samples are intrinsically different.
Although we have not yet quantified the search completeness, we found a similar number of smaller planets, suggesting that they are at least as common as Jupiter-sized ones in the searched period range over $2$--$20\,\mathrm{yr}$. If the sample is complete, our sample implies the occurrence of $\approx 0.4$ planets larger than Neptune per FKG star (Section \ref{ssec:planets_clean_occ}). The radius distribution of planets $>4R_\oplus$ turned out to be indistinguishable from that of the KOIs with $100\,\mathrm{days}\lesssim P \lesssim700\,\mathrm{days}$ (Section \ref{ssec:planets_clean_rad}). We also found, based on the spectroscopic sample, that the host stars of the planets larger than $4\,R_\oplus$ have systematically higher [Fe/H] than the other \kepler\ field stars, while this is not the case for smaller candidates (Section \ref{ssec:planets_clean_feh}). 

\subsection{Comparison with the Doppler Sample}\label{ssec:planets_clean_rv}

Jupiter-sized planets in the period range of our interest have already been detected in long-term Doppler surveys \citep[e.g.][]{2008PASP..120..531C, 2011arXiv1109.2497M}. Here we compare the number of Jupiter-sized planets in our sample against the number expected from the Doppler occurrence.

We use the occurrence rate density modeled as a double power-law function of orbital period and planetary mass by \citet{2018arXiv181205569F}, based on the combined sample of HARPS and CORALIE surveys \citep{2011arXiv1109.2497M}. Specifically, we reproduced the analysis in \citet{2018arXiv181205569F} for periods $3$--$10,000\,\mathrm{days}$ and mass $30$--$6,000\,M_\oplus$ and obtained the posterior sample for the rate density parameters using the {\tt epos} code \citep{gijsmulders_2019_2552594}. Using this function, we assigned planets (i.e., occurrence, mass, and period) to a subset of \kepler\ stars, which we call $\mathcal{S}_{\rm H}$, in the same region of the HR diagram as spanned by the stars in the HARPS volume-limited sample \citep{2011A&A...533A.141S}. This ensures that the comparison can be made for \kepler\ stars with similar properties as those in the HARPS sample. We have not corrected for possible difference in the stellar binarity between the volume-limited HARPS sample and the \kepler\ stars, but this effect is likely minor as long as we focus on giant planets \citep{2018AJ....155..244B}. The planetary masses are converted to radii using a broken power-law fit to the known planets with a break at $100\,M_\oplus$, and the transit duration was computed assuming the Beta distribution for eccentricities \citep{2013MNRAS.434L..51K} and random distributions for the argument of periastron and cosine of the orbital inclination. We then computed the transit signal-to-noise ratio and the corresponding detectability following \citet{2017AJ....154..109F} using Combined Differential Photometric Precision \citep[CDPP;][]{2010ApJ...713L..79K}, and counted the expected number of transit detection for planets larger than $8\,R_\oplus$ as a function of period. Although the process is not totally justified for planets with less than three transits, the detection efficiency curve likely plays a minor role for planets larger than $8\,R_\oplus$ because of large signals. The whole simulation was repeated $1,000$ times for the occurrence density parameters randomly sampled from the {\tt epos} posterior. 

\begin{figure*}
    \centering
    \epsscale{1.15}
    \plotone{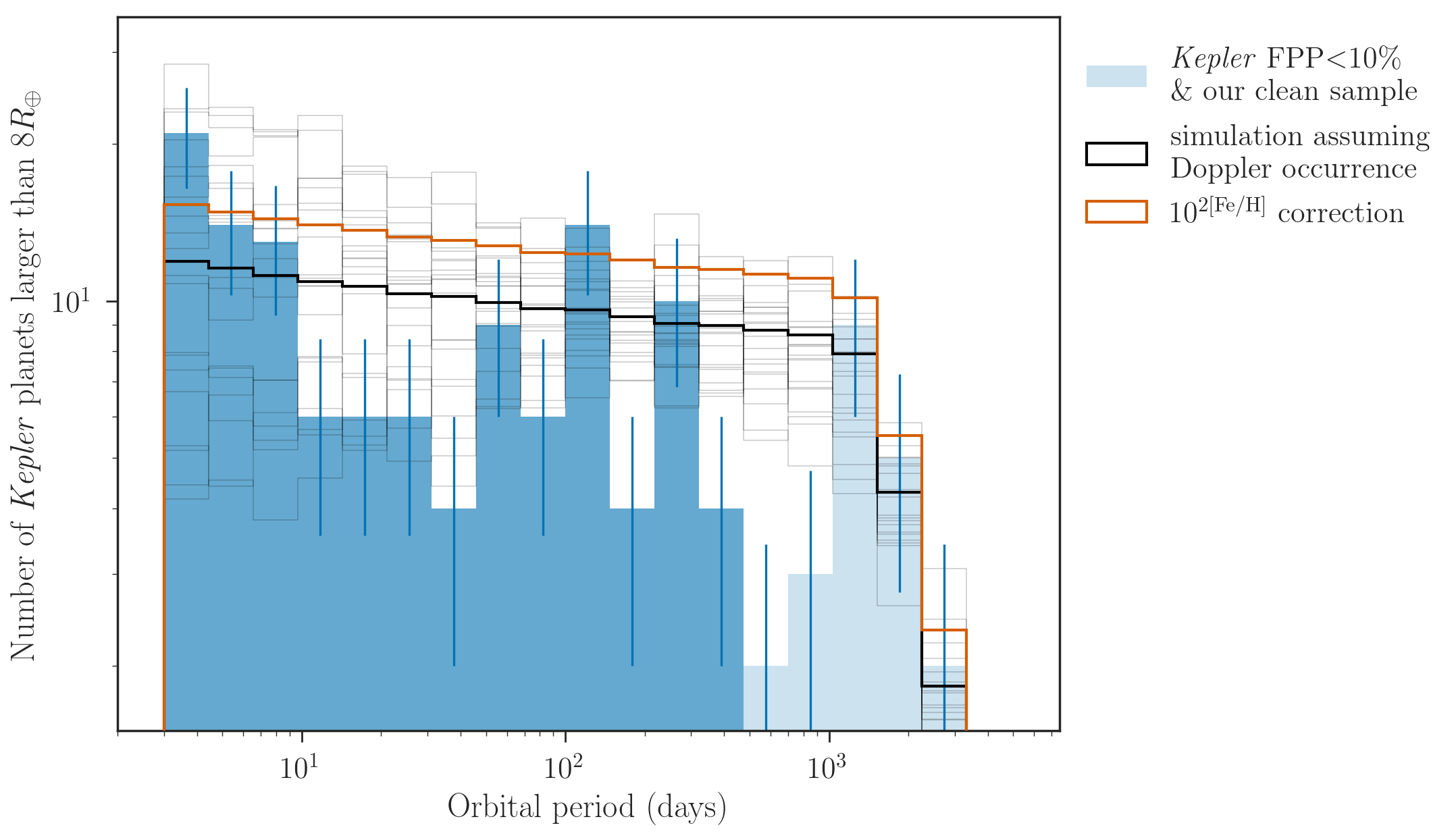}
    \caption{The period distribution of transiting planets larger than $8\,R_\oplus$ from the \kepler\ prime mission. The histogram shows the distribution of the clean KOI sample defined in Section \ref{sec:planets_clean} (thick blue) and our clean sample defined in Section \ref{ssec:planets_prune} (thin blue). The black lines show the simulated distribution based on the Doppler occurrence (thick: mean, thin: random posterior samples). The thick orange line shows the mean of the simulated results corrected for the different mean [Fe/H] between the \kepler\ stars and targets in the Doppler survey. See Section \ref{ssec:planets_clean_rv} for details.}
    \label{fig:epossym}
\end{figure*}

Figure \ref{fig:epossym} compares the result against the subsets of the clean KOIs and our clean sample that are also part of $\mathcal{S}_{\rm H}$. The thick black line shows the mean of the $1,000$ simulations, and thin lines show $20$ random posteriors (i.e., $2\sigma$ samples). Since the mean [Fe/H] of the HARPS sample \citep[$-0.1$;][]{2011A&A...533A.141S} is lower than the mean of the \kepler\ field stars \citep{2014ApJ...789L...3D, 2017ApJ...838...25G}, we also show the mean occurrence corrected for this difference assuming $10^{2\mathrm{[Fe/H]}}$ dependence derived for giant planets out to $4$-$\mathrm{yr}$ orbits \citep{2005ApJ...622.1102F}. We find a generally good agreement except for the region around $10\,\mathrm{days}$ and 1--$2\,\mathrm{yr}$. The former is likely the gap in the occurrence \citep[e.g.][]{2016A&A...587A..64S} that is not taken into account in the adopted power-law model. The origin of the latter tension is unclear; it may indicate that a few planets with three transits are missed in the KOI catalog, and were missed in our search as well.

\subsection{Radius Distribution}\label{ssec:planets_clean_rad}

Let us have a closer look at the radius distribution using the combined clean KOI sample and our long-period sample. Here we focus on giant planets $>4R_\oplus$, for which completeness of the KOI sample is likely high at least for $P\lesssim1\,\mathrm{yr}$ \citep[e.g.][]{2018AJ....155...89P}. 

\begin{figure}
    \centering
    \epsscale{1.15}
    \plotone{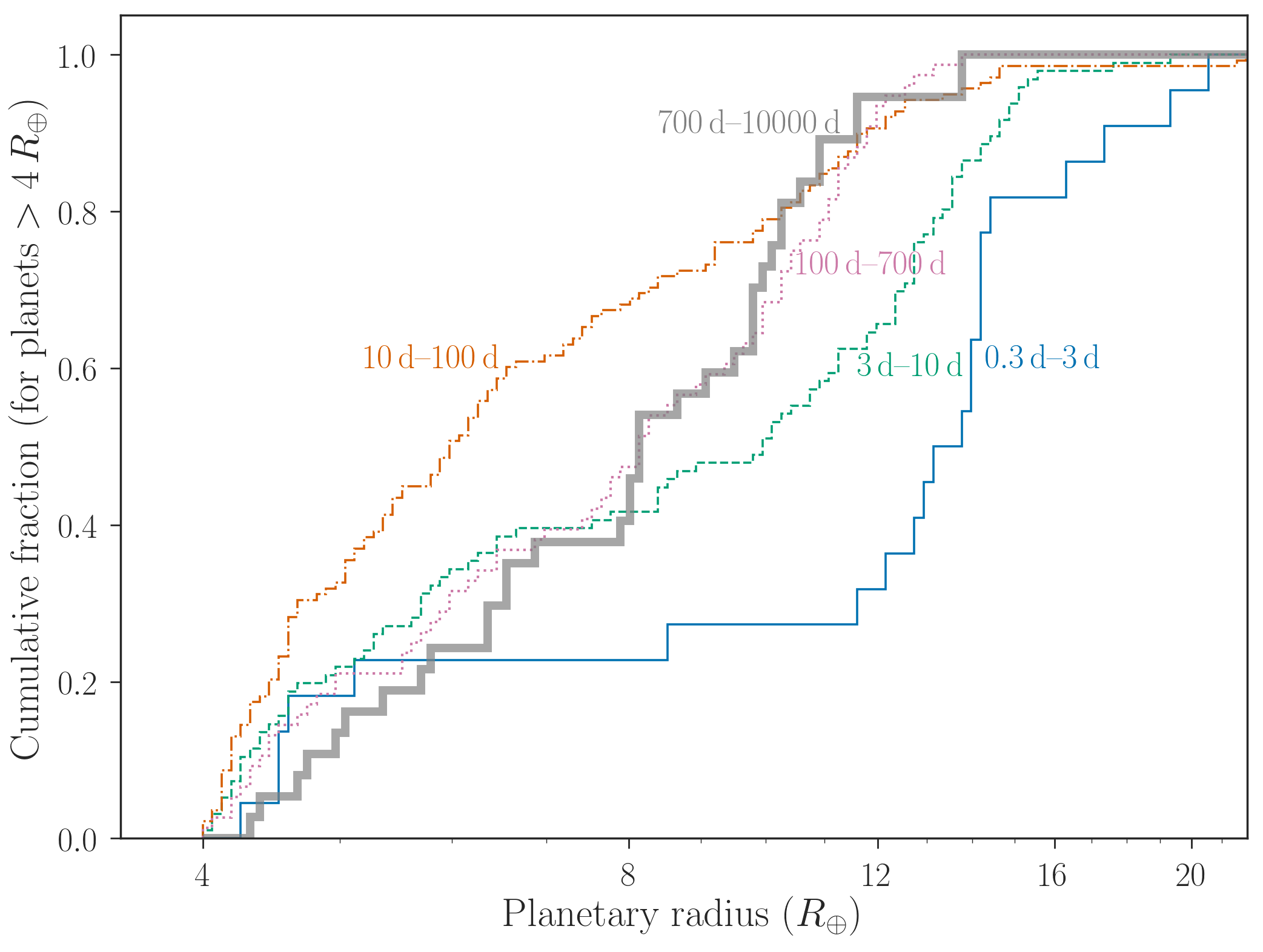}
    \plotone{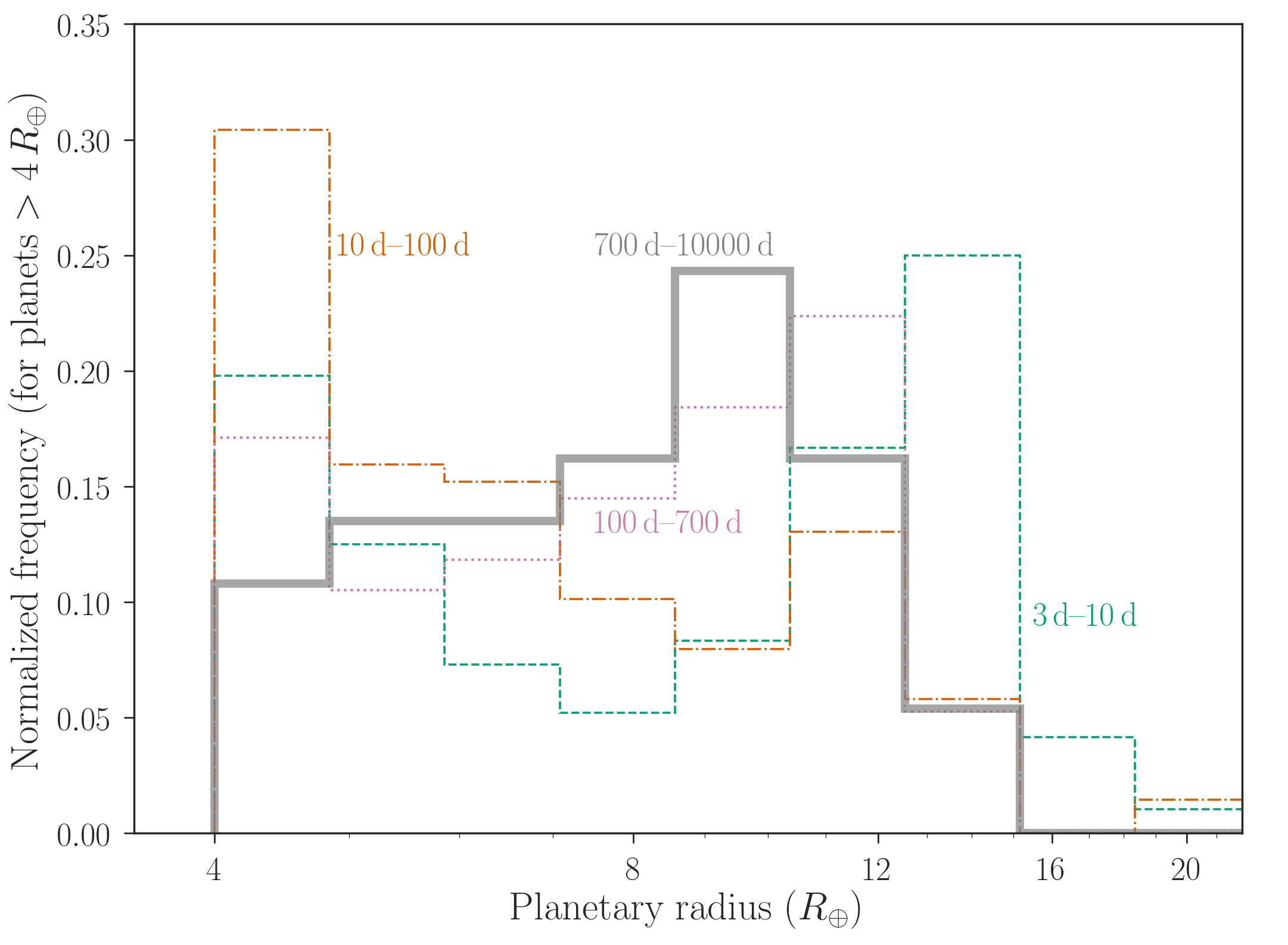}
    \caption{Radius distribution of our samples with $P>700\,\mathrm{days}$ (thick gray line), compared with the shorter-period planets in the clean KOI sample defined in Section \ref{sec:planets_clean}. {\it Upper panel} --- normalized cumulative radius distributions for planets with $r>4\,R_\oplus$. {\it Lower panel} --- normalized radius histograms for planets with $r>4\,R_\oplus$.
    }
    \label{fig:radius_dist}
\end{figure}

Figure \ref{fig:radius_dist} shows the normalized histogram and cumulative distribution for the planetary radius. The radius distribution for $P>700\,\mathrm{days}$ is roughly log-flat down to $4\,R_\oplus$ (bottom panel), suggesting that Neptune-sized planets are at least as common as Jupiter-sized ones in this period range. In addition, we found very similar distributions for $P>700\,\mathrm{days}$ and $100\,\mathrm{days}<P<700\,\mathrm{days}$ (top panel) with the KS $p$-value of 0.9. 
In other words, we did not find any significant change in the radius distribution around the snow line. 
Given that we have not corrected for completeness, two interpretations are possible: the radius distributions do remain to be the same and our search completeness is high for planets down to $4\,R_\oplus$; or there actually exist more small planets in $P>700\,\mathrm{days}$ range than in $100\,\mathrm{days}<P<700\,\mathrm{days}$.

Figure \ref{fig:radius_dist} also shows the radius distributions for planets with $P<100\,\mathrm{days}$. Here the bin edges roughly correspond to the radius values across which the radius distribution was found to show significant changes based on the KS $p$-value. The bin with $P<3\,\mathrm{days}$ is dominated by inflated hot Juptiers larger than $11\,R_\oplus$ and lack smaller planets in the sub-Saturn desert \citep{2011ApJ...727L..44S, 2016NatCo...711201L}. For longer periods, Jupiter-sized planets become less inflated and smaller planets start to dominate. The origin of the difference across $P\sim100\,\mathrm{days}$ is not clear; this 
might be related to the prevalence of lower-density sub-Saturns at longer orbital periods as hinted by systematic analysis of transit timing variations \citep{2017AJ....154....5H}.

\subsection{[Fe/H] of the Host Stars}\label{ssec:planets_clean_feh}

\begin{figure}[h!]
    \centering
    \epsscale{1.15}
    \plotone{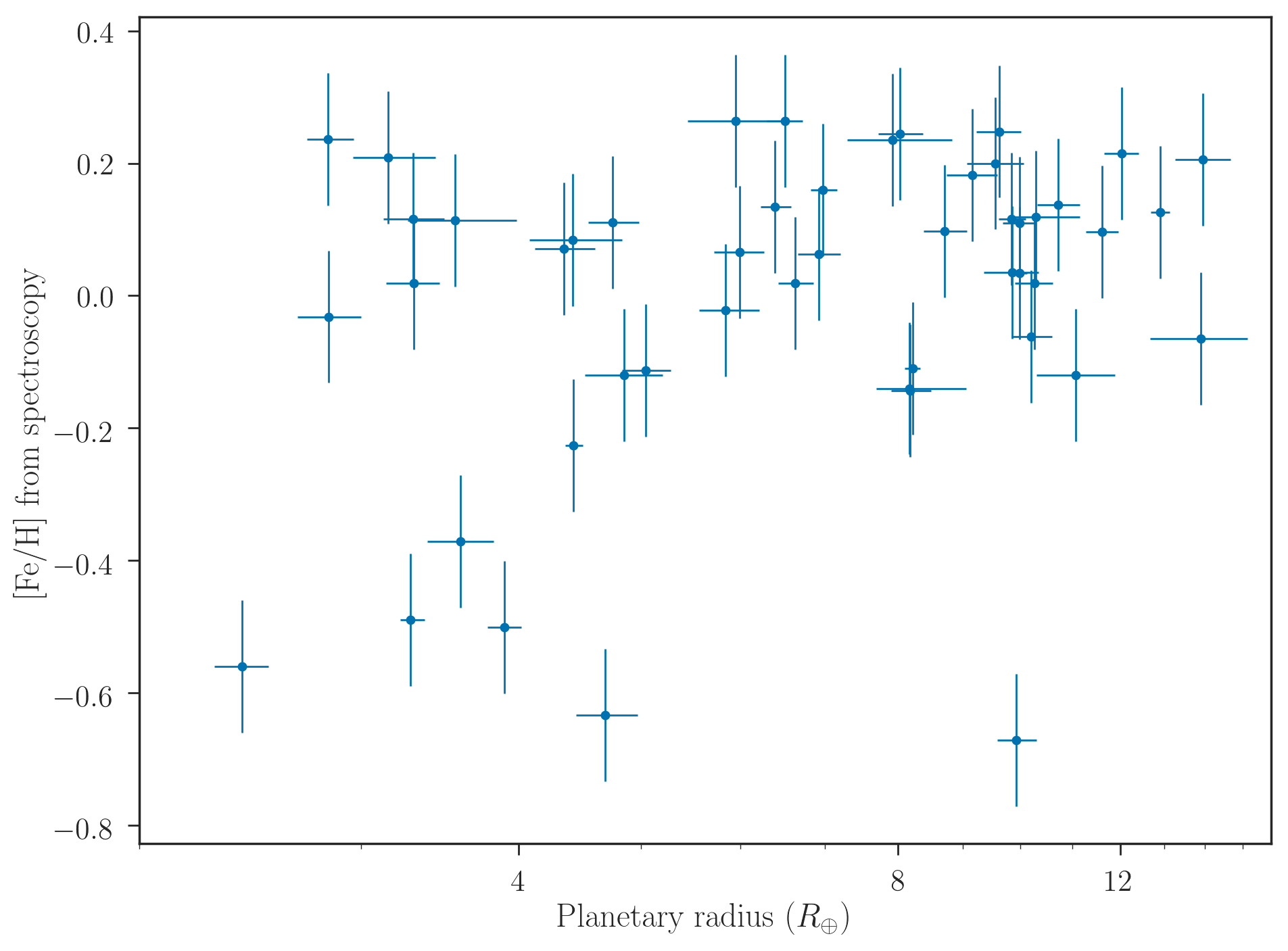}
    \plotone{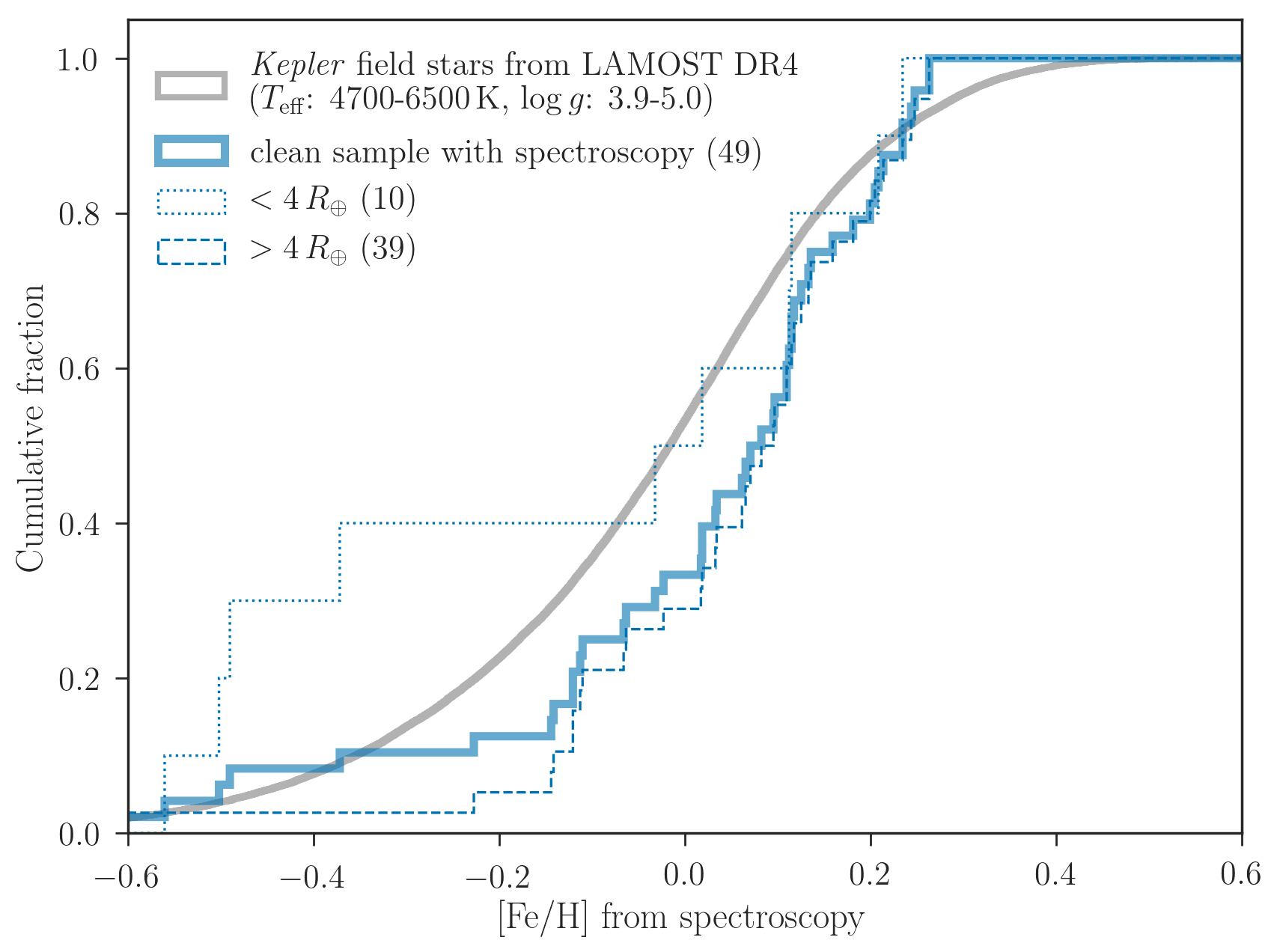}
    \caption{[Fe/H] distribution of the host stars in our clean sample with spectroscopic measurements, compared to the {\it Kepler} field stars. {\it Upper panel} --- [Fe/H] against planetary radii. {\it Lower panel} --- Thick blue line is the cumulative [Fe/H] distribution of the clean sample; dashed and dotted lines are its subsets with $>4\,R_\oplus$ and $<4\,R_\oplus$ planets, respectively; the thick gray line is the [Fe/H] distribution of FGK dwarf \kepler\ stars from LAMOST DR4.
    }
    \label{fig:feh_plrad}
\end{figure}

Figure \ref{fig:feh_plrad} shows stellar [Fe/H] and planetary radius for a subset of the clean sample with spectroscopic parameters. Here we use [Fe/H] values from raw spectroscopy (rightmost column of Table \ref{tab:starparams}), although the following arguments qualitatively remain to be the same for the values from the isochrone fit. 

The planets larger than $4\,R_\oplus$ are found mostly around stars with $\mathrm{[Fe/H]}>-0.15$, while the hosts of smaller planets have a wider range of [Fe/H] (upper panel). The host stars of planets larger than $4\,R_\oplus$ are also systematically more metal rich than the FGK dwarfs ($T_{\rm eff}=4700$--$6500\,$K, $\log g=3.9$--$5.0$) in the \kepler\ field from LAMOST DR4 (lower panel), with the KS $p$-value of $0.8\%$. This indicates that the giant planet--metallicity correlation confirmed for the shorter-period sample \citep{2018AJ....155...89P} also holds for planets with periods longer than two years.

One clear outlier in the plot, KIC 9822143, has the radius of $9.9\,R_\oplus$ and [Fe/H] of $-0.67$ ({\tt SpecMatch-Emp}) or $-0.47$ (isochrone modeling). This candidate might be a brown dwarf, which occurs around stars with a wider range of [Fe/H] \citep{2014MNRAS.439.2781M}.

\subsection{Occurrence Rate}\label{ssec:planets_clean_occ}

Although we have not quantified the detection completeness in our search, here we estimate the occurrence of giant planets $>4\,R_\oplus$ for reference, assuming that the completeness is high.

Defining the transit signal-to-noise $s_{\rm tra}(r)$ by $(r/R_\star)^2/\sigma_{15}$, where $\sigma_{15}$ is the CDPP for the longest $15\,\mathrm{h}$ timescale, we find that our sample has an $s_{\rm tra}$ cutoff at $\approx 10$ and the cutoff is independent of the \kepler\ magnitude. Given this observation, we computed $s_{\rm tra}(4\,R_\oplus)$ for all the \kepler\ stars using the radii from \citet{2018ApJ...866...99B}, and found that $\approx 67000$ \kepler\ stars with $4700\,\mathrm{K}<T_{\rm eff}<6500\,\mathrm{K}$  have $s_{\rm tra}(r)>10$; these are the Sun-like stars (which turned out to be mostly on the main sequence) for which Neptune-sized transiting planets, if present, would have been detected. On the other hand, there are $32$ planets with $r>4\,R_\oplus$ and $P>2\,\mathrm{yr}$ in our sample. 

We follow equations 12 and 13 in \citet{2016AJ....152..206F} \citep[see also the Appendix of][]{2014ApJ...795...64F} and assume that the detection completeness is $100\%$ for any relevant $r$ and $P$ and that the rate density per $\ln P$ is constant in our search range. Then we find the occurrence rate density per $\ln P$ for planets with $r=4$--$14\,R_\oplus$ and $P=2$--$20\,\mathrm{yr}$ to be $0.17\pm0.03$, or the occurrence rate integrated in this range to be $0.39\pm0.07$ per star with $4700\,\mathrm{K}<T_{\rm eff}<6500\,\mathrm{K}$. Here the error bar takes into account the Poisson noise alone.
Interestingly, the result agrees with the value $0.29\pm0.11$ derived by \citet{2016AJ....152..206F} for a similar radius range ($0.4$--$1.0\,R_{\rm Jup}$) but after completeness correction (their table 6).\footnote{Here the $\pi^{1/3}$ error in the transit probability is corrected \citep[see][]{2019arXiv190101974H}.} 

\subsection{Systems with Inner Planets}\label{ssec:planets_clean_koi}

Our clean sample includes 10 planet candidates with confirmed inner transiting planets (Table \ref{tab:clean}). These planets have higher fidelity to be genuine planets \citep{2012ApJ...750..112L}. The fraction of such systems in our sample ($9/67=15\%$) is significantly higher than the fraction of confirmed \kepler\ stars among all the \kepler\ stars ($\sim1\%$). This is likely due to the correlation of planet occurrences and orbital inclinations between inner and outer planets \citep{2016ApJ...822....2U, 2019arXiv190101974H}. It may also be the case that the stars with inner, typically smaller transiting planets detected have systematically less noisy light curves and bias for the detection of longer-period transiting planets.

\subsection{Mass Distribution}

If the empirical mass--radius relation in \citet{2017ApJ...834...17C} is adopted, the radius distribution of our planet sample with $P>700\,\mathrm{days}$ suggests that the occurrence rate density per log planet mass (planet-to-star mass ratio) at $\sim30\,M_\oplus$ ($\sim10^{-4}$) is higher than that around $\sim M_\mathrm{Jup}$ ($\sim10^{-3}$) by almost an order of magnitude. Such a slope is compatible with the value implied from microlensing surveys for planets near the snow line of late dwarfs \citep{2016ApJ...833..145S}.
Quantifying the completeness for smaller planets is essential for more detailed comparisons.

\section{Implications for future direct imaging}\label{sec:imaging}

\begin{figure*}
    \centering
    \epsscale{1.15}
    \plotone{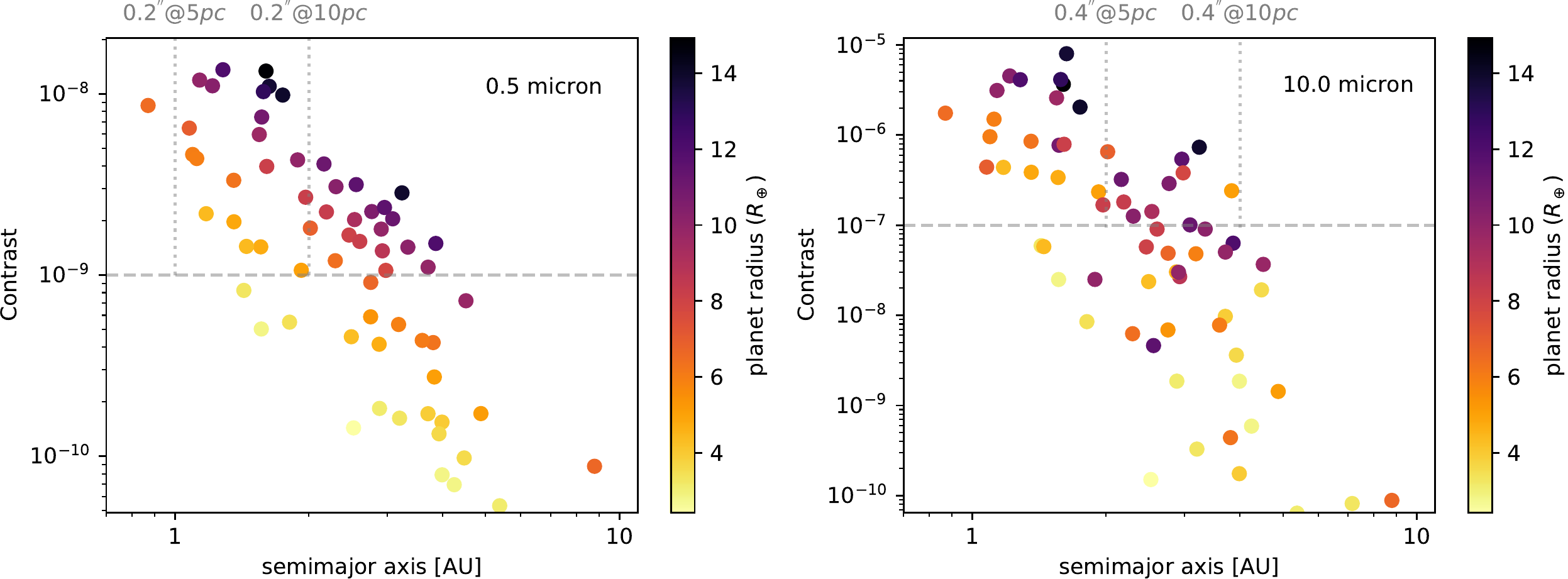}
    \caption{Expected star--planet contrast and semi-major axis of the planets in the clean sample, at 0.5 $\mathrm{\mu m}$ (left) and 10 $\mathrm{\mu m}$ (right). The Bond albedo of 0.4 is assumed for all the planets. The color bar indicates the planet radius. The horizontal dashed lines indicate the baseline technical goal of the WFIRST CGI (left) and a fiducial goal of a ground-based direct imaging instrument in a thermal band (right).  The vertical dotted lines show the semi-major axes corresponding to a fiducial inner working angle (IWA) at distances of 5~pc and 10~pc. The IWA for WFIRST CGI is assumed to be 0.2~arcsec (left) and a fiducial IWA of $2\lambda/D=0.4$~arcsec is adopted for a high-contrast, infrared ($\lambda=10\,\mu$m) instrument on a $D=10$~m telescope (right).}
    \label{fig:di}
\end{figure*}

Abundance and properties of the long-period exoplanets are useful for forecasting the outcome of future direct imaging missions. In Figure \ref{fig:di}, we convert planet and stellar properties in the clean sample to star--planet contrast ratios (hereafter contrast) in visible and infrared bands. The contrast $C$ at the wavelength $\lambda$ was computed as the sum of scattered light and thermal emission from a planet:
\begin{eqnarray}
\label{eq:refplanet}
C (\lambda) = \frac{2}{3} A \, \phi(\beta) \, \frac{r^2}{a^2} + \frac{r^2}{R_\star^2} \displaystyle{ \frac{e^{\frac{h c}{\lambda k_B T_\mathrm{eff}}} - 1 }{e^{\frac{h c}{\lambda k_B T_\mathrm{p}}} - 1 }},
\end{eqnarray}
where $A$ is the Bond albedo, $a$ is the orbital semi-major axis, $h$ is the Planck constant, $c$ is the speed of light, and $k_B$ is the Boltzmann constant.
In the first term (scattered light), 
\begin{eqnarray}
\label{eq:phaselambert}
\phi(\beta) \equiv [\sin{\beta} +  (\pi - \beta) \cos{\beta}]/\pi, 
\end{eqnarray}
is the Lambert phase function,
where $\beta$ is an observer--star--planet phase angle and we adopt $\beta=\pi/2$. Given the values of Bond albedo of Jupiter, Saturn, Uranus, and Neptune are 0.50, 0.34, 0.3, and 0.29, respectively, here we adopt $A=0.4$ for simplicity. 
In the second term (emission light), we adopt the radiative equilibrium temperature
\begin{eqnarray}
\label{eq:radt}
T_\mathrm{p,eq} = \left( \frac{1-A}{4} \right)^{1/4} T_{\rm eff} \sqrt{\frac{R_\star}{a}},
\end{eqnarray}
for the planet temperature $T_\mathrm{p}$. For typical $T_{\rm p}$ for the long-period planets, the scattered light dominates the contrast in the visible band, while the thermal emission is much larger in the infrared band. 

The baseline technical goal of the contrast of the WFIRST coronagraph instrument (CGI) in the visible band is $10^{-9}$ with an inner working angle of 0.2~arcsec, which corresponds to a semi-major axis of 1--2~AU at a distance of 5--10~pc \citep[][see also the website of WFIRST]{2015arXiv150303757S}\footnote{https://wfirst.gsfc.nasa.gov/exoplanets\_direct\_imaging.html}. The left panel of Figure \ref{fig:di} shows that the planets down to $\sim4\,R_\oplus$ in the clean sample have contrasts above the goal of the WFIRST CGI, if those planets are found at 5--10~pc. In addition, those Neptune-sized planets near the snow line around the nearest stars ($\lesssim5$~pc) can be a potential target for ground-based extreme adaptive optics systems in infrared bands to detect planet's thermal emission, such as TIKI \citep{2018SPIE10702E..4AB}, which aims to achieve contrast of $10^{-7}$ for nearby bright stars (the right panel in Figure \ref{fig:di}). Thus, such Neptune-sized planets at a few au, if found around nearby stars, can be promising targets for direct imaging surveys in the next decade.

\acknowledgements

This paper includes data collected by the {\it Kepler} mission. Funding for the {\it Kepler} mission is provided by the NASA Science Mission directorate. We are grateful to Sho Uehara for his help with visual inspection to find some transit events newly reported in the paper. We also thank Masahiro Ikoma, Masataka Aizawa, Tadahiro Kimura, Kentaro Aoki, Akito Tajitsu, Daisuke Suzuki, Chris Packham, Mitsuhiko Honda, and Makiko Nagasawa for helpful and fruitful conversations. We also thank Akihiko Fukui for the inspiration for the name of this catalog. We thank Samuel Yee and Yuan-Sen Ting for their suggestions on the spectroscopic stellar parameters. H.K. is supported by a Grant-in-Aid from JSPS (Japan Society for the Promotion of Science), Nos. JP17K14246, JP18H01247 and JP18H04577. Work by K.M. was performed under contract with the California Institute of Technology (Caltech)/Jet Propulsion Laboratory (JPL) funded by NASA through the Sagan Fellowship Program executed by the NASA Exoplanet Science Institute. This work was also supported by the JSPS  Core-to-Core Program ``Planet$^2$". The authors wish to recognize and acknowledge the very significant cultural role and reverence that the summit of Mauna Kea has always had within the indigenous Hawaiian community.  We are most fortunate to have the opportunity to conduct observations from this mountain.

\vspace{5mm}
\facilities{{\it Kepler}, Subaru/HDS}

\software{scikit-learn \citep{scikit-learn}, matplotlib \citep{Hunter:2007}, lightkurve \citep{lightkurve}, specmatch-emp \citep{2017ApJ...836...77Y}, isochrones \citep{2015ascl.soft03010M}, numpy \citep{2011CSE....13b..22V}, scipy \citep{scipy}, pycuda \citep{kloeckner_pycuda_2012}, python3, celerite \citep{celerite}, emcee \citep{2013PASP..125..306F}, batman \citep{2015PASP..127.1161K}}


\appendix

\section{Algorithm of the GPU-based Trapezoid Least Square}\label{sec:tlsalg}

To minimize equation (\ref{eq:residHLchi}), we use the derivative of $\chi^2$ by $H$,  
\begin{eqnarray}
  \label{eq:residHLchi2}
\sigma^{2} \frac{\partial \chi^2}{\partial H} 
&=& \frac{H}{2}(n_a - n_c + n_e)  + \frac{H W^2}{8 L^{2}} (n_b + n_d) + \sum_{t_i \in a,e,c} x_i - \frac{W}{2 L} \sum_{t_i \in  b, d} x_i+ \frac{2}{L} \left( \sum_{t_i \in  d} x_i t_i - \sum_{t_i \in b} x_i t_i\right) \nonumber \\
&+& \frac{H W}{L^{2}} \left( \sum_{t_i \in b} t_i - \sum_{t_i \in d} t_i \right) + \frac{2 H}{L^{2}} \sum_{t_i \in b,d} t_i^2 = 0.
\end{eqnarray}
where $n_a$ to $n_e$ is the number of the data points in the region of $a$ to $e$. We obtain the height at the minimum $\chi^2$ for each of the set of $L,W,t=t_0$ as 
\begin{eqnarray}
  \label{eq:residHLchimax}
\tilde{H} &=& -B/A \\
A &=& 4 L^2 (n_a - n_c + n_e) + W^2 (n_b + n_d) + 8 W \left( \sum_{t_i \in b} t_i - \sum_{t_i \in d} t_i \right) + 16 \sum_{t_i \in b,d} t_i^2 \\
B &=& 8 L^2 \sum_{t_i \in a,e,c} x_i -  4 W L \sum_{t_i \in b,d} x_i + 16 L \left(\sum_{t_i \in d} x_i t_i - \sum_{t_i \in b} x_i t_i \right).
\end{eqnarray}

\begin{figure}[htbp]
\begin{center}
\includegraphics[width=0.43\linewidth]{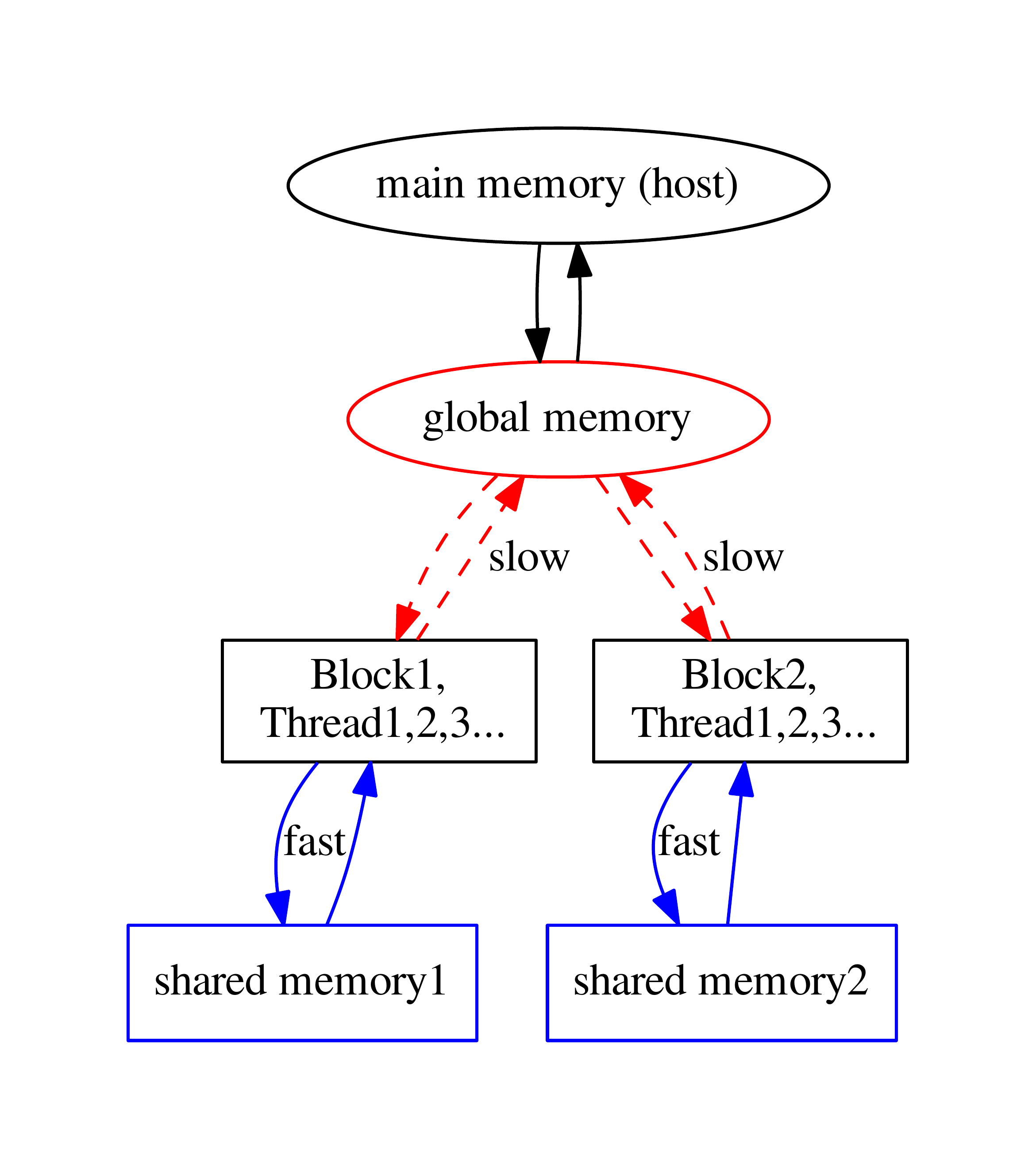}
  \includegraphics[width=0.55\linewidth]{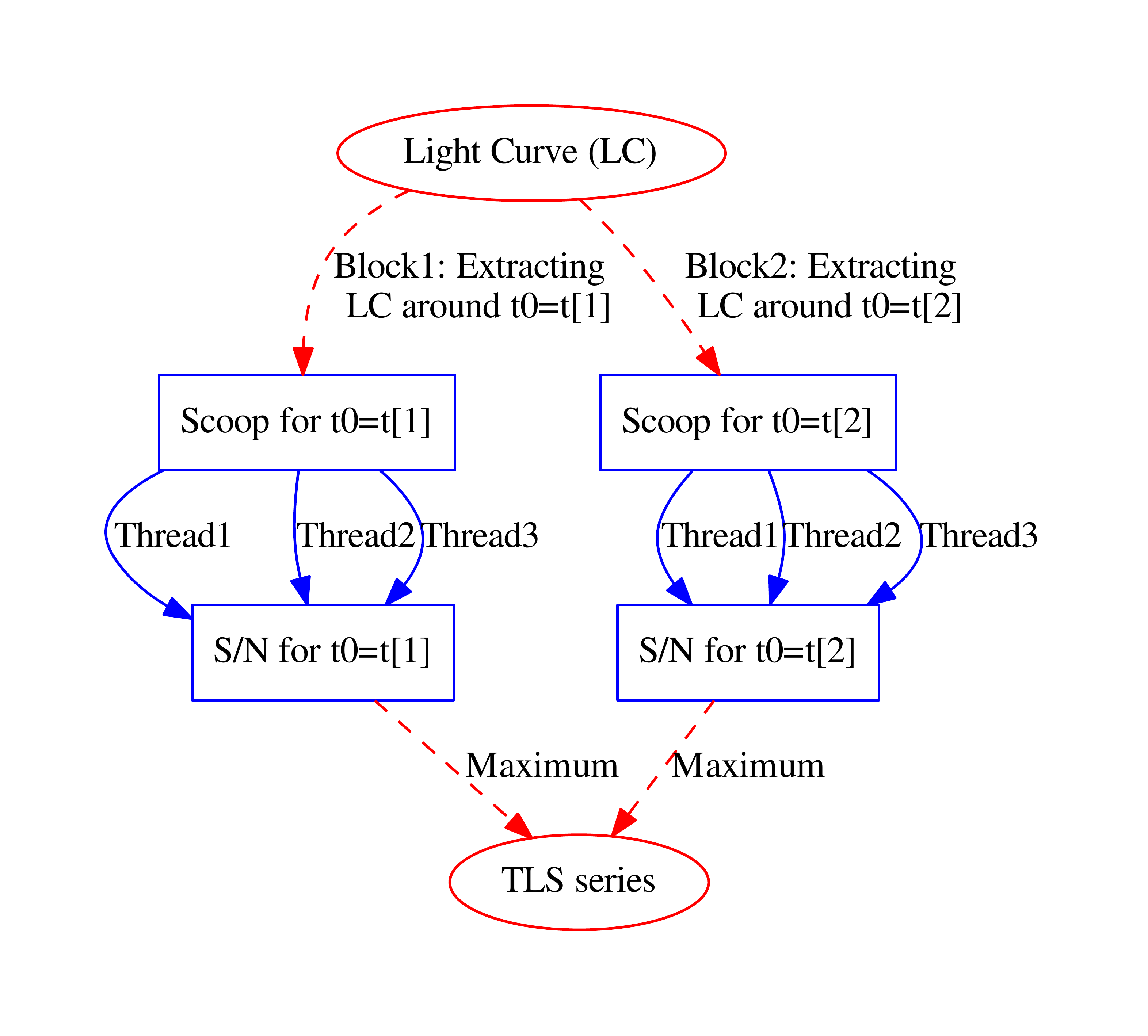}
\caption{The structure of the CUDA-GPU model (left) and the parallelization of the TLS algorithm (right). In the CUDA-GPU model, transfer from shared memory (blue boxes) is faster than from global memory (red circles), although the shared memory size is much smaller than the global memory size. Each block has its own shared memory and consists of a set of threads. In our GPU-based TLS code, we transfer a segment of the light curve around the central time $t_0$ from the entire light curve stored in the global memory to the corresponding shared memories. Each thread in a block computes the residual of the data from the trapezoid model for a different width, $W$.    \label{fig:paraltls}}
\end{center}
\end{figure}

We need to search for the minimum of $\chi^2$ by varying $L$, $W$, and $t_0$. The left panel in Figure \ref{fig:paraltls} shows the structure of the NVIDIA/CUDA model. There are three types of memory: main memory, global memory, and shared memory. CPU uses main memory, which is located outside the GPU unit. The GPU has global memory and shared memory. The former is a common memory for all blocks and has a large size. Each computing block has shared memory. Threads in a block can transfer data from the corresponding shared memory much faster than from global memory. However, the size of shared memory is usually small.

First, we transfer the light curve from the main memory to the global memory. We allocate $t_0$ into blocks. The $i$-th block computes the $\chi^2$ of the trapezoid for $t_0=t[i]$. Then we transfer a small segment around $t_0=t[i]$ ("scoop" in the right panel in Figure \ref{fig:paraltls}) to the shared memory so that the threads can read the data quickly. In addition, the $j$-th thread computes values for different widths $W = W[j]$. Finally, each thread has only a loop to search for the optimal $L$. Each block picks up the maximum value of 
\begin{eqnarray}
  \label{eq:sn}
  S/N(t_0) \equiv \frac{\tilde{H}}{\sqrt{\mathrm{res} (n - 3)}},
\end{eqnarray}
where ${\mathrm{res}}$ is the residual of the fit, and $n-3$ is the degrees of freedom. We call this time series the TLS series.

\section{The list of False Positives}\label{sec:fplist}

Table \ref{tab:fp} provides a list of eclipsing binaries identified in Sections \ref{ss:cvi} and \ref{ss:tls}, as well as the false positives identified in Section \ref{sec:fp}.

\begin{table*}[!tbh]
\begin{center}
\caption{List of the events removed during the visual inspection of stellar-sized signals (ST), the first screening of the targets (SC), 
visual inspection of the pixel data (VIP), search for signals with similar epochs (SE), and centroid shift test (CS). $T_0$ (BKJD) is an approximate position of the dip. ST (flat) is the stellar-sized eclipse with a flat bottom. \label{tab:fp}}
\begin{tabular}{|ccc|ccc|ccc|}
  \hline\hline
  KIC & reason &$T_0$ (BKJD) &  KIC & reason &$T_0$ (BKJD) & KIC & reason &$T_0$ (BKJD) \\
  \hline
  5475628 & ST & 950.5 & 3736610 & VIP & 1418.81 & 9019145 & SE & 142.16   \\  
  6234593& ST & 1148 & 3833007 & VIP & 981.7 & 9019245 & SE &  142.60  \\
  8463272 & ST & 641, 1206.7 & 4744261 & VIP & 478.82 & 9019513 & SE &  141.22  \\
  8508736 & ST & 261.3, 942.2 & 4932576 & VIP & 1272.7 & 9019948 & SE &  139.95  \\
  8570781 & ST & 1535.8 & 5480825 & VIP & 362.7 & 9086943 & SE &  142.65  \\
  9306307 & ST & 1191.3 & 5621767 & VIP & 839.2 & 1717717 & SE & 1439.20\\
  9409599 & ST & 1324.9 & 6947459 & VIP & 1568.8 & 1717722 & CS, SE & 1439.20 \\ 
  4585946 & ST & 878 & 6962233 & VIP & 924.5 & 2162635 & CS & 176.1 \\
  5024447 & ST & 434.5, 1206.5 & 7190443 & VIP & 1200.7 & 3230491 & CS & 315.33 \\
  6757558 & ST & 779 & 7983622 & VIP & 471.7 & 3241604 & CS & 1263.41\\
  9408440 & ST & 1503.5& 8110811  & VIP & 1512.8 & 8082126 & CS & 794.3 \\
  7971363 & ST (Flat) & 675.5  & 10197310 & VIP & 1325.6 & 9970525 & CS & 139.71 \\
  8540376 & SC & 1520, 1552 & 10334763 & VIP & 549.5 & 10058021 & CS & 601.1\\
  5522786 & SC & 283, (1040, 268.5, 872.5) & 10668646 & VIP & 196.2, 1449.1 & 10602068 & CS & 830.81\\
  8489948 & SC & 1578.5 & 3222471 & VIP & 673. & & & \\
   & & & 5480825 & VIP & 362.5 & & & \\
   & & & 9291458 & VIP & 286. & & & \\
  &&& 6681473 & VIP & 471.2, 867.3, 1263.3 & & & \\
  \hline
\end{tabular}
\end{center}
\end{table*}

\section{System Parametesr and Light Curves of the Flagged Systems}

\begin{figure*}
    \epsscale{1.2}
    \plotone{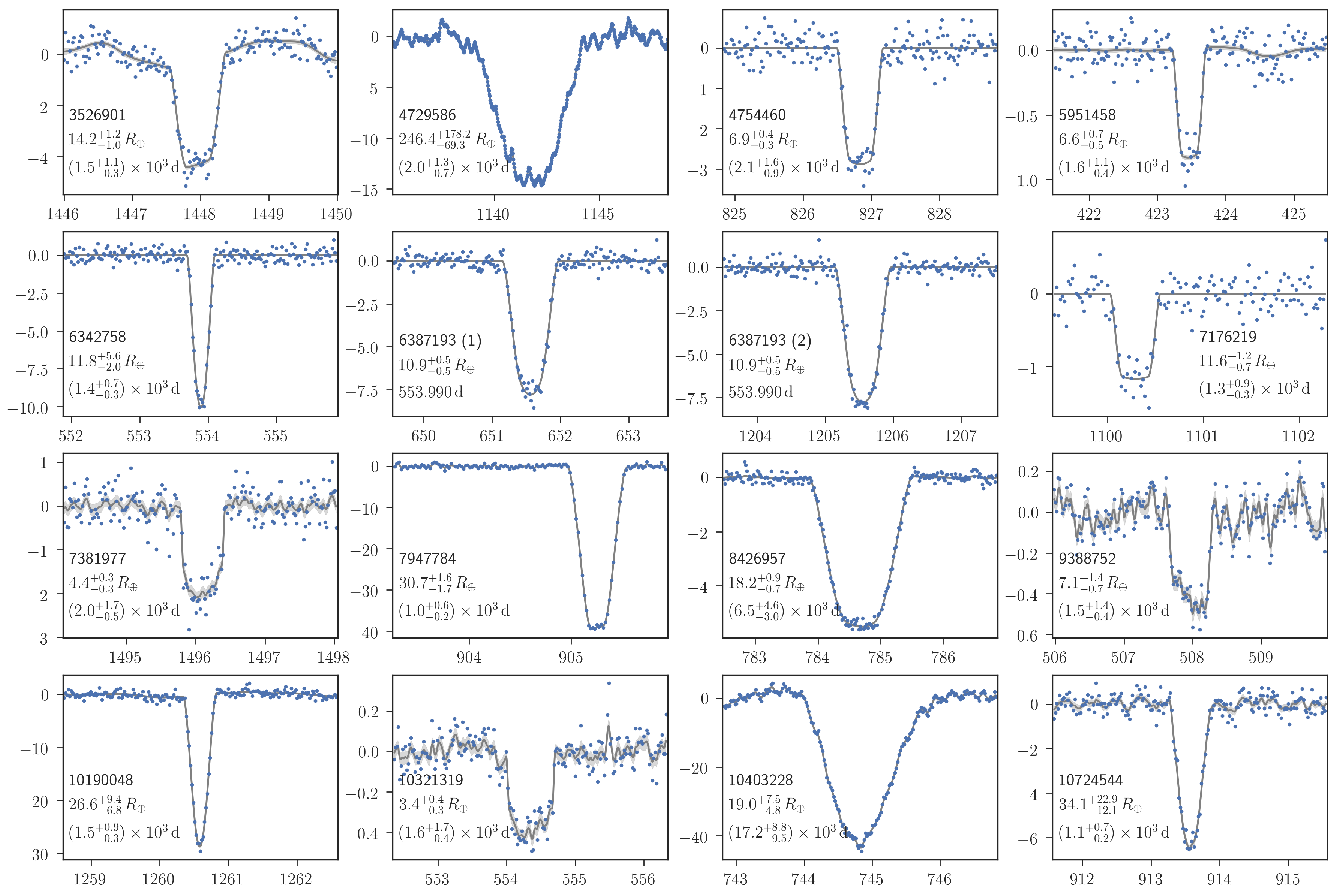}
    \caption{Same as Figure \ref{fig:clean1}, but for potential false positives identified from long ingress/egress duration.}
    \label{fig:lc_sec}
\end{figure*}

\begin{figure*}
    \epsscale{1.2}
    \plotone{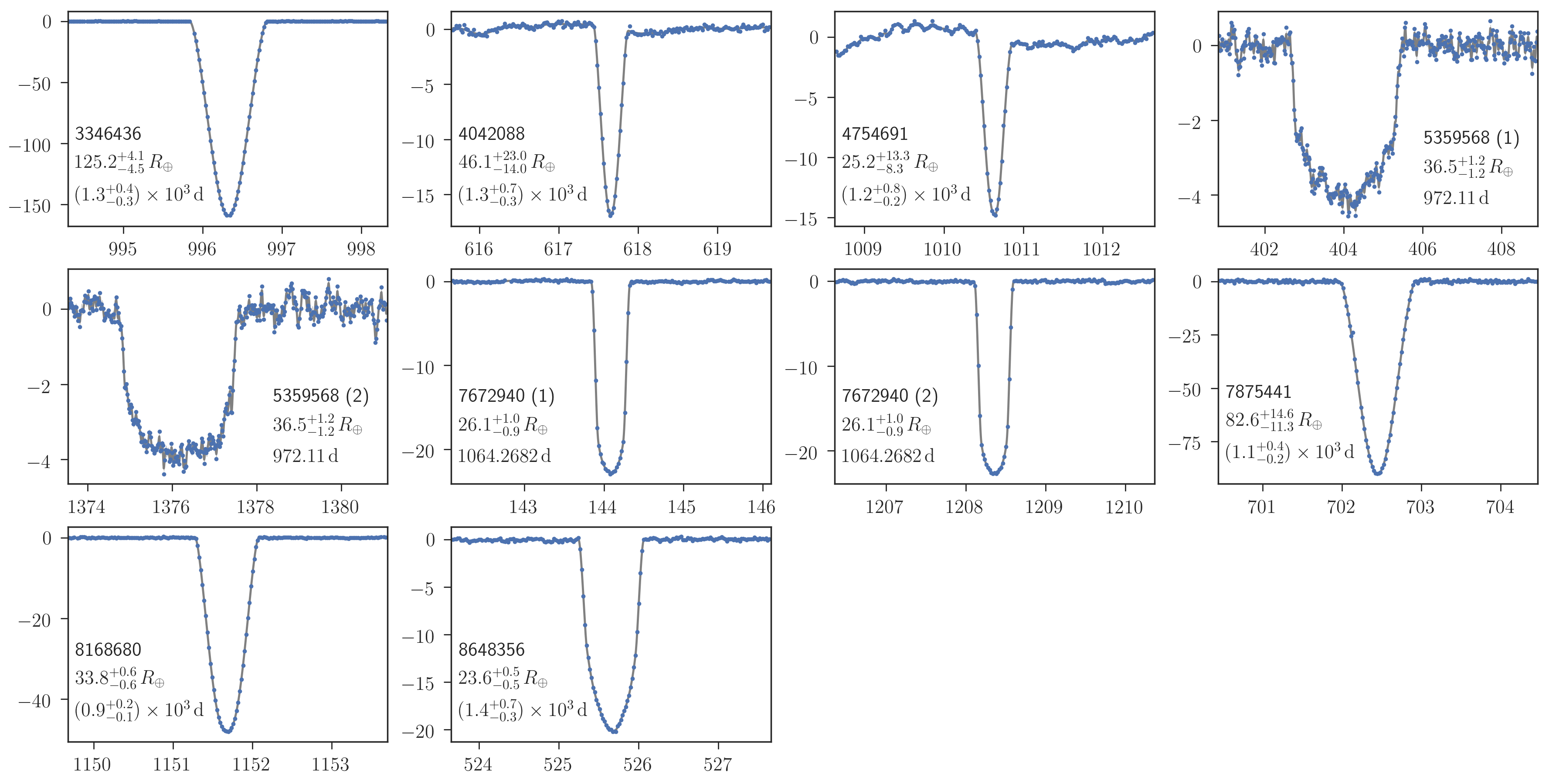}
    \caption{Same as Figure \ref{fig:clean1}, but for stellar-sized candidates.}
    \label{fig:lc_large}
\end{figure*}

Here we show the light curves of the candidates flagged due to long ingress/egress durations (Figure \ref{fig:lc_sec}) or radii too large to be planets (Figure \ref{fig:lc_large}). Table \ref{tab:flagged} reports the parameters for these systems.

\startlongtable
\begin{deluxetable*}{lccccccccccccccccc}
\tablecaption{Parameters of the Flagged Systems.\label{tab:flagged}}
\tablehead{
\colhead{KIC} & \colhead{KOI} & \colhead{$Kp$} & \colhead{$M_\star$ ($M_\odot$)} & \colhead{$R_\star$ ($R_\odot$)} & \colhead{$r$ ($R_\oplus$)} & \colhead{$t_0$ ($\mathrm{BKJD}$)} & \colhead{$P$ ($\mathrm{days}$)} & \colhead{$b$} & \colhead{$e$}
}
\startdata
3346436$^{\dagger}$ & \nodata & $12.4$ & $1.23^{+0.16}_{-0.06}$ & $2.33^{+0.07}_{-0.07}$ & $125^{+4}_{-5}$ & $996.3255^{+0.0001}_{-0.0002}$ & $(1.3^{+0.4}_{-0.3})\times10^3$ & $0.819^{+0.008}_{-0.016}$ & $0.69^{+0.04}_{-0.04}$\\
3526901$^{\dagger}$ & \nodata & $15.4$ & $0.98^{+0.04}_{-0.02}$ & $2.0^{+0.1}_{-0.1}$ & $14^{+1}_{-1}$ & $1447.953^{+0.006}_{-0.006}$ & $(1.5^{+1.1}_{-0.3})\times10^3$ & $0.90^{+0.01}_{-0.02}$ & $0.3^{+0.2}_{-0.2}$\\
4042088$^{\dagger}$ & 6378 (FP) & $13.4$ & $1.25^{+0.04}_{-0.05}$ & $1.43^{+0.03}_{-0.03}$ & $46.1^{+23.0}_{-14.0}$ & $617.6551^{+0.0008}_{-0.0008}$ & $(1.3^{+0.7}_{-0.3})\times10^3$ & $1.1^{+0.2}_{-0.1}$ & $0.54^{+0.10}_{-0.06}$\\
4729586$^{\dagger}$ & \nodata & $12.3$ & $2.65^{+0.08}_{-0.04}$ & $12.0^{+0.6}_{-0.6}$ & $246.4^{+178.2}_{-69.3}$ & $1141.71^{+0.04}_{-0.04}$ & $(2.0^{+1.3}_{-0.7})\times10^3$ & $1.0^{+0.2}_{-0.1}$ & $0.2^{+0.2}_{-0.1}$\\
4754460$^{\dagger}$ & \nodata & $14.9$ & $0.87^{+0.03}_{-0.02}$ & $1.10^{+0.03}_{-0.03}$ & $6.9^{+0.4}_{-0.3}$ & $826.837^{+0.004}_{-0.005}$ & $(2.1^{+1.6}_{-0.9})\times10^3$ & $0.89^{+0.02}_{-0.04}$ & $0.2^{+0.2}_{-0.1}$\\
4754691$^{b}$ & \nodata & $13.3$ & $0.88^{+0.02}_{-0.02}$ & $0.95^{+0.01}_{-0.01}$ & $28.4^{+13.6}_{-9.8}$ & $1010.629^{+0.002}_{-0.002}$ & $(1.2^{+0.9}_{-0.2})\times10^3$ & $1.1^{+0.2}_{-0.1}$ & $0.3^{+0.2}_{-0.1}$\\
5359568$^{\dagger}$ & \nodata & $13.1$ & $1.2^{+0.1}_{-0.2}$ & $5.9^{+0.2}_{-0.2}$ & $37^{+1}_{-1}$ & $404.046^{+0.007}_{-0.007}$ & $972.11^{+0.01}_{-0.01}$ & $0.3^{+0.1}_{-0.1}$ & $0.35^{+0.06}_{-0.05}$\\
5951458$^{\dagger}$ & \nodata & $12.7$ & $1.20^{+0.05}_{-0.03}$ & $1.81^{+0.03}_{-0.04}$ & $6.6^{+0.7}_{-0.5}$ & $423.464^{+0.006}_{-0.006}$ & $(1.6^{+1.1}_{-0.4})\times10^3$ & $0.94^{+0.01}_{-0.02}$ & $0.4^{+0.2}_{-0.2}$\\
6342758$^{\dagger}$ & \nodata & $14.6$ & $0.81^{+0.03}_{-0.03}$ & $0.769^{+0.010}_{-0.009}$ & $12^{+6}_{-2}$ & $553.896^{+0.001}_{-0.001}$ & $(1.4^{+0.7}_{-0.3})\times10^3$ & $0.96^{+0.09}_{-0.06}$ & $0.2^{+0.2}_{-0.1}$\\
6387193$^{\dagger}$ & \nodata & $14.9$ & $0.96^{+0.05}_{-0.04}$ & $1.03^{+0.02}_{-0.02}$ & $10.9^{+0.5}_{-0.5}$ & $651.562^{+0.003}_{-0.003}$ & $553.990^{+0.004}_{-0.003}$ & $0.87^{+0.02}_{-0.03}$ & $0.59^{+0.07}_{-0.06}$\\
7176219$^{}$ & \nodata & $13.8$ & $1.34^{+0.03}_{-0.03}$ & $2.70^{+0.05}_{-0.06}$ & $11.0^{+1.1}_{-0.7}$ & $1100.286^{+0.006}_{-0.006}$ & $(1.3^{+0.8}_{-0.2})\times10^3$ & $0.92^{+0.02}_{-0.05}$ & $0.6^{+0.1}_{-0.2}$\\
7381977$^{\dagger}$ & \nodata & $15.0$ & $0.91^{+0.05}_{-0.04}$ & $0.93^{+0.02}_{-0.02}$ & $4.4^{+0.3}_{-0.3}$ & $1496.095^{+0.007}_{-0.007}$ & $(2.0^{+1.7}_{-0.5})\times10^3$ & $0.6^{+0.2}_{-0.3}$ & $0.3^{+0.2}_{-0.2}$\\
7672940$^{\dagger}$ & 1463 (FP)$^*$ & $12.3$ & $1.36^{+0.04}_{-0.04}$ & $1.67^{+0.06}_{-0.06}$ & $26.1^{+1.0}_{-0.9}$ & $144.0861^{+0.0002}_{-0.0002}$ & $1064.2682^{+0.0003}_{-0.0003}$ & $0.483^{+0.007}_{-0.008}$ & $0.73^{+0.02}_{-0.02}$\\
7875441$^{}$ & \nodata & $15.6$ & $1.35^{+0.08}_{-0.14}$ & $1.9^{+0.1}_{-0.1}$ & $82.6^{+15.8}_{-11.6}$ & $702.4502^{+0.0006}_{-0.0006}$ & $(1.1^{+0.3}_{-0.2})\times10^3$ & $0.87^{+0.10}_{-0.08}$ & $0.48^{+0.09}_{-0.09}$\\
7947784$^{}$ & \nodata & $15.5$ & $1.12^{+0.10}_{-0.09}$ & $1.35^{+0.08}_{-0.07}$ & $30^{+2}_{-2}$ & $905.2548^{+0.0009}_{-0.0009}$ & $(1.0^{+0.6}_{-0.2})\times10^3$ & $0.76^{+0.01}_{-0.01}$ & $0.48^{+0.11}_{-0.09}$\\
8168680$^{\dagger}$ & \nodata & $12.3$ & $1.03^{+0.05}_{-0.05}$ & $1.38^{+0.02}_{-0.02}$ & $33.8^{+0.6}_{-0.6}$ & $1151.6787^{+0.0002}_{-0.0002}$ & $(0.9^{+0.2}_{-0.1})\times10^3$ & $0.776^{+0.003}_{-0.003}$ & $0.22^{+0.06}_{-0.06}$\\
8426957$^{\dagger}$ & \nodata & $13.6$ & $1.33^{+0.06}_{-0.07}$ & $2.05^{+0.04}_{-0.04}$ & $18.2^{+0.9}_{-0.7}$ & $784.677^{+0.005}_{-0.003}$ & $(7^{+5}_{-3})\times10^3$ & $0.900^{+0.009}_{-0.011}$ & $0.2^{+0.2}_{-0.1}$\\
8648356$^{\dagger}$ & \nodata & $11.7$ & $1.70^{+0.05}_{-0.05}$ & $1.77^{+0.03}_{-0.03}$ & $23.6^{+0.5}_{-0.5}$ & $525.6574^{+0.0009}_{-0.0008}$ & $(1.4^{+0.7}_{-0.3})\times10^3$ & $0.28^{+0.04}_{-0.05}$ & $0.49^{+0.09}_{-0.06}$\\
9388752$^{\dagger}$ & \nodata & $11.6$ & $1.51^{+0.09}_{-0.05}$ & $2.75^{+0.04}_{-0.05}$ & $7.1^{+1.4}_{-0.7}$ & $507.97^{+0.01}_{-0.01}$ & $(1.5^{+1.4}_{-0.4})\times10^3$ & $0.93^{+0.03}_{-0.05}$ & $0.3^{+0.2}_{-0.2}$\\
10190048$^{b}$ & \nodata & $15.7$ & $0.81^{+0.03}_{-0.02}$ & $0.81^{+0.02}_{-0.02}$ & $26^{+9}_{-6}$ & $1260.586^{+0.001}_{-0.001}$ & $(1.6^{+0.8}_{-0.3})\times10^3$ & $1.0^{+0.1}_{-0.1}$ & $0.3^{+0.1}_{-0.1}$\\
10321319$^{\dagger}$ & \nodata & $11.9$ & $1.00^{+0.03}_{-0.03}$ & $1.51^{+0.02}_{-0.03}$ & $3.4^{+0.4}_{-0.3}$ & $554.352^{+0.008}_{-0.010}$ & $(1.6^{+1.7}_{-0.4})\times10^3$ & $0.83^{+0.08}_{-0.15}$ & $0.3^{+0.2}_{-0.2}$\\
10403228$^{}$ & 8007 & $16.1$ & $0.59^{+0.01}_{-0.01}$ & $0.574^{+0.009}_{-0.010}$ & $19^{+8}_{-5}$ & $744.84^{+0.02}_{-0.02}$ & $(17^{+9}_{-10})\times10^3$ & $1.0^{+0.2}_{-0.1}$ & $0.5^{+0.2}_{-0.2}$\\
10724544$^{}$ & \nodata & $15.0$ & $1.21^{+0.11}_{-0.07}$ & $1.92^{+0.08}_{-0.08}$ & $32.2^{+26.5}_{-10.8}$ & $913.571^{+0.004}_{-0.005}$ & $(1.1^{+0.8}_{-0.2})\times10^3$ & $1.03^{+0.15}_{-0.08}$ & $0.3^{+0.2}_{-0.2}$\\
\enddata
\tablenotetext{*}{These stars have KOI numbers because of the STEs/DTEs analyzed here; they are the only transiting planet candidates.}
\tablenotetext{\dagger}{\ Stars with spectra.}
\tablenotetext{\square}{\ Flat-bottomed transits with short ingress/egress; see Section \ref{ssec:planets_prune}.}
\tablenotetext{\textit{b}}{\ Flagged as photometric binary by \citet{2018ApJ...866...99B}.}
\tablecomments{The reported values and errors are medians and $15.87$th/$84.13$th percentiles of the marginal posterior.}
\end{deluxetable*}

\section{Transit events in pulsating stars}\label{sec:pulsation}

\begin{figure*}[htbp]
\begin{center}
  \includegraphics[width=0.49\linewidth]{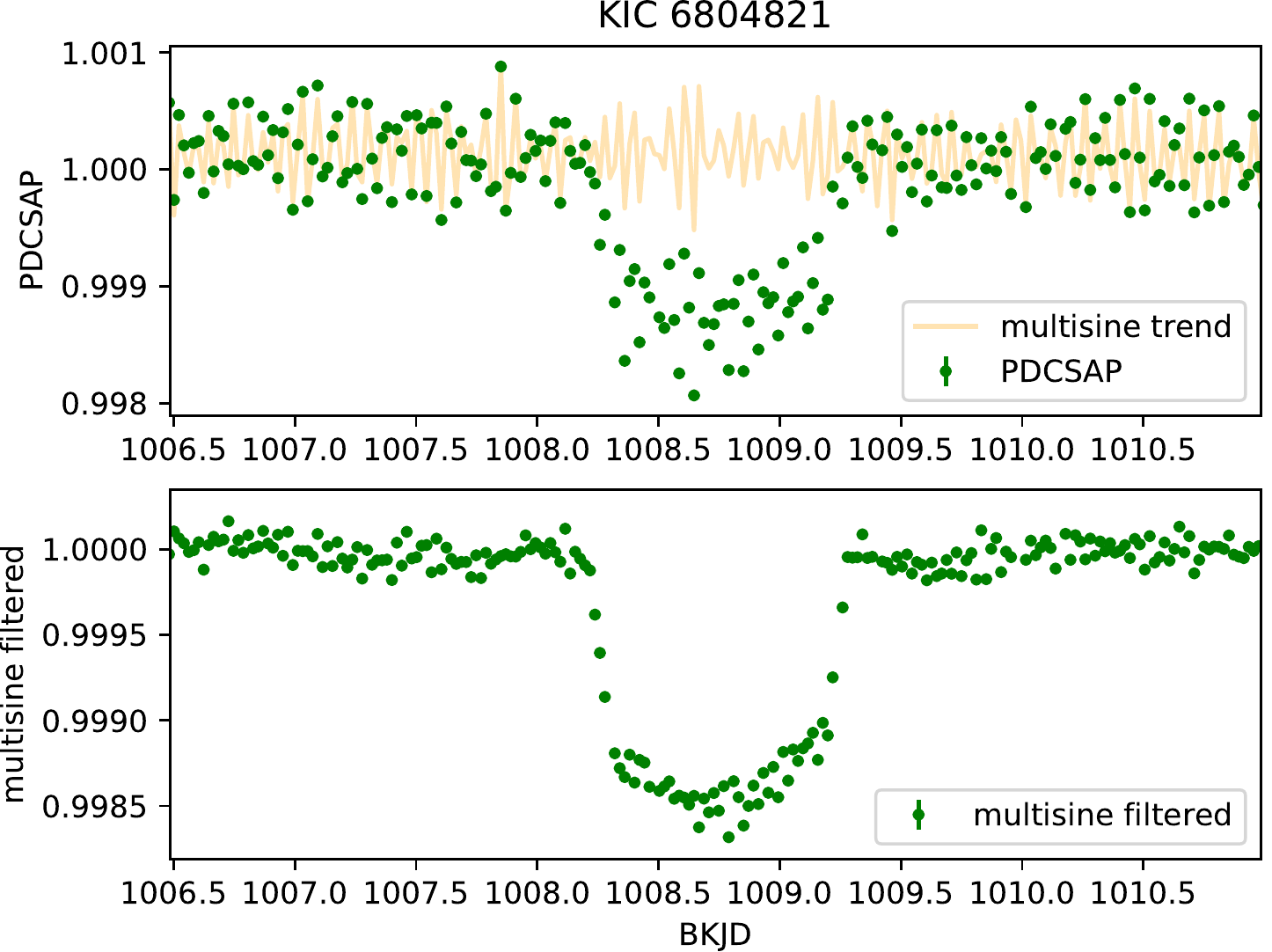}
  \includegraphics[width=0.49\linewidth]{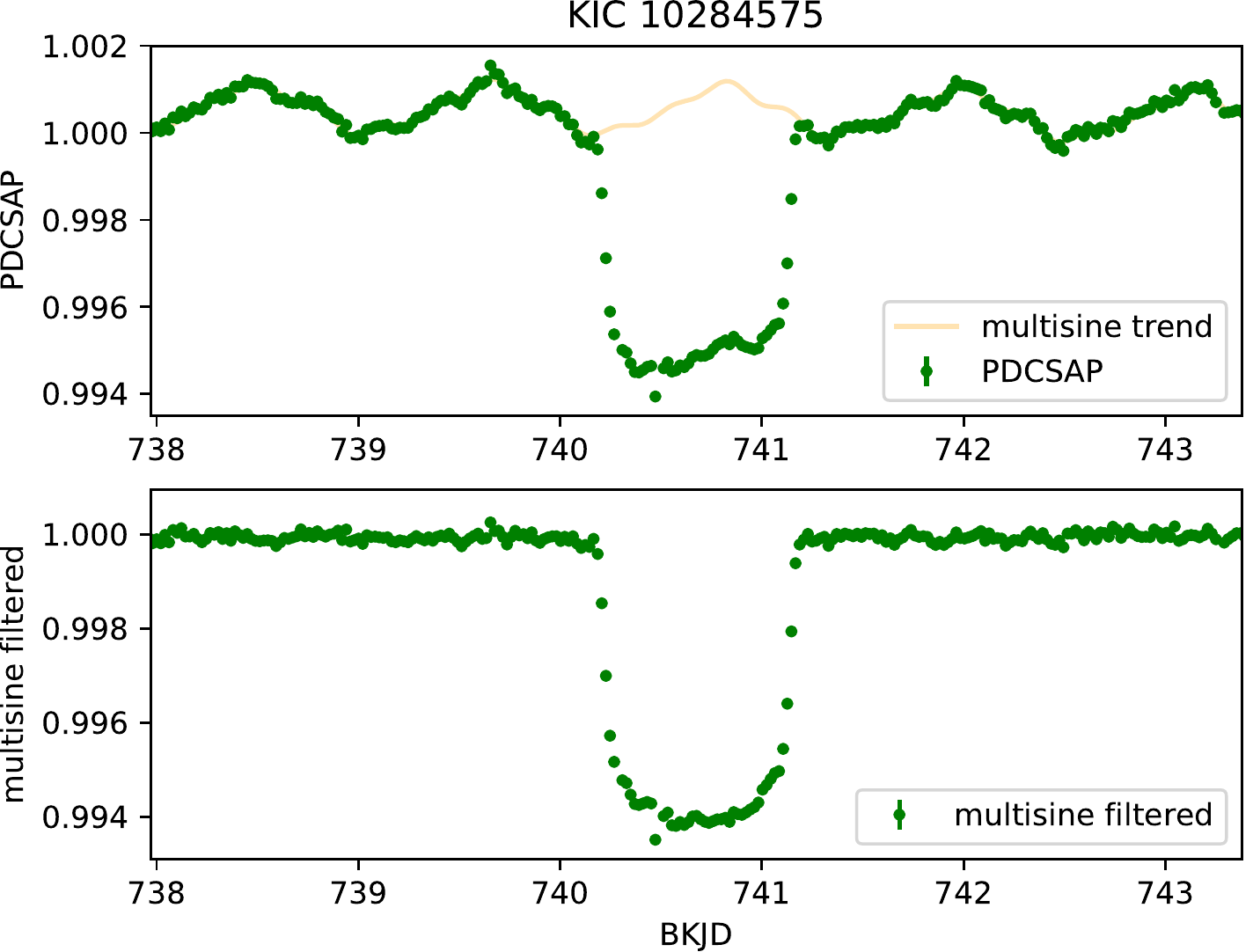}
\caption{STEs in pulsating stars. The upper panels show the original PDCSAP light curves and the multi-sine trend for KIC 6804821 (left) and KIC 10284575 (right). The lower panels are the de-trended light curves. \label{fig:deltas}}
\end{center}
\end{figure*}

KIC 6804821 is a delta scuti star, which is one of the typical variable stars. Oscillation of delta scuti stars consists of dozens of coherent modes. Performing the Fourier analysis of the light curve, we pick up $N=$27 significant Fourier modes ($f_i$, $i=1,...,N$). Fitting the multi-sine function with these modes,
\begin{eqnarray}
  \label{eq:msin}
  \mathrm{trend (t)} = \sum_{i=1}^{N} a_i \sin(2 \pi f_i t + \phi_i) + C
\end{eqnarray}
to the light curve outside the transit phase ($C$ is the offset) around STE, we derive the trend of the star, as shown in the upper panel of Figure \ref{fig:deltas} (left). De-trending the light curve, we obtain a clear signal of the STE of KIC 6804821 (bottom panel in Figure \ref{fig:deltas}). KIC 10284575 and KIC 8648356 also exhibit strong peaks in the power spectra. The mode frequencies of KIC 10284575 are lower than those of KIC 6804821. We show the de-trended curve of KIC 10284575 by the multi-sine de-trending in the right panels of Figure \ref{fig:deltas}. The de-trended light curves for these three stars are used in the transit analysis. The coherent oscillation of the host star gives us information about a companion, if one exists, via the R{\o}mer delay of the frequency \citep{2012MNRAS.422..738S}. Phase modulation has been widely used to find their companions in delta scuti stars \cite[e.g.][]{2014MNRAS.441.2515M,2015MNRAS.450.3999S,2015MNRAS.450.4475M}. A companion in KIC 8648356 was found by \cite{2018arXiv181112659M} when analyzing the phase modulation method. Assuming that STE and phase modulation have the same origin, its mass is consistent with a stellar one.

In our paper, the STEs in these pulsating stars were found in the original PDCSAP light curve, then we de-trend the light curve by sinusoidal fits. \cite{2017MNRAS.467.4663S} searched for transit events in pulsating stars after removing pulsation using a short cadence and found two candidates. The search after de-trending in the pulsating stars of the long cadence data, which increases the sensitivity of detecting the transit events, remains future work.

\end{document}